\documentclass{ps}
\usepackage{dsfont}
\usepackage{hyperref}
\usepackage{graphicx}
\graphicspath{{fig/}}
\usepackage{amsmath,amsthm,amssymb}
\usepackage{caption} 
\usepackage{multirow}
\usepackage{multicol}
\usepackage[scr=boondox]  % heavily sloped
{mathalpha}
\usepackage[ruled,algosection]{algorithm2e}
\usepackage{chngcntr}
\usepackage{xr}
\usepackage{placeins}

\usepackage{xcolor}

\def\frS{\mathfrak{S}}
\def\vec{\mathrm{vec}}
\def\sign{\mathrm{sign}}
\def\pval{\mathrm{pval}}

\def\bD{\mathbf{D}}
\def\bI{\mathbf{I}}
\def\bM{\mathbf{M}}
\def\bP{\mathbf{P}}
\def\T{\mathbf{T}}
\def\y{\mathbf{y}}
\def\z{\mathbf{z}}
\def\m{\mathbf{m}}
\def\t{\mathbf{t}}

\def\C{\mathcal{C}}

\def\N{\mathcal{N}}

\def\cT{\mathcal{T}}
\def\V{\mathcal{V}}

\def\R{\mathbb{R}}

\def\E{\mathbb{E}}

\def\bone{\boldsymbol{1}}
\def\bzero{\boldsymbol{0}}
\def\scrM{\boldsymbol{\mathscr{M}}}
\def\scrm{\boldsymbol{\mathscr{m}}}
\def\X{\boldsymbol{X}}
\def\x{\boldsymbol{x}}
\def\Z{\boldsymbol{Z}}
\def\obZ{\boldsymbol{\overline{Z}}}
\def\bz{\boldsymbol{z}}
\def\B{\boldsymbol{B}}
\def\D{\boldsymbol{D}}
\def\Y{\boldsymbol{Y}}

\def\bI{\boldsymbol{I}}
\def\be{\boldsymbol{e}}
\def\bv{\boldsymbol{v}}

\def\bc{\boldsymbol{c}}
\def\obc{\boldsymbol{\overline{c}}}
\def\bbeta{\boldsymbol{\beta}}
\def\bkappa{\boldsymbol{\kappa}}
\def\boeta{\boldsymbol{\eta}}
\def\bmu{\boldsymbol{\mu}}
\def\bSigma{\boldsymbol{\Sigma}}
\def\bDelta{\boldsymbol{\Delta}}
\def\bGamma{\boldsymbol{\Gamma}}

\def\1{\mathds{1}} 
\begin{document}
%%-----------------------------
%%      the top matter
%%-----------------------------
\title{Selective inference after convex clustering\\ with $\ell_1$ penalization}
\runningtitle{Selective inference after convex clustering}

\thanks{This work was supported by the Projects GAP (ANR-21-CE40-0007) and BACKUP (ANR-23-CE40-0018-01) of the French National Research Agency (ANR), and by the MITI at CNRS through the DDisc project}
%\thanks{...}% At most 5 thanks

\author{Francois Bachoc}
\address{Institut de Mathématiques de Toulouse; UMR5219,
	Université de Toulouse; CNRS, 
	UPS, F-31062 Toulouse Cedex 9,
	Institut universitaire de France (IUF),
	France\\
    \email{francois.bachoc@math.univ-toulouse.fr}}

\author{Cathy Maugis-Rabusseau}
\address{Institut de Mathématiques de Toulouse; UMR5219,
	Université de Toulouse; CNRS, 
	INSA, F-31077 Toulouse, France\\
	\email{cathy.maugis@insa-toulouse.fr}}

\author{Pierre Neuvial}
\address{Institut de Mathématiques de Toulouse; UMR5219,
	Université de Toulouse; CNRS, 
    UPS, F-31062 Toulouse Cedex 9, France\\
	\email{pierre.neuvial@math.univ-toulouse.fr}}

%
%\date{\textcolor{magenta{...}}
%

\begin{abstract}
Classical inference methods notoriously fail when applied to data-driven test hypotheses or inference targets.
Instead, dedicated methodologies are required to obtain statistical guarantees for these selective inference problems.
Selective inference is particularly relevant post-clustering, typically when testing a difference in mean between two clusters. 
In this paper, we address convex clustering with $\ell_1$ penalization, by leveraging related selective inference tools for regression, based on Gaussian vectors conditioned to polyhedral sets.
In the one-dimensional case, we prove a polyhedral characterization of obtaining given clusters, that enables us to introduce a test procedure with statistical guarantees.
This characterization also allows us to provide a computationally efficient regularization path algorithm. 
Then, we extend the above test procedure and guarantees to multi-dimensional clustering with $\ell_1$ penalization, and also to more general multi-dimensional clusterings that aggregate one-dimensional ones. 
With various numerical experiments, we validate our 
statistical guarantees and study the power of our methods to detect differences in mean between clusters.
Our methods are implemented in the \texttt{R} package \texttt{poclin}.
\end{abstract}

\begin{resume} 
Les méthodes d'inférence classiques ne sont pas applicables lorsque les hypothèses nulles testées ou les cibles de l'estimation dépendent des données.
Des développements spécifiques sont nécessaires pour obtenir des garanties statistiques pour ces problèmes d'inférence sélective.
C'est le cas pour l'inférence après une procédure de classification non supervisée (clustering), où il s'agit généralement de tester une différence de moyenne entre deux classes.
Dans cet article, nous abordons le clustering convexe avec une pénalisation $\ell_1$, en exploitant des outils d'inférence sélective développés pour la régression, basés sur des vecteurs gaussiens conditionnés à des ensembles polyédraux.
Dans le cas unidimensionnel, nous prouvons une caractérisation polyédrale de  classes obtenues par clustering convexe, ce qui nous permet de définir une procédure de test avec des garanties statistiques.
Cette caractérisation nous permet également de fournir un algorithme de chemin de régularisation efficace. 
Ensuite, nous étendons cette procédure de test, et les garanties associées, au clustering multidimensionnel avec une pénalisation $\ell_1$, et aussi à des clusterings multidimensionnels plus généraux qui agrègent les clusterings unidimensionnels. 
A l'aide de plusieurs expériences numériques, nous validons nos garanties statistiques et nous étudions
la puissance de nos méthodes pour détecter les différences de moyenne entre les classes.
Nos méthodes sont implémentées dans le package
 \texttt{R} \texttt{poclin}.
\end{resume}
\subjclass{62F03, 62H30}
\keywords{
Selective inference;
clustering;
regularization path;
hypothesis test;
truncated Gaussian
}
\maketitle
%%-----------------------------
%%      your text
%%-----------------------------
\section{Context and objectives} \label{section:context:and:objectives}

The problem of \textbf{selective inference} occurs when the same dataset is used (i) to detect a statistical signal and (ii) to evaluate the strength of this signal~\cite{taylor2015statistical}. 
In this article, we focus on the problem of post-clustering testing, where step (i) corresponds to a clustering of the input data, and step (ii) to an hypothesis test stemming from the clustering step. In such a situation, the naive application of a test that does not account for the data-driven clustering step is bound to violate type I error control \cite{gao2022}. 

This problem occurs in several applications. For instance, it is well-identified in the analysis of single-cell RNA-seq data (see \cite{Lahnemann2020}) where the genes expression is measured for several cells:  we want to test if each gene has a differential expression between two cells clusters, which are determined beforehand with a clustering procedure on the same expression matrix. 
This practical question has motivated numerous recent statistical developments to address this post-clustering testing problem.

A data splitting strategy has been studied by \cite{zhang2019}, where the $n$ observations are split into two samples. From the clustering of the first sample, they use a supervised approach to predict labels on the second sample before the test procedure. But this assignment is not taken into account in the correction, leading to a possible loss of the type I error control.
A conditional testing approach has been proposed by \cite{gao2022} for the problem of the difference in mean between two clusters. The authors condition by the event ``the two compared clusters are obtained by the random clustering'' and also by an additional event which is necessary for computational reasons (fixing some elements of the decomposition of Gaussian vectors). In particular, their $p$-values can be exactly computed in some cases of agglomerative hierarchical clustering and an extension to k-means clustering is proposed in \cite{chen2023selective} under a stronger conditioning. 
This approach has been extended to the test of the difference in mean between two clusters for each fixed variable \cite{hivert2024post}. A strategy to aggregate these $p$-values, and another approach using tests of multimodality (without statistical guarantees) are also suggested in \cite{hivert2024post}. In the context of single-cell data analysis, a count splitting approach under a Poisson assumption \cite{neufeld2024inference} and a more flexible Negative Binomial assumption \cite{neufeld2023negative} have recently been proposed. 
Finally, data thinning (or data fission) strategies are explored in \cite{leiner2023data-fission,dharamshi2024generalized,neufeld2024data}: they consist in generating two (or more) independent random datasets in such a way that, by applying a non-random operation (e.g the sum) to them, the initial dataset can be recovered.
This idea can be applied to various distributions belonging to the exponential family.

%Finally, data thinning (or data fission) strategies are explored in \cite{leiner2023data-fission,dharamshi2024generalized,neufeld2024data}: they consist in generating two (or more) independent random matrices that sum to the initial data matrix. This idea can be applied to various distributions belonging to the exponential family.  

The present contribution takes a different route from the above references and builds on \cite{Lee16}, where a Gaussian linear model is considered, and test procedures are provided, together with associated guarantees post-selection of variables based on the Lasso.
The nature of the Lasso optimization problem is carefully analyzed in \cite{Lee16}, and conditionally valid test procedures are obtained, based on properties of Gaussian vectors conditioned to polyhedral sets. 

We will extend this approach and its statistical guarantees to clustering procedures based on solving a convex optimization problem with $\ell_1$ penalization. 
We note that this optimization problem is a special case of a generalized Lasso problem \cite{TibshiraniTaylor11}. There has been a fair amount of recent contributions to inference post-generalized Lasso \cite{hyun2018exact,le2022more,chen2023more}. Our contribution focuses on a special case of the generalized Lasso (see Equations \eqref{eq:conv-clust:multi-dim} and \eqref{eq:conv-clust:j} below). We exploit this specificity to provide efficiency of implementation (see in particular the end of Section \ref{subsection:regularization:path:one-dimensional} and Section~\ref{sec:complexity}).

Let us now describe the setting of the paper in more details. 
wWe observe, for $n$ observations of $p$ variables (or features), a matrix $\Y = (Y_{ij})_{i\in [|n|], j\in [|p|] }$, where $[|u|]:=\{1,\ldots,u\}$ for any positive integer $u$.
We assume that $\vec(\Y)$ is a $np$-dimensional Gaussian vector with mean vector $\bbeta$ and  $np \times np$ covariance matrix $\bGamma$, where $\vec(.)$ denotes the vectorization by column of a matrix.
The vector $\bbeta$ is unknown but the matrix $\bGamma$ is assumed to be known (as in several of the articles cited above, we will discuss  this hypothesis in Section~\ref{sec:discussion-known-gamma}). 
Note that this setup covers in particular the case considered e.g. in \cite{gao2022}, where $\Y$ follows the matrix normal distribution $\mathcal{MN}_{n\times p}(\mathbf{u},\bSigma,\bDelta)$ where $\mathbf{u}$ is the $n\times p$ mean matrix, $\bSigma$ is the $n\times n$ covariance matrix among rows and $\bDelta$ is the $p\times p$ covariance matrix among variables.
Indeed, this matrix normal setup is equivalent (by definition) to that  $\vec(\Y)$ is a $np$-dimensional Gaussian vector with mean vector $\bbeta:=\vec(\mathbf{u})$ and  $np \times np$ covariance matrix $\bGamma=\bDelta\otimes \bSigma$, where $\otimes$ denotes the Kronecker product.

Under this framework, as announced, we will develop test procedures that extend the line of analysis of \cite{Lee16} to a clustering counterpart of the Lasso in linear models. Thus we consider the convex clustering problem \cite{hocking2011,lindsten2011clustering,pelckmans2005}
which consists in solving the following optimization problem	
\begin{equation} \label{eq:conv-clust:multi-dim} 
	\widehat{\B}(\Y)
	\in 
	\underset{\B = (\B_{1.}^\top,\ldots,\B_{n.}^\top)^\top \in \R^{n \times p }}{\mathrm{argmin}}
	~ ~
	\frac{1}{2}
	||\B - \Y||_F^2
	+
	\lambda 
	\sum_{\substack{i,i'=1 \\ i < i'}}^n
	||\B_{i'.} - \B_{i.}||_1
\end{equation}
where $||\cdot||_F$ is the Frobenius norm and $\B_{i.}$ denotes the $i$-th row of $\B$. 
The quantity $\lambda>0$ is a tuning parameter that we consider fixed here (as for the covariance matrix $\bGamma$, this assumption is further discussed in Section \ref{sec:discussion-known-gamma}).
We can immediately notice that Problem  \eqref{eq:conv-clust:multi-dim} is separable, and can be solved by addressing, for $j \in [|p|]$, the one-dimensional problem 
\begin{equation} \label{eq:conv-clust:j} 
	\widehat{\B}_{.j}(\Y_{.j})
	\in 
	\underset{\B_{.j} = (B_{1j},\ldots,B_{nj})^\top \in \R^{n }}{\mathrm{argmin}}
	~ ~
	\frac{1}{2}
	||\B_{.j} - \Y_{.j}||_2^2
	+
	\lambda 
	\sum_{ \substack{ i,i'=1 \\ i < i'}}^n
	|B_{i'j} - B_{ij}|,
\end{equation}
where $\B_{.j}$ is the $j$-th column of $\B$. 
It is worth pointing out that if the norm $\| \cdot \|_1$ is replaced by another norm $\| \cdot \|_q$, $q \in (0,\infty) \backslash \{ 1 \}$ in \eqref{eq:conv-clust:multi-dim}, then the optimization problem is no longer separable.
Hence, it becomes more challenging from a computational perspective.
This topic has been the object of a fair amount of recent work, see \cite{chi2015,sun2021,wang2018,weylandt2020,zhou2020} and our discussions
at the end of Section \ref{subsection:regularization:path:one-dimensional} and
in Section \ref{subsection:choice:lone:norm}.

The solution $ 	\widehat{\B}_{.j}(\Y_{.j})$ of \eqref{eq:conv-clust:j} naturally provides a one-dimensional clustering $\C^{(j)}$ of the observations for the variable $j$, by affecting $i$ and $i'$ to the same cluster if and only if $\widehat{\B}_{ij} = \widehat{\B}_{i'j}$.
Similarly, the solution of \eqref{eq:conv-clust:multi-dim} provided by the matrix $\hat{\B} = (\widehat{\B}_{1.}^\top,\ldots,\widehat{\B}_{n.}^\top)^\top$  naturally yields a multi-dimensional clustering of the observations, by affecting $i$ and $i'$ to the same cluster if and only if $\widehat{\B}_{i.} = \widehat{\B}_{i'.}$.
In this article, we will consider more general multi-dimensional clusterings that can be obtained by aggregation of the one-dimensional clusterings  $\C^{(1)} , \ldots, \C^{(p)}$ (see Section \ref{subsection:test:after:aggregation}).
A clustering of the rows of $\Y$ in $K$ clusters will be denoted by $\C = \C(\Y) = \left( \C_1(\Y), \ldots, \C_K(\Y) \right) $.
Of course these clusters and the number of clusters $K$ are random (depending on $\Y$).

Our goal is to provide test procedures for a (data-dependent) hypothesis of the form
\[
\bkappa^\top \bbeta = 0, 
\]
where $\bkappa = \bkappa(\C(\Y))$ is a deterministic function of the clustering $\C(\Y)$ and where we recall that $\bbeta$ is the $np \times 1$ mean vector of $\vec(\Y)$.  We refer to Section \ref{subsection:benefits:test} for further discussions on the merits and interpretations of the tests considered in this paper. 
\begin{xmpl}[feature-level two-group test]
	\label{ex:hyp} 
	The following typical example of a choice of $\bkappa$ enables to compare, for a variable $j_0 \in [|p|]$, the average signal difference between two clusters $\C_{k_1}$ and $\C_{k_2}$, $k_1,k_2 \in [|K|]$, $k_1 \neq k_2$. We write, for $i \in [|n|]$ and $j \in [|p|]$,
	\begin{equation}\label{eq:kappa}
		\bkappa_{i+(j-1)n} =
		\left(
		\frac{\1_{i \in  \C_{k_1}}}{  |\C_{k_1}| }
		-
		\frac{\1_{i \in \C_{k_2}}}{  |\C_{k_2}| }
		\right)\  \1_{ j = j_0 } ,
	\end{equation}
	where $|A|$ denotes the cardinality of any finite set $A$. This yields 
	\begin{equation} \label{eq:example:comparison:signal:multiD}
		\bkappa^\top \bbeta
		=
		\frac{1}{|\C_{k_1}|}
		\sum_{i \in \C_{k_1}} \beta_{i + (j_0 -1)n}
		-
		\frac{1}{|\C_{k_2}|}
		\sum_{i \in \C_{k_2}} \beta_{i + (j_0-1)n}. 
	\end{equation}
	In the particular matrix normal setup discussed above,
	\begin{equation*}
		\bkappa^\top \bbeta=
		\frac{1}{|\C_{k_1}|}
		\sum_{i \in \C_{k_1}} \mathbf{u}_{ij_0}
		-
		\frac{1}{|\C_{k_2}|}
		\sum_{i \in \C_{k_2}} \mathbf{u}_{ij_0}.
	\end{equation*}
	Rejecting this hypothesis corresponds to deciding that Clusters $\C_{k_1}$ and $\C_{k_2}$ have a discriminative power for the variable $j_0$, since their average signal indeed differs.
\end{xmpl}

The separation of Problem \eqref{eq:conv-clust:multi-dim} into $p$ one-dimensional optimization problems in \eqref{eq:conv-clust:j}  will be key for the testing procedures we develop in this paper. In Section \ref{section:one:dimensional case}, we will thus develop our methodology and theory related to the one-dimensional Problem \eqref{eq:conv-clust:j}. A test procedure is proposed and its statistical guarantees are established in Section~\ref{subsection:test:one:dimensional}. In Section \ref{subsection:regularization:path:one-dimensional}, a discussion of the existing optimization procedures to solve Problem \eqref{eq:conv-clust:j} is given and an original regularization path algorithm is also provided, specifically for this problem (obtained by leveraging our theoretical results in Section~\ref{subsection:polyhedral:caracterization:}).
In Section~\ref{section:p:dimensional case}, the proposed test procedure and its guarantees are extended to the $p$-dimensional framework.
Numerical experiments are presented in Sections \ref{sec:numerical-experiments-1d} for $p=1$ and \ref{sec:numer-exper-p} for $p > 1$.
In Section \ref{sect-discuss}, we provide a detailed overview of our contributions, together with various conclusive discussions regarding them and remaining open problems.
The proofs are postponed in Appendix (Sections~\ref{section:technical:lemmas:and:proofs} to \ref{appendix:proof:multiD}). Section~\ref{sec:complexity} contains 
additional material regarding computational aspects of convex clustering, in particular with our suggested regularization path. Section~\ref{apx:addit-illustr} contains additional numerical illustrations.

\section{The one-dimensional case} \label{section:one:dimensional case}

\subsection{Setting and notation}

In this section, for notational simplification, we consider a single Gaussian vector $\X$ of size $n \times 1$, with unknown mean vector $\bmu$ and known covariance matrix $\bSigma$.  This vector $\X$ should be thought of as an instance of $\Y_{.j}$ in \eqref{eq:conv-clust:j} for some fixed $j \in [|p|]$.

We consider the convex clustering procedure (as Problem \eqref{eq:conv-clust:j}) obtained for a given $\lambda > 0$ by  
\begin{equation} \label{eq:conv-clust} 
	\widehat{\B}(\X)
	\in 
	\underset{\B = (B_1,\ldots,B_n) \in \R^{n }}{\mathrm{argmin}}
	~ ~
	\frac{1}{2}
	||\B - \X||_2^2
	+
	\lambda 
	\sum_{ \substack{ i,i'=1 \\ i < i'}}^n
	|B_{i'} - B_{i}|.
\end{equation}
Solving this optimization problem defines a clustering of the $n$ observations, each cluster corresponding to a distinct value of $\widehat{\B}(\X)$. This mapping is formalized by the following definition.

\begin{dfntn} \label{def:clustering}
	%For $\B= (B_1,\ldots,B_n) \in \mathbb{R}^n$, the clustering given by $\B$ is obtained by the equivalence relation $\sim$ on $[|n|]$ given by $i \sim i'$ if and only if $B_i = B_{i'}$. Its classes are the equivalence classes of $\sim$. 
	For $\B= (B_1,\ldots,B_n) \in \mathbb{R}^n$, let $b_1 > b_2 > \dots > b_K$ be the sorted distinct values of the set $\{B_i: i \in [|n|]\}$. The clustering associated to $\B$ is $\C = (\C_k)_{k \in [|K|]}$, where
	$\C_k = \{i; B_i = b_k\}$ for $k \in [|K|]$.
\end{dfntn}
Note that, indifferently, we address clusterings of a set of elements $(x_1,\ldots,x_n)$ (for instance scalars or vectors) either with clusters that are subsets of $(x_1,\ldots,x_n)$ or subsets of $[|n|]$.
It is convenient to point out the following basic property of the optimization of Problem \eqref{eq:conv-clust}, implying in particular that the clusters are composed by successive scalar observed values, which is very natural. 

\newpage
\begin{lmm}
	\label{lm:increasing-hat-B}
	Consider a fixed $\x = (x_1 , \ldots , x_n) \in \mathbb{R}^n$.
	Consider $\widehat{\B}=\widehat{\B}(\x)$ given by Problem \eqref{eq:conv-clust}.
	%, with $\X$ replaced by $\x$. 
	Then, for $i,i'\in [|n|], i \neq i'$,
	\begin{enumerate}
		\item $x_i = x_{i'}$ implies $\widehat{B}_i = \widehat{B}_{i'}$
		\item $x_i \geq x_{i'}$ implies $\widehat{B}_i \geq \widehat{B}_{i'}$. 
	\end{enumerate}
\end{lmm}

Similarly as discussed in Section \ref{section:context:and:objectives}, for the clustering $\C = 
\C(\X) = \left( \C_1(\X), \ldots, \C_K(\X) \right) $ obtained from \eqref{eq:conv-clust}, we
will provide a valid test procedure for an hypothesis of the form  $\boeta^\top \bmu = 0$, 
where $\boeta = \boeta(\C(\X))$. 

\subsection{Polyhedral characterization of convex clustering in dimension one}
\label{subsection:polyhedral:caracterization:}

As in \cite{Lee16}, we will suggest a test procedure (see Section \ref{subsection:test:one:dimensional})
based on analyzing Gaussian vectors conditioned to polyhedral sets. At first sight, one could thus aim at showing that the observation vector $\X$ yields a given clustering with \eqref{eq:conv-clust} if and only if it belongs to a corresponding polyhedral set. However, this does not hold in general. Hence, we will characterize a more restricted event with a polyhedral set. This event is that (i) a given clustering is obtained and (ii) the scalar observations are in a given order. The same phenomenon occurs in \cite{Lee16}, where variables are selected in a linear model. There, it does not hold that a given set of variables is selected by the Lasso if and only if the observation vector belongs to a given polyhedral set. Nevertheless, the event that can be characterized with a polyhedral set is that (i) a given set of variables is selected and (ii) the signs of the estimated coefficients for these variables take a given sequence of values. We refer to Section \ref{section:cond:by:order} for further discussion on conditioning also by the observations' order. 

Before stating the polyhedral characterization, let us provide some notation.
We let $\frS_n$ be the set of permutations of $[|n|]$. Consider observations $x_1, \ldots, x_n$, ordered as $x_{\sigma(1)} \geq \dots \geq x_{\sigma(n)}$ for $\sigma \in \frS_n$. When these observations are clustered into $K$ clusters of successive values, the clustering is in one-to-one correspondence with the positions of the cluster right-limits $t_1,\ldots,t_K$, where $ 0 =t_0 < t_1  \cdots  < t_K = n$, and where for $k \in [|K|]$,  cluster $\C_k$ is composed by the indices $\sigma(t_{k-1}+1),\ldots,\sigma(t_k)$. This corresponds to the following definition. 
\begin{dfntn}
	\label{def:clust-seg}
	For $n\in\mathbb{N}$ and $K\in[|n|]$, let 
	$$
	\mathcal{T}_{K,n} := \left\{(t_k)_{0\leq k \leq K};\ 0=t_0<t_1< \dots < t_{K}=n\right\}.
	$$
	For any $\sigma \in \frS_n$ and any vector $\mathbf{t} \in \mathcal{T}_{K,n}$, the clustering associated to $(\mathbf{t},\sigma)$ is defined as $\C(\t, \sigma)=\{\C_1, \dots, \C_K\}$, where for $k \in [|K|]$, $n_k = t_k - t_{k-1}$ and 
	$\C_k = \{\sigma(t_{k-1}+i)\}_{i \in [| n_k|]}$.
\end{dfntn}
In particular, let us consider the clustering $\C=(\C_k)_{k \in [|K|]}$ obtained from Definition~\ref{def:clustering} by solving Problem~\eqref{eq:conv-clust} for a given $\x \in \mathbb{R}^n$.
This clustering can be written as $\C(\t, \sigma)$, for any $\sigma$ such that $x_{\sigma(1)} \geq \dots \geq x_{\sigma(n)}$, $t_0=0$ and $t_k = \sum_{j \in [|k|]} |\C_j|$ for $k \in [|K|]$.

\begin{figure}[p]
\centering
\includegraphics{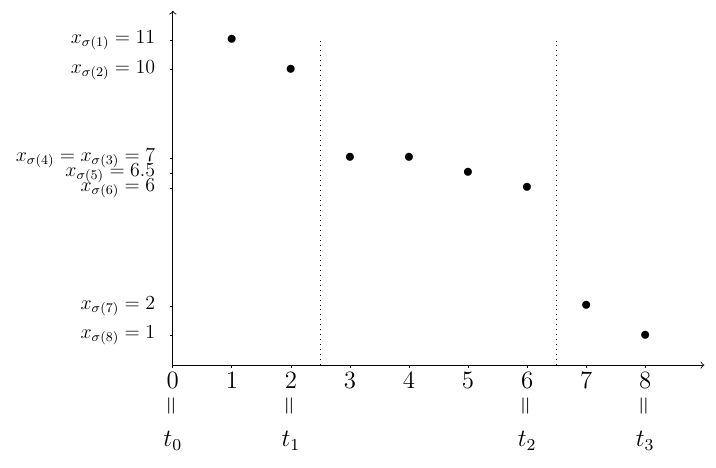}
\caption{Illustration of Definition \ref{def:clust-seg} for a clustering with $K=3$ clusters of the observed values $\x=(2,6,11,10,7,1,6.5,7)$.}\label{fig:Toy}
\end{figure}

\begin{figure}[p]
\centering	
\includegraphics{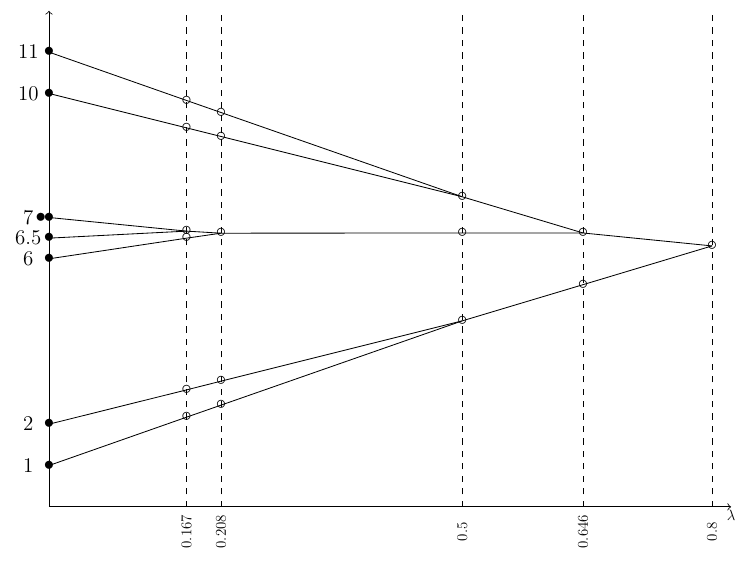}
\caption{Regularization path (see Section \ref{subsection:regularization:path:one-dimensional}) associated to the convex clustering problem for the observed values $\x=(2,6,11,10,7,1,6.5,7)$.}\label{fig:regpathToy}
\end{figure}

\begin{xmpl}\label{extoy}
	To illustrate Definition \ref{def:clust-seg} and Lemma \ref{lm:increasing-hat-B}, let $\x=(2,6,11,10,7,1,6.5,7)$ be observed data. A permutation reordering the values of $\x$ by decreasing order is  
	$$\sigma:(1,\ldots,n=8)\mapsto (3,4,5,8,7,2,1,6).$$ 
	For the clustering $\C=(\C_1,\C_2,\C_3)$ with $\C_1=\{11,10\}$, $\C_2=\{7,7,6.5,6\}$ and $\C_3=\{2,1\}$, the associated vector $\t$ is $t_0=0$, $t_1=2$, $t_2=6$ and $t_3=8$, as shown in 
	Figure \ref{fig:Toy}. Note that the clustering $\C$ of observations is equivalent to the clustering of indices $\C_1=\{\sigma(1),\sigma(2)\}=\{3,4\}$, $\C_2=\{\sigma(3),\sigma(4),\sigma(5),\sigma(6)\}=\{5,8,7,2\}$ and $\C_3=\{\sigma(7),\sigma(8)\}=\{1,6\}$. The regularization path (see Section \ref{subsection:regularization:path:one-dimensional}) associated to the convex clustering problem on the observed values $\x$ is represented in Figure~\ref{fig:regpathToy}. The vertical line at $x = \lambda$ intersects the regularization path at $y = \hat B_i$.  
	The order property between $x_i$ and $\hat B_i$ stated in Lemma \ref{lm:increasing-hat-B} is observed all along the regularization path. 
	For $\lambda=0.5$, we find the clustering in three clusters where the $\hat B_i$ values take three distinct values $\hat b_k$ ($\hat b_1=7.5$, $\hat b_2=6.625$ and $\hat b_3=4.5$).  
\end{xmpl}

Next, we can provide the announced polyhedral characterization of obtaining a given clustering, together with a given order of the observations. 

\begin{thrm} \label{theorem:equivalence:polyhedral:one:d}
	Let $\mathbf{t}$ be a fixed vector in $\mathcal{T}_{K,n}$ with $K \in [|n|]$, and let $\sigma \in \frS_n$ be a fixed permutation of $[|n|]$.
	Let $\C=\C(\t, \sigma)$ be the clustering obtained from Definition~\Rref{def:clust-seg}, with cluster cardinalities $n_1,\ldots,n_K$.
	Consider a fixed $\x = (x_1 , \ldots , x_n) \in \mathbb{R}^n$. 
	Let $\widehat{\B}=\widehat{\B}(\x)$ be the solution of Problem \eqref{eq:conv-clust} for some fixed $\lambda>0$, with $\X$ replaced by $\x$.
	From Definition \Rref{def:clustering}, $\widehat{\B}$ yields a clustering.
	Then the set of conditions
	\begin{align} 
		&
		\text{$\C(\t, \sigma)$ is the clustering given by $\widehat{\B}$},
		\label{eq:conditions:un} \\
		& 
		x_{\sigma(1)} \geq x_{\sigma(2)} \geq \dots \geq x_{\sigma(n)}
		\label{eq:conditions:deux} 
	\end{align}
	is equivalent to the set of the three following conditions 
	\begin{align} 
		& \notag
		\text{for} ~ k \in [|K-1|]: \\ 
		&
		\quad \quad
		\frac{1}{n_k}
		\sum_{i=1}^{n_k} x_{\sigma(t_{k-1} + i)} - 
		\frac{1}{n_{k+1}}
		\sum_{i=1}^{n_{k+1}} x_{\sigma(t_{k} + i)} 
		> 
		\lambda (t_{k+1} - t_{k-1}),
		\label{eq:equi:conditions:un}
		\\
		& \notag
		\text{for} ~ k \in [|K|] ~\text{such that $n_k \geq 2$},~\text{for} ~ \ell \in [| n_k-1|]: ~ ~\\
		& 
		\quad \quad 
		\frac{1}{n_k}
		\sum_{i=1}^{n_k} x_{\sigma(t_{k-1} + i)} 
		-
		\frac{1}{\ell}
		\sum_{i=1}^{\ell} x_{\sigma(t_{k-1} + i)} 
		\geq  
		\lambda (  \ell - n_k),
		\label{eq:equi:conditions:deux} \\
		& 
		x_{\sigma(1)} \geq x_{\sigma(2)} \geq 
		\dots \geq x_{\sigma(n)}.   
		\label{eq:equi:conditions:trois}
	\end{align}
	Finally, when \eqref{eq:conditions:un} and \eqref{eq:conditions:deux} hold, then for $i \in [|n|]$, for $k \in [| K |]$ with $i \in \C_k$, we have  
	\begin{equation} \label{eq:hat:B:optimum}
		\hat{B}_i =
		\frac{1}{n_k}
		\sum_{i' \in \C_{k}} 
		x_{i'}
		+
		\lambda 
		\sum_{k' = 1}^{k-1}
		n_{k'}
		- 
		\lambda 
		\sum_{k'  = k+1}^{K}
		n_{k'}. 
	\end{equation}
\end{thrm}

In \eqref{eq:hat:B:optimum}, note that by convention $\sum_{k'=a}^b \cdots = 0$ for $a,b \in \mathbb{Z}$, $a > b$. We will use this convention in the rest of the paper.
Note also that, apart from the polyhedral characterization given by \eqref{eq:equi:conditions:un} to \eqref{eq:equi:conditions:trois}, Theorem \ref{theorem:equivalence:polyhedral:one:d} also provides the explicit expression of the optimal $\widehat{\B}$, solution of Problem \eqref{eq:conv-clust}. This expression depends of the optimal clustering, so it cannot be directly computed to optimize \eqref{eq:conv-clust} in practice. Nevertheless, Theorem \ref{theorem:equivalence:polyhedral:one:d} is the basis of a regularization path algorithm provided in Section~\ref{subsection:regularization:path:one-dimensional}. 

Next, the following lemma provides a formulation of \eqref{eq:equi:conditions:un} to \eqref{eq:equi:conditions:trois} in Theorem \ref{theorem:equivalence:polyhedral:one:d} as an explicit polyhedral set. 
In this lemma and in the rest of the paper, for $a \in \mathbb{N}$, we let $ \bzero_{a} $ be the $a \times 1$ vector composed of zeros.

\begin{lmm}\label{corPolyh}
	Consider the setting of Theorem \Rref{theorem:equivalence:polyhedral:one:d}.
	Let $\bP_\sigma$ be the $n\times n$ permutation matrix associated to $\sigma \in \frS_n$: $\bP_{\sigma} \x = (\x_{\sigma(1)}, \dots, \x_{\sigma(n)})^\top$, for a $n 
	\times 1 $ vector $\x$.
	Then, Conditions \eqref{eq:equi:conditions:un}, \eqref{eq:equi:conditions:deux} and \eqref{eq:equi:conditions:trois} can be written as
	\begin{equation}
		\label{eq:polyh}
		\{\bM(\t) \bP_\sigma \x \leq \lambda\ \m(\t)\}
	\end{equation}
	where $\bM(\t) \in\R^{2(n-1) \times n}$ and $\m(\t) \in\R^{2(n-1)}$ are given by:
	$$\bM(\t)=\left(\begin{array}{c} \bM_1\\ \bM_2(\t)\\ \bM_3(\t)\end{array}\right) \textrm{ and } \m(\t)=\left(\begin{array}{c} \m_1\\ \m_2(\t)\\ \m_3(\t)\end{array}\right),$$
	with
	$\bM_1 \in\R^{n-1 \times n}$, $\bM_2(\t) \in \R^{K-1 \times n}$ and $\bM_3(\t) \in \R^{n-K \times n}$, explicitly expressed in Section~\ref{section:proof:one:dimensional} (Equations \eqref{eq:exprM1}, \eqref{eq:exprM2} and \eqref{eq:exprM3} respectively); $\m_1=\bzero_{n-1}$, $\m_2(\t)\in\R^{K-1}$ and $\m_3(\t)\in\R^{n-K}$, explicitly expressed in Section~\Rref{section:proof:one:dimensional} (Equations  \eqref{eq:exprm2} and \eqref{eq:exprm3} respectively).  Furthermore, the inequality $\bM_2(\t) \bP_{
		\sigma} \x \leq \lambda \m_2(\t)$ is strict in \eqref{eq:polyh}.
\end{lmm}

\subsection{Test procedure and its guarantees}
\label{subsection:test:one:dimensional}

In this section, we construct the test procedure and provide its theoretical guarantees, based on Theorem~\ref{theorem:equivalence:polyhedral:one:d} and Lemma \ref{corPolyh}. Since the polyhedral characterization has been shown from these two results, the construction and guarantees here are obtained similarly as in \cite{Lee16}. We nevertheless provide the full details, for the sake of self-completeness. 

\subsubsection{Construction of the test procedure} \label{subsection:construction:test}

We want to test $\boeta^\top  \bmu = 0,$
where $\boeta = \boeta(\C(\X))$ and  $\C(\X)$ is obtained from Problem \eqref{eq:conv-clust} and Definition \ref{def:clustering}. 
The test statistic is naturally
$\boeta^\top  \X$,
and we will construct a $p$-value from it, based on the polyhedral lemma (Lemma 5.1) of \cite{Lee16}, that we restate in our setting for convenience. In the next statement, $\bI_a$ is the identity matrix in dimension $a \in \mathbb{N}$ and we use the conventions that the minimum over an empty set is $+ \infty$ and the maximum over an empty set is $- \infty$.

\begin{prpstn}[Polyhedral lemma, adapted from \cite{Lee16}]
	\label{prop:polyhedral:lemma}
	Let $\mathbf{t}$ be a fixed vector in $\mathcal{T}_{K,n}$ with $K \in [|n|]$.
	Let $\sigma \in \frS_n$ be a fixed permutation of $[|n|]$, and $\bP_\sigma$ be the $n\times n$ associated permutation matrix. \\
	Let $\X\sim\N(\bmu,\bSigma)$ with $\bSigma$ invertible and let $\boeta$ be a fixed non-zero $n \times 1$ vector (allowed to depend on $\t$ and $\sigma$). \\
	Let $\Z:=\Z(\X):=[ \bI_n - \bc  \boeta^\top] \X$ with $\bc=\bSigma \boeta (\boeta^\top \bSigma \boeta)^{-1}$. 
	Let $\bM:=\bM(\t)$ and $\lambda \m:= \lambda \m(\t)$ defined in \eqref{eq:polyh}.
	%\PN{$\lambda$ is defined only implicitly via \eqref{eq:polyh}}
	Then, for any fixed $\lambda>0$,  we have the following properties: 
	\begin{itemize}
		\item $\Z$ is uncorrelated with, and hence independent of, $\boeta^\top \X$.
		\item The conditioning set can be written as follows
		\begin{equation} \label{eq:conditioning:set}
			\{\bM \bP_\sigma \X \leq \lambda\  \m\} = \{ \V^{-}(\Z)\leq \boeta^\top \X \leq \V^{+}(\Z), \V^{0}(\Z)\geq 0\}
		\end{equation}
		where
		\begin{itemize}
			\item $\V^{-}(\Z) := \underset{l:(\bM \bP_\sigma \bc)_l<0}{\max}\ \frac{\lambda m_l - (\bM \bP_\sigma \Z)_l}{(\bM \bP_\sigma \bc)_l}$
			\item  $\V^{+}(\Z) := \underset{l:(\bM \bP_\sigma \bc)_l>0}{\min}\ \frac{\lambda m_l - (\bM \bP_\sigma \Z)_l}{(\bM \bP_\sigma \bc)_l}$ 
			\item $\V^{0}(\Z):=\underset{l:(\bM \bP_\sigma \bc)_l=0}{\min} \lambda m_l - (\bM \bP_\sigma \Z)_l$.
		\end{itemize}
	\end{itemize}
	Note that $\V^{-}(\Z)$, $\V^{+}(\Z)$ and $\V^{0}(\Z)$ are independent of $\boeta^\top \X$. Finally, when the event in \eqref{eq:conditioning:set} has non-zero probability, conditionally to this event, the probability that $ \V^{-}(\Z) =  \V^{+}(\Z)$ is zero.
\end{prpstn}

%For $\nu \in \mathbb{R}$, $\tau >0$ and $a<b$, let $F_{\nu, \tau^2}^{[a,b]}(.)$ denote the cumulative distribution function (cdf) of a Gaussian distribution $\N(\nu,\tau^2)$ truncated on the interval $[a,b]$ that is, for $a \leq x \leq b$, 
%$$
%F_{\nu, \tau^2}^{[a,b]}(x) = \frac{\Phi( (x-\nu)/\tau)   - \Phi( (a-\nu)/\tau) }{\Phi( (b-\nu)/\tau)   - \Phi( (a-\nu)/\tau) }
%$$
%where $\Phi$ is the cdf of the standard Gaussian distribution. 
%

From Proposition \ref{prop:polyhedral:lemma}, it is shown in \cite{Lee16} that, for any fixed $\z_0$ with $\V^{-}(\z_0) < \V^{+}(\z_0)$ and $\V^0(\z_0) \geq 0$, under the null hypothesis $\boeta^\top \bmu = 0$, conditionally to $\left\{\bM \bP_{\sigma} \X \leq \lambda\  \m, \Z = \z_0\right\}$, the test statistic $\boeta^\top  \X$ follows a truncated Gaussian distribution: for all $t \in \mathbb{R}$
\begin{equation} \label{eq:the:invariant:stat:oneD}
\mathbb{P}_{\boeta^\top  \bmu = 0}
\left( 
\left.
\boeta^\top  \X \le t
\right| 
\bM \bP_{\sigma} \X \leq \lambda\  \m, \Z = \z_0
\right)
= 
F_{0, \boeta^\top \bSigma \boeta}^{[\V^{-}(\z_0), \V^+(\z_0)]} (t),
\end{equation}
where $F_{\nu, \tau^2}^{[a,b]}(.)$ is the cumulative distribution function (cdf) of a Gaussian distribution $\N(\nu,\tau^2)$ truncated on the interval $[a,b]$. 

%The test statistics is finally defined by, for any $n \times 1$ observation vector $\x$, 
%\[
%T( \x, \t,\sigma) : =  F_{0, \boeta^\top \bSigma  \boeta}^{[\V^{-}(\z_0), \V^+(\z_0)]} \left(\boeta^\top \x\right).
%\]

Now, let us define the $p$-value when having observed $\X = \x$ for a $n \times 1$ observation vector $\x$, and corresponding to considering two-sided alternative hypotheses to $\boeta^\top \bmu = 0$. 
Let $\z_0 = [ \bI_n - \bc  \boeta^\top] \x$ as in Proposition \ref{prop:polyhedral:lemma}.
The  $p$-value is
\begin{equation}
\label{eq:cond-p-value-1d}
	\pval( \x, \t,\sigma)  = 
 \mathbb{P}_{W} \left( 
| W |
\ge 
| \boeta^\top \x |
 \right),
\end{equation}
where $W$ has cdf $F_{0, \boeta^\top \bSigma \boeta}^{[\V^{-}(\z_0), \V^+(\z_0)]}$.
As is standard, the $p$-value is interpreted as the (conditional) probability of the test statistic to be more ``extreme'' than its observed value, under the null hypothesis.
Note that the two definitions
\eqref{eq:the:invariant:stat:oneD} and \eqref{eq:cond-p-value-1d}
require $\V^{-}(\z_0) < \V^+(\z_0)$, which holds almost surely conditionally to  $\{\bM \bP_{\sigma} \X \leq \lambda\ \m\}$, as stated in Proposition \ref{prop:polyhedral:lemma}.

The next lemma provides the explicit expression of $\pval( \x, \t,\sigma)$. 

\begin{lmm}
\label{lemma:expression:pvalue}
Let $\Phi$ be the cdf of the Gaussian distribution $\N(0,\boeta^\top \bSigma \boeta)$. The $p$-value $\pval( \x, \t,\sigma)$ can be expressed as follows, in three different cases.
\begin{itemize}
    \item {\bf Case 1.1 (one sided, $\V^-$):} $|\boeta^\top \x| >  \min (|\V^{-}(\z_0)| , |\V^{+}(\z_0)| )$ with $|\V^{-}(\z_0)| > |\V^{+}(\z_0)|$.
We have
\[
\pval( \x, \t,\sigma)
=
\frac{\Phi(\boeta^\top \x)
-
\Phi(\V^-(\z_0))
}{\Phi(\V^+(\z_0))
-
\Phi(\V^-(\z_0))
}.
\]
\item {\bf Case 1.2 (one sided, $\V^+$):}
$|\boeta^\top \x| >  \min (|\V^-(\z_0)| , |\V^+(\z_0)| )$ with $|\V^+(\z_0)| > |\V^-(\z_0)|$. We have
\[
\pval( \x, \t,\sigma)
=
\frac{
\Phi(\V^+(\z_0))
- \Phi(\boeta^\top \x)
}{
\Phi(\V^+(\z_0))
-
\Phi(\V^-(\z_0))
}.
\]
\item {\bf Case 2 (two sided):}
$|\boeta^\top \x| \leq  \min (|\V^-(\z_0)| , |\V^+(\z_0)| )$. We have
\[
\pval( \x, \t,\sigma)
= 
\frac{
\Phi(-|\boeta^\top \x|)
-
\Phi(\V^-(\z_0))
}{
\Phi(\V^+(\z_0))
-
\Phi(\V^-(\z_0))
}
+
\frac{
\Phi(\V^+(\z_0))
- \Phi(|\boeta^\top \x|)
}{
\Phi(\V^+(\z_0))
-
\Phi(\V^-(\z_0))
}.
\]
\end{itemize}
\end{lmm}

In Lemma \ref{lemma:expression:pvalue}, the wording ``one sided, $\V^-$'' (resp. ``one sided, $\V^+$'') means that the set $\{  w \in \mathbb{R}; |w| \ge  |\boeta^\top \x| \}$ is a single segment containing $\V^-(\z_0)$ (resp. $\V^+(\z_0)$). 
%The wording ``one sided, $\V^+$'' means that this set is a single segment containing $\V^+(\z_0)$. Finally, 
The wording ``two-sided'' means that this set is composed of two distinct segments (unless $\boeta^\top \x = 0$), containing $\V^-(\z_0)$ and $\V^+(\z_0)$ respectively. This is also made explicit in the proof of the lemma in Section \ref{section:proof:one:dimensional}.

\subsubsection{Conditional level} \label{subsubsection:conditional:valid}

Next, we show that the suggested test is conditionally valid. That is, conditionally to a clustering and a data order, when the null hypothesis (that is fixed by the clustering) holds, the $p$-value is uniformly distributed.
In particular, the probability of rejection is equal to the prescribed level.
Conditional validity naturally yields unconditional validity, as shown in Section \ref{subsubsection:unconditional:valid}. Hence conditional validity is mathematically a stronger property than unconditional validity. A statistical benefit of conditional validity is that the null hypothesis is fixed after conditioning; in particular $\boeta^\top \bmu$ becomes a fixed target of interest, which is beneficial for interpretability. In the related context of linear models, for instance, the tests obtained from the confidence intervals of \cite{bachoc2020uniformly,berk2013valid} are unconditionally valid while the tests provided in \cite{Lee16,Tibshirani2018uniform} are conditionally (and unconditionally) valid. 
The interpretability benefit we discuss above is also discussed in \cite{Lee16}.

\begin{prpstn} \label{prop:level:conditional}
	Let $\mathbf{t}$ be a fixed vector in $\mathcal{T}_{K,n}$ with $K \in [|n|]$.
	Let $\sigma \in \frS_n$ be a fixed permutation of $[|n|]$, and $\bP_\sigma$ be the $n\times n$ associated permutation matrix. 
	Let $\C=\C(\t, \sigma)$ be the clustering obtained from Definition~\Rref{def:clust-seg}, with cluster cardinalities $n_1,\ldots,n_K$.
	Let $\X\sim\N(\bmu,\bSigma)$ with $\bSigma$ invertible.
	Consider a fixed $n \times 1$ non-zero vector $\boeta \in \mathbb{R}^n$ (that is only allowed to depend on $(\mathbf{t},\sigma)$). 
	Assume that
	\[
	\boeta^\top \bmu = 0.
	\]
	Let $\widehat{\B} = \widehat{\B}(\X)$ from Problem \eqref{eq:conv-clust} for some fixed $\lambda>0$.
	Assume that with non-zero probability, the event 
	\[
	E_{\t,\sigma} ~ ~ ~ 
	:= 
	~ ~ ~
	\left\{ \text{$\C(\t, \sigma)$ is the clustering given by $\widehat{\B}$}, ~ ~
	X_{\sigma(1)} \geq X_{\sigma(2)} \geq \dots \geq X_{\sigma(n)}
	\right\}
	\]
	holds. 
	Then, conditionally to $E_{\t,\sigma}$, $\pval( \X, \t,\sigma)$ is uniformly distributed on $[0,1]$:
	$$
	\mathbb{P}_{\boeta^\top \bmu=0}
	\left(
	\pval( \X, \t,\sigma) \leq t \big\vert E_{\t,\sigma}
	\right) = t \quad \quad
	\forall t \in [0,1].
	$$
\end{prpstn}

\subsubsection{Unconditional level} \label{subsubsection:unconditional:valid}

We now show that $\pval( \X, \t,\sigma)$ is unconditionally uniformly distributed, which we call unconditional validity. Here, ``unconditionally'' means that the clustering is not fixed, but it is still necessary to condition by the fact that the null hypothesis $\boeta^\top \bmu = 0$ is well-defined and true. Regarding well-definiteness, the vector $\boeta = \boeta (\C(\X))$ may indeed not be well-defined for all clusterings $\C(\X)$. In the next proposition, we thus introduce the set $\mathcal{E}$ of clusterings, indexed by an ordering $\sigma$ and a sequence of right-limits $\t$ as in Definition \ref{def:clust-seg}, that make $\boeta$ well-defined.

For instance, in the case of the two-group test of Example \ref{ex:hyp}, $\boeta$ can be defined similarly as in \eqref{eq:example:comparison:signal:multiD}, with  
\begin{equation}  \label{eq:example:comparison:signal:singleD}
	\boeta^\top \bmu
	=
	\frac{1}{|\C_{k_1}(\X)|}
	\sum_{i \in \C_{k_1}(\X)} \mu_{i }
	-
	\frac{1}{|\C_{k_2}(\X)|}
	\sum_{i \in \C_{k_2}(\X)} \mu_{i }.
\end{equation}
In this case, $\mathcal{E}$ is the set of clusterings for which the number of clusters is larger than or equal to $\max(k_1,k_2)$, enabling $\boeta$ to be well-defined. 
When $k_1 =1 $ and $k_2=2$,
this definition is possible for all clusterings, except the one with only one cluster. In this case, $\mathcal{E}$ should thus be defined as restricting $\t$ to have at least $3$ elements $0=t_0 < t_1 < t_2 = n$, that is to correspond to a clustering with at least two clusters.

Then, Proposition \ref{proposition:unconditional:level} shows that conditionally to $\mathcal{E}$ and to $\boeta^\top \bmu = 0$, the $p$-value is uniformly distributed, which we call unconditional validity, in the sense that we do not condition by a single clustering, as commented above. 

\begin{prpstn}
	\label{proposition:unconditional:level}
	Let $\mathcal{E}$ be a subset of the set of all possible values of $
	(\t,\sigma)$ in Proposition \Rref{prop:level:conditional}. 
	Consider a deterministic function $\boeta : \mathcal{E} \to \mathbb{R}^n$, outputing a non-zero column vector. 
	Assume that $\bSigma$ is invertible. Let $\widehat{\B}$ be as in \eqref{eq:conv-clust}.   Let $S = S(\X)$ be a random permutation obtained by reordering $\X$ as: $X_{S(1)} \geq  \cdots \geq X_{S(n)}$ (uniquely defined with probability one). Let $\C(\X) = \C$ be the random clustering given by $\widehat{\B}$ (Definition \Rref{def:clustering}),  of random dimension $K(\X) = K$. 
	Let $\T(\X) = \T \in \cT_{K,n}$ be the random vector, such that $\T$ and $S$ yield $\C$ as in Definition \Rref{def:clust-seg}. 
	
	Assume that
	\[
	\mathbb{P} 
	\left(
	(\T,S) \in \mathcal{E}
	,
	\boeta(\T,S)^\top \bmu
	= 0
	\right) 
	>0.
	\]
	Then, conditionally to the above event, $\pval( \X, \T,S)$ is uniformly distributed on $[0,1]$:
	$$
	\mathbb{P}
	\left(
	\pval( \X, \T, S) \leq t \big\vert (\T,S) \in \mathcal{E},\boeta(\T,S)^\top \bmu= 0
	\right) = t \quad \quad
	\forall t \in [0,1].
	$$
\end{prpstn}

Note that unconditional validity implies that, for any $\alpha \in (0,1)$, the test procedure constructed by rejecting when $	\pval( \X, \T, S) \le \alpha$ has the following property:
We upper bound by $\alpha$ the probability that at the same time the (random) hypothesis is true (and well-defined) and that the test rejects it.

Proposition \ref{proposition:unconditional:level} proves that conditional validity (as in Proposition \ref{prop:level:conditional}) 
implies unconditional validity for our constructed $p$-value $	\pval( \X, \T, S)$. The proof of Proposition \ref{proposition:unconditional:level} equally applies to any conditionally valid $p$-value as in Proposition \ref{prop:level:conditional}. Hence, all (conditionally valid) selective inference procedures are also unconditionally valid, as mentioned in Section~\ref{subsubsection:conditional:valid}. 
Note that there exist selective inference procedures that are unconditionally valid but not conditionally so, in particular from the PoSI literature \cite{bachoc2020uniformly,berk2013valid,Kuchibhotla2020}.

%%%%%%%%%%%%%%%%%%%%%%%%%%%%%%%%
\subsection{Regularization path} \label{subsection:regularization:path:one-dimensional}

At first sight, \eqref{eq:conv-clust} is a convex optimization problem, whose (unique) minimizer does not have any explicit expression, and thus \eqref{eq:conv-clust} requires numerical optimization to approximate its solution. Furthermore, this numerical optimization would be repeated for different values of $\lambda$. 
However, thanks to the polyhedral characterization of Theorem \ref{theorem:equivalence:polyhedral:one:d}, we can provide a regularization path for solving \eqref{eq:conv-clust}. This regularization path is an algorithm, only performing elementary operations, that provides the entire sequence of exact solutions to \eqref{eq:conv-clust}, for all values of $\lambda$.  
This algorithm is exposed in Algorithm \ref{alg:regularization:path}. Then, Theorem \ref{thm:regularization:path} shows that this algorithm is well-defined and indeed provides the set of solutions to Problem \eqref{eq:conv-clust}.

\SetKwProg{Init}{Initialization}{}{}
\SetKwProg{ForAll}{For all}{}{}

\begin{algorithm}[tbp] 
	\KwIn{$\x = (x_1 , \ldots , x_n) \in \mathbb{R}^n$}
	
	\Init{}{
		$r\gets 0$;
		$\lambda^{(0)} \gets 0$\;
		$\tilde{x}_1 > \dots > \tilde{x}_{K^{(0)}}$:  the $K^{(0)}$ distinct values in $\x$\;
		$\C^{(0)} = (\C_1^{(0)},\ldots , \C_{K^{(0)}}^{(0)}) \gets$ clustering of $[|n|]$ where $\C^{(0)}_k = \{  i \in [|n|] : x_i = \tilde{x}_k \}$\; 
		$n^{(0)}_k \gets |\C^{(0)}_k|$ for $k\in  [|K^{(0)}|]$\;
		$\hat{b}^{(0)}_k(\lambda^{(0)}) \gets \tilde{x}_k$ for $k \in [|K^{(0)}|]$\;	
		$\hat{B}^{(0)}_i(\lambda^{(0)})\gets \hat{b}^{(0)}_k(\lambda^{(0)})$ if $i \in \C^{(0)}_k$  ($k$ is unique) for $i \in [|n|]$\;
	}	
	\While{$K^{(r)} \geq 2$}{
		\ForAll{$\lambda \geq \lambda^{(r)}$ we define}{
			\begin{flalign}\label{eq:hat:b:algo}
				&\hat{b}^{(r)}_k(\lambda) := \hat{b}^{(r)}_k(\lambda^{(r)})    + \left( \lambda -  \lambda^{(r)} \right)\left( \sum_{k' = 1}^{k-1}n^{(r)}_{k'} - \sum_{k'  = k+1}^{K^{(r)}}n^{(r)}_{k'}\right)\ \forall k\in [|K^{(r)}|] &
			\end{flalign}
			$\hat{B}^{(r)}_i(\lambda) :=  \hat{b}_k^{(r)}(\lambda)$ if $i \in \C^{(r)}_k$ ($k$ is unique) for $i \in [|n|]$\; 	
		}
		%	\begin{itemize}
		%	\item For $k\in [|K^{(r)}|]$ and for $\lambda \geq \lambda^{(r)}$, define 
		\begin{flalign}  \label{eq:lambda:equal:inf}
			& \lambda^{(r+1)} \gets \lambda^{(r)} +  \min_{k \in [|  K^{(r)}-1 |] }\frac{\hat{b}^{(r)}_k(\lambda^{(r)})-\hat{b}^{(r)}_{k+1}(\lambda^{(r)})}{n^{(r)}_{k} +n^{(r)}_{k+1}};&
		\end{flalign}
		%	
		%	\item For $i \in [|n|]$, let $\hat{B}^{(r)}_i(\lambda ) =  \hat{b}_k^{(r)}(\lambda)$, where $k$ is the unique index in $[| K^{(r)} |]$ such that $i \in \C^{(r)}_k$.
		%$K^{(r+1)}$: number of distinct values of $\hat{b}_1^{(r)}(\lambda^{(r+1)}) ,\ldots,\hat{b}_{K^{(r)}}^{(r)}(\lambda^{(r+1)})$\;
		$(\hat{b}^{(r+1)}_k(\lambda^{(r+1)}))_{k\in[|K^{(r+1)}|]} \gets$ distinct values of $(\hat{b}_k^{(r)}(\lambda^{(r+1)}))_{k\in[|K^{(r)}|]}$, sorted decreasingly\; 
		%\begin{gather*}
		%\hat{b}^{(r+1)}_1(\lambda^{(r+1)}) > \ldots > \hat{b}^{(r+1)}_{K^{(r+1)}}(\lambda^{(r+1)})
		%\end{gather*}
		$\C^{(r+1)}  \gets$ clustering of $[|n|]$ obtained from $\left( \hat{B}_i^{(r)}(\lambda^{(r+1)}) \right)_{i \in [|n|]}$ by Definition \ref{def:clustering}\;
		$n^{(r+1)}_k \gets |\C^{(r+1)}_k|$ for $k\in  [|K^{(r+1)}|]$\;
		%	 \end{itemize}
		$r\gets r+1$\;
	}
	\caption{Regularization path for one-dimensional convex clustering}\label{alg:regularization:path}
\end{algorithm}

\begin{thrm} \label{thm:regularization:path}
	Algorithm \ref{alg:regularization:path} stops at a final value of $r$ that we write $r_{\max}$, such that $r_{\max} \leq n-1$ and we have $K^{(0)} > \dots > K^{(r_{\max})} = 1$. 	
	Let $\lambda^{(r_{\max} +1)} = + \infty$ by convention. For $r \in  \{ 0 , \ldots , r_{\max} \}$ and $\lambda \in [ \lambda^{(r)} , \lambda^{(r+1)} )$, $(\hat{B}_i^{(r)}(\lambda))_{i \in [|n|]}$ minimizes Problem \eqref{eq:conv-clust}.
\end{thrm}

In Theorem \ref{thm:regularization:path}, we remark that 
even if Algorithm \ref{alg:regularization:path} stops at $r = r_{\max}$, we can still define $(\hat{B}_i^{(r_{\max})}(\lambda))_{i \in [|n|]}$ there, with \eqref{eq:hat:b:algo}, with the convention that $\sum_{k'=1}^0 n_{k'}^{(r_{{\max}})} = 0$ and $\sum_{k'=2}^1 n_{k'}^{(r_{{\max}})} = 0$. This vector has all its components equal to $\sum_{i=1}^n x_i / n$ (see the proof of Theorem \ref{thm:regularization:path} in Section \ref{section:proof:one:dimensional}).

By way of illustration, Algorithm \ref{alg:regularization:path} was applied  to the observations of Example \ref{extoy}, and the resulting regularization path is shown in Figure \ref{fig:regpathToy}. In Algorithm \ref{alg:regularization:path},  since $r \mapsto K^{(r)}$ is strictly decreasing during the execution, there are at most $n-1$ induction steps. A straightforward implementation of \eqref{eq:lambda:equal:inf} can lead to a time complexity of order $\mathcal{O}( K^{(r)} )$  for each step, and thus a total time complexity of order $\mathcal{O}( n^2 )$ in the worst case. The space complexity is linear ($\mathcal{O}(n)$). Indeed, in order to recover the entire regularization path, it is sufficient to record at each step $r$ the labels of the clusters merged at this step. 
We have implemented this algorithm in the open source \texttt{R} package \texttt{poclin} (which stands for ``post convex clustering inference''), which is available from \url{https://plmlab.math.cnrs.fr/pneuvial/poclin}. The empirical time complexity of our implementation is substantially below $\mathcal{O}(n^2)$ for $n \leq 10^5$, as illustrated in Section~\ref{sec:complexity}. In this section, we also explain that the time complexity of Algorithm \ref{alg:regularization:path} could be further decreased to $\mathcal{O}(n \log(n))$ without compromising the linear space complexity by storing merge candidates more efficiently using a min heap.
\FloatBarrier

\begin{rmrk}[Final value of the regularization parameter]
	As a consequence of Theorem~\ref{theorem:equivalence:polyhedral:one:d} (see in particular~\eqref{eq:equi:conditions:deux}), the final value of $\lambda$ in Algorithm~\ref{alg:regularization:path} is obtained analytically as:
	\begin{equation}
		\lambda^{(r_{\max})} = \max_{i \in [|n-1|]}
		\frac{
			\frac{1}{i} \sum_{i'=1}^{i} x_{(i')}    
			-
			\frac{1}{n} \sum_{i'=1}^{n} x_{(i')} 
		}{n - i}.
		\label{eq:lambda-max}
	\end{equation}
	It corresponds to the smallest value of $\lambda$ for which the convex clustering yields exactly one cluster.  The range of values for which there are two or more clusters has also been studied by \cite{tan2015} for convex clustering procedures that include Problem \eqref{eq:conv-clust}. We note that in the specific case of Problem \eqref{eq:conv-clust}, $\lambda^{(r_{\max})}$ can be computed using \eqref{eq:lambda-max} in linear time after an initial sorting of the input vector. Our numerical experiments below make use of~\eqref{eq:lambda-max} to choose $\lambda$ in a non data-driven way, see also Section~\ref{sec:calibration-lambda}.
\end{rmrk}

\noindent
{\bf Relation to other existing regularization path algorithms.}
Algorithm~\ref{alg:regularization:path} has similarities with the following two more general regularization path algorithms, that can be applied to Problem  \eqref{eq:conv-clust}. 
First, for the generalized Lasso, a penalization term of the form $
\| \D \B \|_1
$ is studied in \cite{TibshiraniTaylor11}, for a general matrix $\D$. 
It is then simple to find a  $n(n-1)/2 \times n$ (sparse) matrix $\D$ leading to the penalization term $
\lambda 
\sum_{ \substack{ i,i'=1, i < i'}}^n
|B_{i'} - B_{i}|$ of \eqref{eq:conv-clust}.
The benchmarks that we have conducted in Section~\ref{sec:complexity} show that the procedure based on the generalized Lasso has a very large memory footprint and is very slow (more than 10 seconds for $n=50$), as it relies on the matrix $\D$, whose total number of entries is $\mathcal{O}(n^3)$. 
Second, the fused Lasso signal approximator (FLSA) suggested by \cite{hoefling2010path} can handle a penalization term of the form
$
\lambda 
\sum_{ \substack{ i,i'=1, (i,i') \in E}}^n
|B_{i'} - B_{i}|$, where $E$ is a set of pairs of indices. Similarly as before, taking $E$ as the complete set of pairs recovers  the penalization term of \eqref{eq:conv-clust}. 
The theoretical time complexity of the regularization path for FLSA has been shown in \cite{hocking2011} to be $\mathcal{O}(n \log(n))$ in this case.
The benchmarks that we have conducted in Section~\ref{sec:complexity} show that the procedure based on FLSA is much more efficient than the one based on the generalized Lasso.
Nevertheless, our implementation of Algorithm~\ref{alg:regularization:path}
remains preferable, as it can address larger dataset sizes (see Figure~\ref{fig:computational-time}).

On top of these numerical performances, the benefit of Algorithm \ref{alg:regularization:path}, relatively to these two general procedures, is that its description and proof of validity (Theorem \ref{thm:regularization:path}) are self-contained and specific to the one-dimensional convex clustering problem  \eqref{eq:conv-clust}. Furthermore,
the proof of validity exploits the specific analysis of \eqref{eq:conv-clust} given by Theorem \ref{theorem:equivalence:polyhedral:one:d}. 
Finally, the efficiency of our regularization path algorithm is a benefit of our post-clustering test procedure, compared to other procedures that address the generalized Lasso problem more broadly \cite{hyun2018exact,le2022more,chen2023more} (see also Section \ref{sec:comp-gen-lasso}).

\subsection{Numerical experiments}
\label{sec:numerical-experiments-1d}

In order to illustrate the behaviour of our post-clustering testing procedure, we have performed the following numerical experiments in the one-dimensional framework.
The code to reproduce these numerical experiments and the associated figures is available from \url{https://plmlab.math.cnrs.fr/pneuvial/poclin-paper}.

We consider a Gaussian sample $\X=(X_1,\ldots,X_n)^\top$ with mean vector $\bmu= (\nu \bone_{n/2}^\top, \bzero^\top_{n/2})^\top$ and known covariance matrix $\bSigma= \bI_n$.
Here and in the rest of the paper, for $a \in \mathbb{N}$, we let $ \bone_{a} $ be the $a \times 1$ vector composed of ones. We recall that $\bzero_{a}$ is similarly composed of zeros.

We set $n=1000$ and $\lambda = 0.0025$. This value of $\lambda$ has been chosen to ensure that with high probability, the convex clustering finds at least two clusters under the null hypothesis. 
The procedure that we have used in our numerical experiments to achieve this property relies on \eqref{eq:lambda-max} and is described in Section~\ref{sec:calibration-lambda}.
Let $\C=(\C_k)_{k \in [|K|]}$ be the result of the one-dimensional convex clustering obtained from Algorithm~\ref{alg:regularization:path} with $\lambda = 0.0025$. If $K>2$, we merge adjacent clusters in to obtain a 2-class clustering of the form $\overline{\C}_1 := \C_1 \cup  \dots \cup \C_q, \overline{\C}_2 := \C_{q+1} \cup \dots \cup \C_K$, where $q$ is chosen so that the sizes of $\overline{\C}_1$ and $\overline{\C}_2$ are as balanced as possible.
We then apply the test procedure introduced in Section~\ref{subsection:construction:test} to compare the means of $\overline{\C}_1$ and $\overline{\C}_2$, as in Example~\ref{ex:hyp}.
Note that this yields $\eta_i = \1_{i \in \overline{\C}_1} / |\overline{\C}_1| - \1_{i \in \overline{\C}_2} / |\overline{\C}_2|$ for $i \in [|n|]$, which is indeed a deterministic function of $\C_1,\ldots,\C_K$ and thus in the scope of the guarantees obtained in Section \ref{subsection:test:one:dimensional}.
For each signal value $\nu\in\{0, 1, 2, 3, 4, 5\}$, we retain $N=1000$ numerical experiments for which $K \geq 2$.
Note that the event $K \geq 2$ corresponds to the set $\mathcal{E}$ in Proposition~\ref{proposition:unconditional:level}.

\begin{figure}[htp!]
	\includegraphics[height=6.5cm]{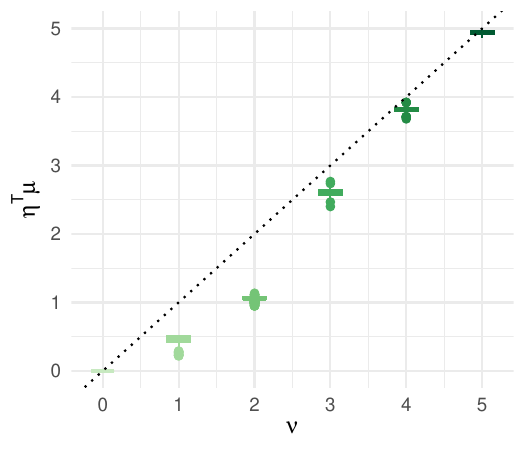}
	\includegraphics[height=6.5cm]{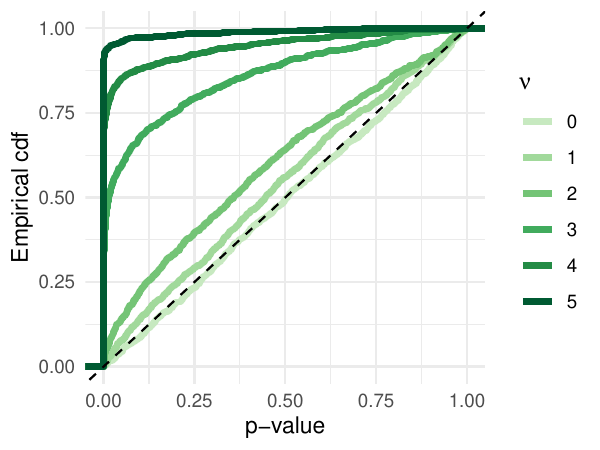}
	\caption{Left: empirical density of $\boeta^\top \bmu$ for each $\nu$. Right: empirical cumulative distribution functions of the $p$-value of the test of equality between the means of two clusters.}
	\label{fig:num-exp-1d-H1}
\end{figure}

Figure~\ref{fig:num-exp-1d-H1} (left) gives the empirical density of $\boeta^\top \bmu$, the difference between the true means of the estimated clusters, for each value of $\nu$ considered. This plot quantifies the performance of the clustering step: for a perfect clustering, we would have $\boeta^\top \bmu = \nu$, corresponding to the diagonal line. As expected, the larger the signal ($\nu$ increases), the easier the clustering step. 

Figure~\ref{fig:num-exp-1d-H1} (right) shows the empirical  distribution of the proposed $p$-value (see \eqref{eq:cond-p-value-1d}).
For $\nu=0$ (no signal), the curve illustrates the uniformity of the distribution of the $p$-values: it shows that the level of the test is appropriately controlled. Another simulation to control the level of the test is available in Section~\ref{apx:numerical-experiments-1d-level}. As expected, the power of the test is an increasing function of the distance between the null and the alternative hypotheses (as encoded by the parameter $\nu$). Our conditional test is able to detect the signal only for $\nu > 1$.

Because of the computational limitations of procedures based on the generalized Lasso (both at the clustering and inference steps) when specialized to the convex clustering problem, we have not been able to include them in the numerical experiments reported in this section, where $n=1000$. 
For completeness, in Section \ref{sec:comp-gen-lasso}, we have included  a comparison to the \texttt{genlasso} procedure introduced in \cite{hyun2018exact}. 
In a nutshell, when reducing the number of observations to $n=40$, the \texttt{genlasso} procedure could be applied but was essentially powerless, whereas our procedure showed reasonable power for this number of observations.

\section{The \texorpdfstring{$p$}{p}-dimensional case} 
\label{section:p:dimensional case}

\subsection{Aggregating one-dimensional clusterings} \label{subsection:test:after:aggregation}

Consider the $p$-dimensional setting of Section \ref{section:context:and:objectives}.  
For $j \in [|p|]$, consider the one-dimensional clustering $\C^{(j)} = \C^{(j)}(\Y_{.j}) = (\C^{(j)}_1(\Y_{.j}) , \ldots, \C^{(j)}_{K^{(j)}}(\Y_{.j}))$ obtained by computing $\hat{\B}_{.j}$ by solving \eqref{eq:conv-clust:j} and with Definition~\ref{def:clustering}. 
We consider a $p$-dimensional clustering $\C$ obtained by aggregation of the one-dimensional clusterings  $\C^{(1)} , \ldots, \C^{(p)}$ as follows.

For $i \in [|n|]$ and $j \in [|p|]$, let $\tilde{Y}_{ij}$ be the class index of $Y_{ij}$ in the clustering $\C^{(j)}$, rescaled from $\{1, 2, \dots, K^{(j)}\}$ to $\{0, 1/(K^{(j)}-1), \ldots, 1\}$ (to $\{ 0 \}$ if $K^{(j)} = 1$).
We obtain a $p$-dimensional clustering $\C$ by applying a clustering procedure to the rows of the $n \times p$ matrix $\tilde{\Y}$, for instance a hierarchical clustering \cite{murtagh2012algorithms} with the Euclidean distance.
We are then in a position to test an hypothesis $\bkappa^\top \bbeta = 0$, where  $\bkappa = \bkappa(\C)$, as motivated in Section~\ref{section:context:and:objectives}. 
In particular, we can test the signal difference for the column $j_0$ between two clusters $\C_{k_1}$ and $\C_{k_2}$ in the multi-dimensional clustering $\C$, as in \eqref{eq:example:comparison:signal:multiD} in Example \ref{ex:hyp}.

\begin{rmrk}
	\label{rem:clustering-aggregation}
	Above, we focus on a specific aggregation using the hierarchical clustering with the Euclidean distance for simplicity.
	However, we can construct more general $p$-dimensional clusterings $\C$ by more general aggregations of $\C^{(1)} , \ldots, \C^{(p)}$.
	Indeed, our statistical framework (see Section \ref{subsection:test:procedure:multiD}) encompasses any case where $\bkappa = \bkappa(\C)$, as long as $\C$ is a function of the one-dimensional clusterings and orderings.
	In particular, one could also consider the hierarchical clustering with the Hamming distance, or the ``unanimity'' clustering, 
	($i$ and $i'$ are in the same cluster of $\C$ if and only if they are in the same cluster for each $\C^{(j)}$). This latter clustering is actually the one provided by Problem \eqref{eq:conv-clust:multi-dim}. 
	For more background on clustering aggregation, we refer for instance to \cite{Gionis07clustering,nguyen2007consensus,WANG2009668} and references therein.
\end{rmrk}

\subsection{Test procedure and its guarantees}
\label{subsection:test:procedure:multiD}

\subsubsection{Construction of the test procedure} \label{subsubsubsection:construction:p:dim}

The test procedure for the hypothesis $\bkappa^\top \bbeta = 0$ is constructed similarly as in Section \ref{subsection:construction:test}. We consider $p$ permutations $\sigma^{(1)}, \ldots , \sigma^{(p)}$ that provide the orderings of the columns of the $n \times p$ observation matrix $\Y$. As in Definition \ref{def:clust-seg}, we identify the $p$ clusterings $\C^{(1)} , \ldots, \C^{(p)}$ by their numbers of classes $K^{(1)} , \ldots , K^{(p)} \in [|n|]$ and by the right-limit sequences $\t^{(j)} \in \cT_{K^{(j)},n}$ for $j \in [|p|]$.

For $j\in [|p|]$, we consider the matrix $\bM(\t^{(j)}) \bP_{\sigma^{(j)}}$ of size $2(n-1) \times n$
and the vector $\lambda \m (\t^{(j)})$ of size $n$, defined in Lemma \ref{corPolyh}. Recall from Section \ref{section:one:dimensional case} that, if only the variable $j$ and its clustering $\C^{(j)}$ and order $\sigma^{(j)}$ were considered, then the conditioning event would be $\left\{\bM(\t^{(j)}) \bP_{\sigma^{(j)}} \Y_{.j} \leq \lambda \m(\t^{(j)})\right\}$.  

We then explicit the conditioning constraints in dimension $p$, corresponding to all the clusterings $\C^{(1)},\ldots,\C^{(p)}$ and orders $\sigma^{(1)},\ldots,\sigma^{(p)}$. 
We define the matrix $\scrM$ of size $2(n-1)p \times np$ in the following block-wise fashion. There are $p^2$ rectangular blocks (corresponding to dividing the rows into $p$ groups and the columns into $p$ groups). The block indexed by row-group $j$ and column-group $j'$ has size $2(n-1) \times n$. It is zero if $j \neq j'$ and it is equal to $\bM(\t^{(j)})$ if $j = j'$. 
Define also $\bD_{\sigma} $ as the $np \times np$ block diagonal matrix with $p$ diagonal blocks and block $j$ equal to $ \bP_{\sigma^{(j)}}$, for $j \in [|p|]$.
With these definitions, we have 
\begin{equation} \label{eq:polyhedral:clustering:multiD}
\scrM \bD_{\sigma}  \vec(\Y) 
=
\begin{pmatrix}
	\bM(\t^{(1)}) \bP_{\sigma^{(1)}} \Y_{.1} 
	\\
	\vdots \\
	\bM(\t^{(p)}) \bP_{\sigma^{(p)}} \Y_{.p}
\end{pmatrix}.
\end{equation}
We let $\lambda \scrm$ be the vector obtained by stacking the column vectors $\lambda \m(\t^{(j)}) $, $j \in [|p|]$, one above the other. 
The conditioning constraints in dimension $p$ are then $\left\{\scrM \bD_{\sigma}  \vec(\Y)  \leq \lambda \scrm\right\}$.

Consider a column vector $\bkappa$ of size $np$, that is allowed to depend on $(\t^{(j)},\sigma^{(j)}), j \in [|p|]$.
This includes the setting $\bkappa = \bkappa(\C^{(1)} , \ldots, \C^{(p)}) $ of Section \ref{subsection:test:after:aggregation}, with the additional mathematical flexibility that $\bkappa$ is allowed to depend on the orderings of the columns, besides their clusterings. 

Recall that $\bGamma$ is the $np \times np$ covariance matrix of $\vec(\Y)$. 
Note that in the definition of $\pval( \x,\t,\sigma)$ in Section \ref{subsection:construction:test} (one-dimensional case), the values of $\X$, $\bSigma$, $\boeta$, $\bM \bP_{\sigma}$ and $\lambda \m$ are sufficient to determine
%the invariant statistic $T( \X, \t,\sigma)$ in \eqref{eq:the:invariant:stat:oneD} and
the $p$-value $\pval( \x, \t,\sigma)$
(see Equation \eqref{eq:cond-p-value-1d} and Lemma \ref{lemma:expression:pvalue}).
Thus we can define the test statistic $\bkappa^\top \vec(\Y)$, 
%then the invariant statistic $T(\Y) = T(\Y,\t^{(1)} , \ldots , \t^{(p)},\sigma^{(1)} , \ldots , \sigma^{(p)})$ in the same way as $T(\X,\t,\sigma)$ in  \eqref{eq:the:invariant:stat:oneD} 
and consequently the
$p$-value $\pval(\y)$, for a $n \times p$ realization $\y$ of $\Y$, in the same way as $\pval( \x, \t,\sigma)$ in \eqref{eq:cond-p-value-1d}.
The explicit correspondence between the notation of the one-dimensional case and the present notation is 
given in Table~\ref{tab:corresp-notation-p}.

In the frame of Section \ref{subsection:test:after:aggregation}, it is interesting to highlight the impact of the aggregation of the one-dimensional clusterings $\C^{(1)} , \ldots, \C^{(p)}$ into the $p$-dimensional clustering $\C$. This aggregation has an impact on the vector $\bkappa$ defining the null hypothesis
$\bkappa^\top \bbeta = 0$ and the corresponding estimator
$\bkappa^\top \vec(\Y)$ of $\bkappa^\top \bbeta$. For instance, when $\bkappa$ is as in \eqref{eq:kappa}, 
then the final aggregation impacts the two clusters $\C_{k_1}$ and $\C_{k_2}$ which impact the vector $\bkappa$. Note that, in contrast, $\bkappa$ or $\C$ do not impact the polyhedral characterization of the conditioning $\left\{\scrM \bD_{\sigma}  \vec(\Y)  \leq \lambda \scrm\right\}$ in \eqref{eq:polyhedral:clustering:multiD}.
Indeed, the individual matrices $\bM(\t^{(j)}) \bP_{\sigma^{(j)}}$, $j \in [|p|]$, depend on the individual clusterings for each column of the data matrix $\Y$ (and their orders) but not on the final aggregation by hierarchical clustering. The same holds for the vector $\lambda \scrm$. 

The next section provides additional explanations on the computation of the $p$-value $\pval(\y)$, in the special case of independent variables, for the sake of exposition.

% The vector $\x$ of size $n$ of Section \ref{subsection:construction:test} corresponds to the vector $\vec(\y)$ of size $np$ here.
% The $n \times n$ matrix $\bSigma$ of  Section \ref{subsection:construction:test} corresponds to the $np \times np$ matrix $\bGamma$ here. 
% The vector $\boeta$ of Section \ref{subsection:construction:test} corresponds to $\bkappa$ here. 
% The matrix $\bM \bP_{\sigma}$ of Section \ref{subsection:construction:test} corresponds to the matrix $\bM \bD_{\sigma} $ here.
%  The vector $\lambda \m$ of Section \ref{subsection:construction:test} corresponds to the vector $\lambda \m$ here. 
\renewcommand{\arraystretch}{1.5}
\begin{table}
	\begin{tabular}{c|c|c|c|c|c|c|c|c|c}
		\hline
		$p=1$
		& $\x $ 
		& $\X$
		& $\bSigma $
		&  $\boeta $
		& $\bP_{\sigma} $
		& $\bM $
		& $\lambda \m $
		& $\Z$ 
		& $\bc$ \\ 
		size
		& $n$
		& $n$
		& $n \times n$
		&  $n$
		& $ n \times n$
		& $2(n-1) \times n$
		& $n$ 
		& $n$ 
		& $n$ \\
		\hline 
		$p>1$ 
		& $\vec(\y)$   
		& $\vec(\Y)$  
		& $\bGamma $
		& $\bkappa $
		& $\bD_{\sigma} $
		& $\scrM  $
		& $\lambda \scrm $
		& $\vec(\obZ)$ 
		& $\vec(\obc)$ \\
		size
		& $np$ 
		& $np$ 
		& $np \times np$
		& $np$
		& $np \times np$
		& $2(n-1)p \times np$
		& $np$ 
		& $np$ 
		& $np$ \\
		\hline
	\end{tabular}
	\caption{Correspondence between the notation of Section \ref{subsection:construction:test} (dimension one) and the notation of Sections~\ref{subsubsubsection:construction:p:dim} and \ref{subsubsubsection:example:p:dim} (dimension $p$).
	The notation $\vec(\obZ)$ and $\vec(\obc)$ is used in the proof of Proposition~\ref{proposition:example:p:dim} in Section~\ref{appendix:proof:multiD}.}
	\label{tab:corresp-notation-p}
\end{table}

\subsubsection{A detailed example: testing the signal difference along a variable $j_0$ with independent variables} 
\label{subsubsubsection:example:p:dim}

Consider testing the signal difference for the column $j_0$ between two clusters $\C_{k_1}$ and $\C_{k_2}$ in the multi-dimensional clustering $\C$, as in Example \ref{ex:hyp}. 
It is interesting to explicit the construction of the $p$-value in the special case of the matrix normal distribution (see Section \ref{section:context:and:objectives}) where $\bDelta$ is diagonal, that is the $p$ $n$-dimensional observation vectors corresponding to the $p$ variables are independent. For the sake of simplicity, let us even consider that $\bDelta = \bI_p$.

Observe first that the test statistic satisfies $\bkappa^\top \vec(\Y) = \boeta^\top \Y_{.j_0}$, where $\eta_i = \1_{ i \in \C_{k_1} } / |\C_{k_1}| -
 \1_{ i \in \C_{k_2} } /
 \linebreak[1]
  |\C_{k_2}|$. That is, the test statistic is constructed as it would be in the one-dimensional case (Section \ref{subsection:construction:test}), except that the one-dimensional clustering  $\C^{(j_0)}$ is replaced by the aggregated one $\C$. The variance of the test statistic (unconditional to the clusterings and orders of observations) is thus $ \boeta^\top \bSigma \boeta $ and is as in the one-dimensional case (up to the distinction between $\C^{(j_0)}$ and $\C$). Then, the next proposition specifies the computation of the $p$-value.

\begin{prpstn} \label{proposition:example:p:dim}
	In the context of Section \Rref{subsubsubsection:example:p:dim}, computing the $p$-value as described in Section \Rref{subsubsubsection:construction:p:dim} is equivalent to proceed as described in Section \Rref{subsection:construction:test} (one-dimensional case), with $\boeta $ defined by $\eta_i = \1_{ i \in \C_{k_1} } / |\C_{k_1}| - \1_{ i \in \C_{k_2} } / |\C_{k_2}|$ for $i \in [| n|]$, with $\x$ replaced by $\Y_{.j_0}$
	and with the conditioning set $\left\{\bM \bP_\sigma \X \leq \lambda\  \m\right\}$ replaced by $\left\{\bM(\t^{(j_0)}) \bP_{\sigma^{(j_0)}} \Y_{.j_0} \leq \lambda \m (\t^{(j_0)})\right\}$. 
\end{prpstn}

In Proposition \ref{proposition:example:p:dim}, the observations corresponding to the variables $j \neq j_0$, for which the average signal difference is not tested, have an impact on the clusterings $\C^{(j)}$, $j \neq j_0$, and thus have an impact on the multi-dimensional clustering $\C$ and thus on $\boeta$. Besides $\boeta$, these observations have no other influence on the construction of the $p$-value, which is computed only from $\Y_{.j_0}$ and its conditioning set
$\{ \bM(\t^{(j_0)}) \bP_{\sigma^{(j_0)}} \Y_{.j_0} \leq \lambda \m (\t^{(j_0)}) \}$ as in the one-dimensional case.
This fact can be interpreted in light of the general properties of conditioning and independence. Indeed, we are studying events of the form $E_j$ on $\Y_{.j}$, $j \in [|p|]$ and we are studying
a test statistic $\boeta(E_1,\ldots,E_p)^\top \Y_{.j_0}$ conditionally to these events. 
Here $E_j$ encodes the event corresponding to \eqref{eq:conditions:un} and \eqref{eq:conditions:deux} in Theorem \ref{theorem:equivalence:polyhedral:one:d} for variable $j$.
By independence of $\Y_{.j}$, $j \in [|p|]$,
the events $E_j$, $j \neq j_0$ simply have an influence on $\boeta$, while the event $E_{j_0}$ also has an impact on the conditional distribution of $\Y_{.j_0}$ given $E_{j_0}$.

\subsubsection{Conditional level}

For $j \in [|p|]$, let $\widehat{\B}_{.j}$ be obtained from \eqref{eq:conv-clust:j}. The next proposition is similar to Proposition \ref{prop:level:conditional} and proves that the $p$-value $\pval(\Y)$ in Section \ref{subsubsubsection:construction:p:dim}
is uniformly distributed, conditionally to the one-dimensional clusterings and orders, when the null hypothesis is true. 
We remark that in the context of Section~\ref{subsection:test:after:aggregation}, this implies that the $p$-value is also uniformly distributed conditionally to the $p$-dimensional clustering obtained by aggregation, when the null hypothesis is true.

\begin{prpstn} \label{prop:level:conditional:multidim}
	Consider $p$ fixed permutations $\sigma^{(1)}, \ldots , \sigma^{(p)}$ of $[|n|]$. Let $K^{(1)} , \ldots , K^{(p)} \in [|n|]$. 
	For $j \in [|p|]$, let $\t^{(j)} \in \cT_{K^{(j)},n}$ and consider the clustering $\C^{(j)}$ associated to $(\t^{(j)},\sigma^{(j)})$ by Definition~\Rref{def:clust-seg}.
	
	Consider a fixed non-zero vector $\bkappa \in \mathbb{R}^{np}$ (that is only allowed to depend on $(\t^{(j)},\sigma^{(j)}), j\in [|p|]$). 
	Assume that 
	\[
	\bkappa^\top \bbeta = 0.
	\]
	Assume that with non-zero probability, the event 
	\[
	E ~ ~ ~ 
	:= 
	~ ~ ~
	\left\{ \text{for $j\in [|p|]$, $\C^{(j)}$ is the clustering given by $\widehat{\B}_{.j}$ and
		$Y_{\sigma^{(j)}(1)j}  \geq \dots \geq Y_{\sigma^{(j)}(n)j} $}
	\right\}
	\]
	holds. 
	Assume also that the $np \times np$ matrix $\bGamma$ is invertible.
	Then, conditionally to $E$, $\pval( \Y)$ is uniformly distributed on $[0,1]$ under the null hypothesis.
\end{prpstn}

\subsubsection{Unconditional level}

The unconditional guarantee is similar to that of Proposition \ref{proposition:unconditional:level} for the one-dimensional case. In particular, here we also introduce the subset $\mathcal{E}$ on which the null hypothesis is well-defined. 

\begin{prpstn} \label{proposition:unconditional:level:multi-dim}
	Let $\mathcal{E}$ be a subset of the set of all possible values of $
	(\t^{(j)},\sigma^{(j)})_{j \in [|p|]}$ in Proposition~\Rref{prop:level:conditional:multidim}. 
	Consider a deterministic function $\bkappa : \mathcal{E} \to \mathbb{R}^{np}$, outputing a non-zero column vector. 
	Assume that $\bGamma$ is invertible. For $j \in [|p|]$, let $\widehat{\B}_{.j}$ be obtained from \eqref{eq:conv-clust:j}.  Let also $S^{(j)} = S^{(j)}(\Y_{.j})$ be the random permutation obtained by the order of $\Y_{.j}$: $Y_{S^{(j)}(1)j} \geq  \cdots \geq Y_{S^{(j)}(n)j}$ (uniquely defined with probability one). Let $\C^{(j)}(\Y_{.j}) = \C^{(j)}$ be the random clustering given by $\widehat{\B}_{.j}$ (Definition~\Rref{def:clustering}). 
	Let $\T^{(j)}(\Y_{.j}) = \T^{(j)} \in \cT_{K^{(j)},n}$ be the random vector (with random $K^{(j)}(\Y_{.j}) = K^{(j)}$), such that $(\T^{(j)},S^{(j)})$ yields $\C^{(j)}$ as in Definition \Rref{def:clust-seg}. 
	
	Assume that
	\[
	\mathbb{P} 
	\left(
	(\T^{(j)},S^{(j)})_{j \in [|p|]} \in \mathcal{E}
	,
	\bkappa( (\T^{(j)},S^{(j)})_{j \in [|p|]})^\top \bbeta
	= 0
	\right) 
	>0.
	\]
	
	Then, conditionally to the above event, $\pval( \Y)$ is uniformly distributed on $[0,1]$. 
\end{prpstn}

\subsection{Numerical experiments}
\label{sec:numer-exper-p}

In this section, we describe the numerical experiments that we have performed in order to illustrate the behaviour of our post-clustering testing procedure for $p>1$.
The code to reproduce these numerical experiments and the associated figures is available from \url{https://plmlab.math.cnrs.fr/pneuvial/poclin-paper}.

We consider the specific case where $\Y$ is distributed from a matrix normal distribution $\mathcal{MN}_{n\times p}(\mathbf{u},\linebreak[1]\bSigma,\bDelta)$ (see Section \ref{section:context:and:objectives}) with
$p=3$, 
$\mathbf{u} = \left(
\begin{array}{c c c}
	\nu \bone_{n/2}
	& \bzero_{n/2}
	&   \bzero_{n/2} \\
	-\nu \bone_{n/2}
	& \bzero_{n/2}
	&   \bzero_{n/2}
\end{array}
\right)$ with $\nu\in\{0,1,2,5\}$, $\bSigma=\bI_n$, and
$\bDelta=\left(
\begin{array}{c c c}
	1 & 0 & \rho\\
	0 & 1 & 0\\
	\rho & 0 & 1
\end{array}\right)$ with $\rho\in\{0,0.1,0.5\}$.
%$\rho\in\{0,0.3,0.5\}$.

\begin{figure}[htb]
	\includegraphics[width=0.95\textwidth]{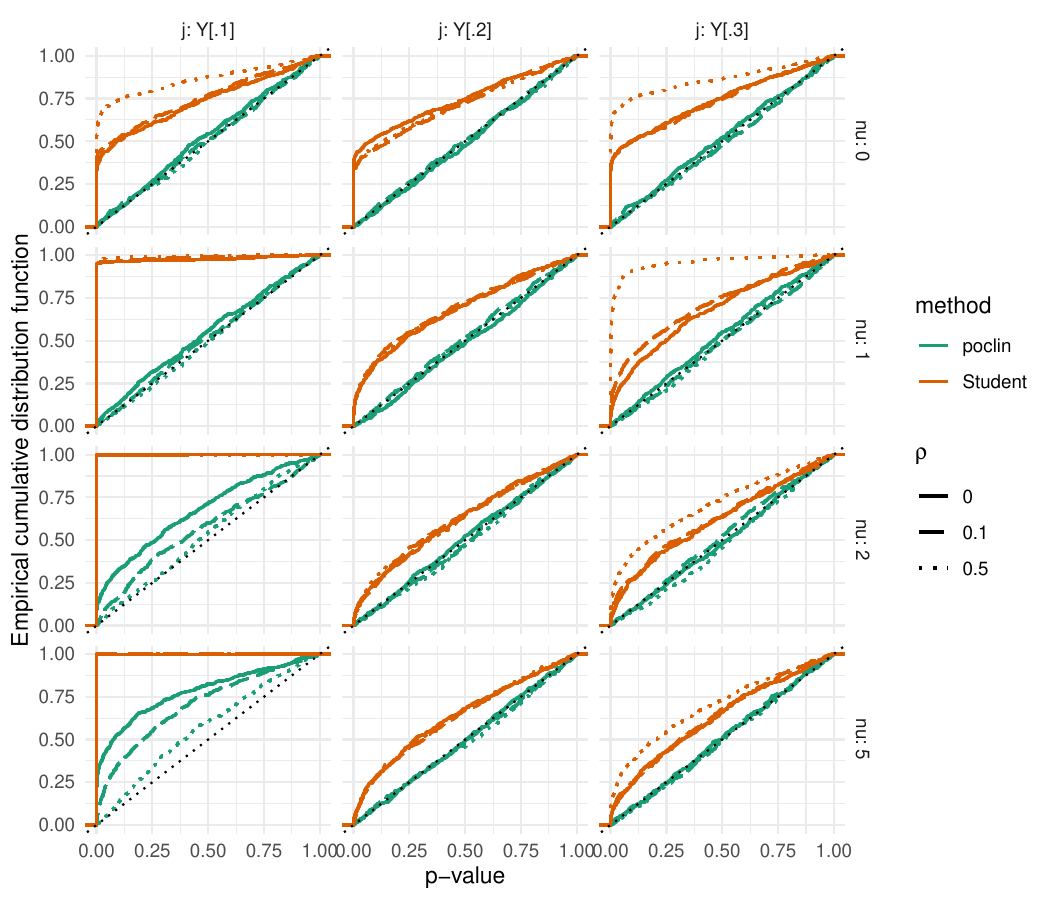}
	\caption{The empirical cumulative distribution function of the $p$-values across $500$ experiments for $n=100$ with our method \texttt{poclin} (in green) and the Student test (in orange). Each column corresponds to a variable $j$, each row to a value of $\nu$ and each line type to a value of $\rho$.}
	\label{fig:cdf-scenario1-n100}
\end{figure}

\begin{figure}[htb]
	\includegraphics[width=0.95\textwidth]{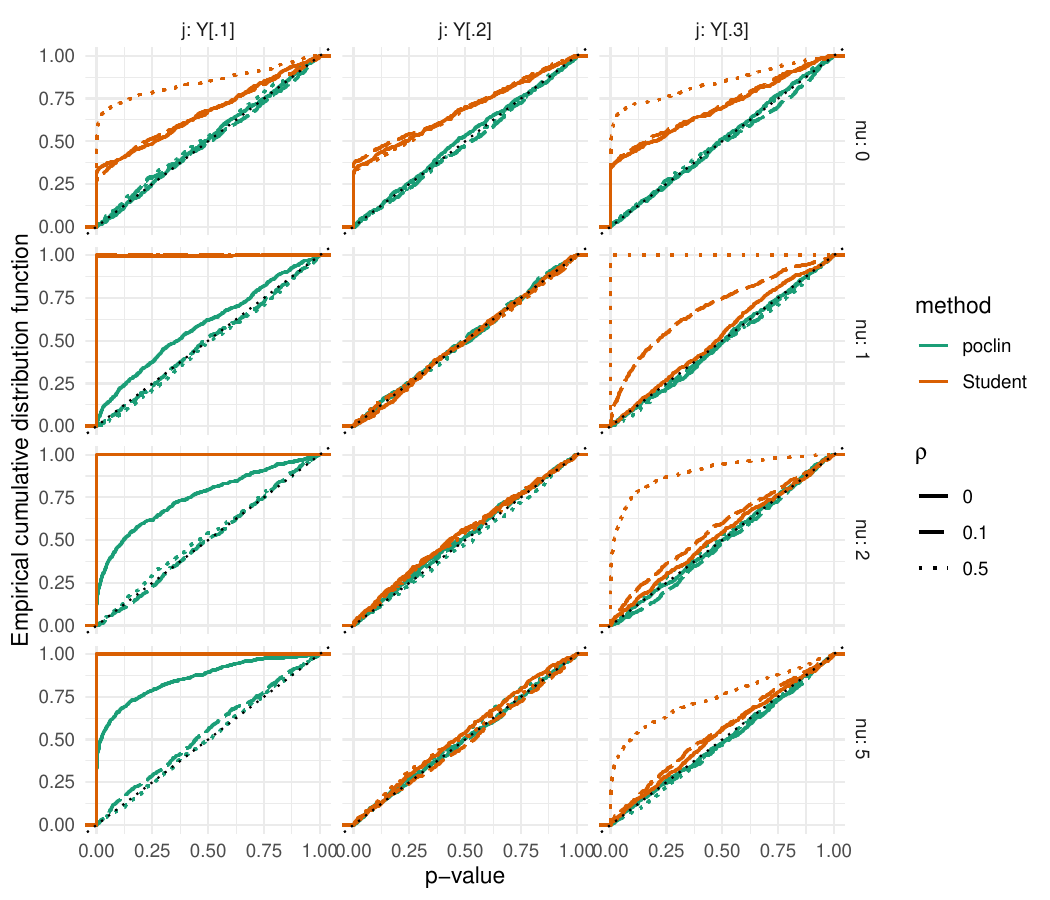}
	\caption{The empirical cumulative distribution function of the $p$-values across $500$ experiments for $n=1000$ with our method \texttt{poclin} (in green) and the Student test (in orange). Each column corresponds to a variable $j$, each row to a value of $\nu$ and each line type to a value of $\rho$.}
	\label{fig:cdf-scenario1-n1000}
\end{figure}

We obtain $K=2$ clusters by aggregating one-dimensional convex clusterings obtained for a given value of $\lambda$, as explained in Section~\ref{subsection:test:after:aggregation}. 
For each variable $j\in\{1,2,3\}$, we want to compare the means of the two clusters. This corresponds to the test of the null hypothesis $\bkappa^\top \bbeta = 0$, where $\bkappa$ is defined by \eqref{eq:kappa} (see Example \ref{ex:hyp}).
We compare our procedure with $\lambda=0.016$ (resp. $\lambda=0.0025$) for $n=100$ (resp. $n=1000$) and the two-sample Student and Wilcoxon tests, as implemented in the \texttt{R} functions \texttt{t.test} and \texttt{wilcox.test}. Since the results of the Student and Wilcoxon tests were very similar, only the results for the Student test are shown below. This choice of $\lambda$ ensures to have at least two clusters under the null hypothesis with high probability, as explained in Section \ref{sec:numerical-experiments-1d} and Section~\ref{sec:calibration-lambda}.
The empirical cumulative distribution function of the $p$-values $\pval(\y)$ across $500$ experiments is represented for different values of the simulation parameters in Figures~\ref{fig:cdf-scenario1-n100} and~\ref{fig:cdf-scenario1-n1000} for $n=100$ and $n=1000$, respectively.
For each parameter combination, the $p$-value distribution of the proposed method (in green) is compared to that of the two-sample Student test (in orange) for all three variables $\Y_{.j}$, for $j=1,2,3$ (in columns). Each row corresponds to a value of $\nu$ and each line type corresponds to a value of $\rho$. 


First, the clustering procedure described in Section~\ref{subsection:test:after:aggregation} works reasonably well in this setting.
Indeed, for the variable $\Y_{.1}$, the absolute value of the difference between the true means of the estimated clusters (obtained as $\bkappa^\top \bbeta$) is generally close to the true value of the signal (that is $2 \nu$), see Figure~\ref{fig:signal-scenario1} in Section~\ref{apx:p:dimensional case}.

The proposed test controls the type I error rate: in all situations where there is no signal (that is, for $\nu=0$ or $j \in \{2, 3\}$), the empirical $p$-value distribution is close to the uniform distribution on $[0,1]$ ($y=x$).
Under the alternative hypothesis (i.e. for $j=1$ and $\nu>0$), our proposed test is able to detect some signal for $\nu \geq 2$. For $\nu=1$ the signal is too small to be detected. 

In contrast, the naive Student test yields severely anti-conservative $p$-values in absence of signal.
This test is naturally much more sensitive than our proposed test.
However, one cannot compare the power of the two tests, since the Student test fails to control the type I error.
It should also be noted that while our proposed test is reasonably powerful when $\rho=0$ (corresponding to independent variables, as studied in Section \ref{subsubsubsection:example:p:dim}), it loses power as $\rho$ increases.
Our proposed method gains some power for $\rho=0$ as $n$ increases from $100$ to $1000$.

However, when $\rho > 0$, the power of our test deteriorates as $n$ increases.
Our numerical investigations indicate that this is certainly due to the impact of the third variable on the conditioning event $\left\{\scrM \bD_{\sigma}  \vec(\Y)  \leq \lambda \scrm\right\}$. 
More precisely, the truncation interval $[\V^{-}, \V^{+}]$
(see Section \ref{subsection:construction:test} and its extension to the multidimensional case in Section \ref{subsubsubsection:construction:p:dim}) is substantially reduced due to the order constraint in \eqref{eq:equi:conditions:trois} for the third variable. 
Note that when $\rho=0$ and we test for the first variable, the third variable does not impact the conditioning and thus does not impact $[\V^{-}, \V^{+}]$, as explained in Section \ref{subsubsubsection:example:p:dim}. This is why this specific issue arises for $\rho > 0$.

For $n=1000$, the Student test is able to distinguish the signal from the noise when $\rho=0$ and actually becomes well-calibrated for $\Y_{.2}$ when $\nu \neq 0$.
However, due to the correlation between $\Y_{.1}$ and $\Y_{.3}$, the Student test is anti-conservative for $\Y_{.3}$ when $\rho > 0$.

\section{Discussion}\label{sect-discuss}

We first provide an overview of our contributions, and then we discuss various specific aspects of them and various remaining open questions.

\subsection{Overview of the contributions}

Selective inference, in the post-clustering context, is a challenging problem and statistical guarantees could be obtained for it only in the recent years, see the references  provided in Section \ref{section:context:and:objectives}. 
In this paper, we suggest a solution based on exhibiting polyhedral conditioning sets for Gaussian vectors, extending a line of work that has proved to be very successful in other statistical contexts, especially for regression models. This line of work was pioneered by \cite{Lee16} and then developed by \cite{panigrahi2022approximate,Tibshirani2018uniform}, among others.

Nevertheless, extending the existing approaches from regression models to clustering models is challenging. As such, the proofs we provide require innovations (for instance for Theorems \ref{theorem:equivalence:polyhedral:one:d} and \ref{thm:regularization:path}). Furthermore, obtaining polyhedral conditioning sets is made possible by focusing on intermediate one-dimensional convex clustering optimization problems based on $\ell_1$ penalties (see \eqref{eq:conv-clust:j}).
In the end, we provide the following workflow for selective inference post-clustering. 

(a) We characterize a one-dimensional clustering by polyhedral constraints on the observation vector (Section~\ref{subsection:polyhedral:caracterization:}). 

(b) As a by-product, we provide a regularization path algorithm to implement this clustering (Section~\ref{subsection:regularization:path:one-dimensional}). The computational efficiency of this algorithm is demonstrated numerically, also in comparison with other existing procedures.

(c) Following \cite{Lee16}, from the polyhedral constraints, we obtain a test procedure which is conditionally and unconditionally valid post-clustering (Section \ref{subsection:test:one:dimensional}). The procedure enables to test the nullity of any linear combination of the unknown mean vector, provided this combination only depends on the clustering (and on the order of the observations). In particular, it is possible to test for the significance of the signal difference between two  clusters as in Example~\ref{ex:hyp} (see \eqref{eq:example:comparison:signal:singleD}). 
Although we do not develop it in this paper, 
confidence intervals for the above linear combination
can be constructed from our test procedure, similarly as in \cite{Lee16}. Numerical experiments (Section \ref{sec:numerical-experiments-1d}) confirm the validity of the test procedure, and indicate that it has power to detect cases where the clustering on the observation vector was able to cluster the unknown mean vector as well into inhomogeneous groups. Section \ref{sec:comp-gen-lasso} also highlights the benefit of the test procedure, compared to broader ones based on the generalized Lasso.

(d) We suggest to aggregate one-dimensional clusterings to form a single multi-dimensional clustering for the data matrix.
Our above contributions can thus be naturally leveraged to obtain a valid test procedure, posterior to this  multi-dimensional clustering (Section \ref{subsection:test:procedure:multiD}). 
In particular, we can test the significance of the signal difference between clusters along a specific variable, as in Example \ref{ex:hyp}. This feature could be beneficial in potential applications to single-cell RNA-seq data, since in this context, testing along a specific variable enables to study genes expressions individually~\cite{hivert2024post,neufeld2024inference,neufeld2023negative}. 
%It is also a welcome complement to related references, in particular \cite{gao2022}, that focuses on testing the global nullity of the signal mean difference vector across two clusters, rather than considering individual components (i.e. variables). 

\noindent This workflow (a)-(d) depends on a regularization parameter $\lambda$ that should not be data-driven (see Section~\ref{sec:discussion-known-gamma} below). From a practical point of view, we provide a procedure to choose $\lambda$ in a non data-driven way, from a choice of the covariance matrix, see Sections \ref{sec:numerical-experiments-1d} and \ref{sec:numer-exper-p}, and Section~\ref{sec:calibration-lambda}. 

In (d), similarly as in the one-dimensional case, we provide numerical experiments (Section \ref{sec:numer-exper-p}) highlighting that inference post-clustering is challenging. Statistical procedures that do not account for the data-driven nature of the clustering are strongly anti-conservative. Indeed, the standard Student test wrongly indicates signal differences across clusters in many cases where there is actually no difference. 
In contrast, these numerical experiments confirm the validity of our test procedure. We also study its power when the clustering procedure successfully yields clusters with significant signal difference for individual variables. 
While the power is satisfactory for independent variables, we have found that dependence can overly restrict the conditioning set, yielding a loss of power.
It is an important perspective for future work to attempt to alleviate this issue.

Note that the numerical experiments are focused on the hierarchical-clustering-based aggregation of one-dimensional clusterings, as described in Section \ref{subsection:test:after:aggregation}. In future investigations, it would be relevant to quantify the benefit brought by alternative aggregation methods. Indeed, a flexibility of our framework is that our statistical guarantees hold for any aggregation procedure.

\subsection{Benefits of the test procedure in well- and misspecified clustering problems} \label{subsection:benefits:test}

For simplicity, let us focus on the one-dimensional case of Section \ref{section:one:dimensional case}, with the observation vector $\X \sim \mathcal{N} ( \bmu , \bSigma  )$. The discussion of the multi-dimensional case of Section \ref{section:p:dimensional case} would be similar. The clustering problem can be considered as well-specified if there are clusters of indices for the mean vector $\bmu$ with equal values, corresponding to a Gaussian mixture setting (see for instance \cite{heinrich2018strong,laurent2016non,McLachlanfinite,nguyen2013convergence} for expositions and recent contributions on mixture models). In the well-specified case, there are thus intrinsic classes of the observations and it is natural to aim at recovering them.  

Consider for the sake of discussion that $n/2$ components of $\bmu$ are zero and the other $n/2$ components are one (there are two intrinsic classes) and that the clustering procedure yields two clusters $\C_1$ and $\C_2$ of equal size. Then if the null hypothesis $(2/n )\sum_{i \in \C_1} \mu_i = (2/n ) \sum_{i \in \C_2} \mu_i$ is rejected by our test procedure, it means that one empirical cluster contains a strict majority of individuals from one intrinsic class, and vice versa for the second cluster. If our test procedure is extended to yield a confidence interval on $(2/n ) \sum_{i \in \C_1} \mu_i - (2/n )\sum_{i \in \C_2} \mu_i$ showing that with high probability this quantity is larger than some $\delta \in (0,1)$, then one can see that the first empirical cluster contains at least $n (\delta+1)/4$ observations from an intrinsic class (corresponding to mean one; and conversely for the second cluster). Hence, generally speaking, for a well-specified clustering problem with intrinsic classes, our test procedure is relevant to recover these classes, similarly as statistical procedures that are dedicated to finite mixture problems, see the references given above.

On the other hand, the clustering problem can be considered as misspecified when the $n$ components of $\bmu$ are two-by-two distinct. In this case one can consider that there are no intrinsic classes.
Nevertheless, providing tests or confidence intervals on the same quantity $(2/n ) \sum_{i \in \C_1} \mu_i - (2/n )\sum_{i \in \C_2} \mu_i$ as before enables to assess if the clustering procedure was able to cluster the unknown mean vector, besides the random/noisy observations. Hence, a benefit of the post-clustering framework considered here is that it is meaningful both in well- and misspecified settings. 
A similar discussion can be made in the related context of selective inference in regression settings, see in particular \cite{bachoc2019valid,berk2013valid,buja2019models}. 

\subsection{Known covariance matrix and fixed $\lambda$}
\label{sec:discussion-known-gamma}

As pointed out above, we assume the covariance matrix ($\bSigma$ in Section \ref{section:one:dimensional case} and $\bGamma$ in Section \ref{section:p:dimensional case}) to be known and the tuning parameter $\lambda$ to be fixed. These two assumptions are necessary for our statistical guarantees in Sections \ref{section:one:dimensional case} and \ref{section:p:dimensional case}. Indeed, the obtention of these guarantees relies first on exhibiting a Gaussian vector constrained to a  polyhedron. Then, the Gaussian vector is decomposed into a linear combination (corresponding to the statistical hypothesis to test) and an independent remainder. This two-step strategy corresponds in particular to Lemma \ref{corPolyh} and Proposition \ref{prop:polyhedral:lemma} in the one-dimensional case. It was previously suggested by \cite{Lee16} in the related context of post-selection inference for the Lasso model selector, with Gaussian linear models. 

Obtaining a polyhedron in the first step relies on $\lambda$ not depending on the data, and computing the decomposition in the second step relies on knowing the covariance matrix. 
Broadly speaking, in the selective inference context, it is relatively common to assume independent individuals ($\bSigma = \bI_n$) and known covariance $\bDelta$ for the variables (see the matrix normal case in Section~\ref{section:context:and:objectives}), or fixed tuning parameters, in order to obtain rigorous mathematical guarantees.
This is indeed the case in  \cite{Lee16} mentioned above, but also for instance in \cite{gao2022}.
In this latter reference, the covariance matrix $\bDelta$ is assumed to be known and spherical ($\bDelta = \sigma^2 \bI_p$) for most of the theoretical results, except for the
asymptotic results in Section 4.3 for the case of a conservative variance estimator. 
In \cite{yun2023selective}, an adaptation of \cite{gao2022} is proposed to avoid covariance estimation, in the specific case of unknown spherical covariance.
Also, data thinning procedures, for instance in \cite{neufeld2024data}, require full knowledge of the data distribution in order to produce independent parts, where the independence property enables valid statistical inference. The negative impact of the misspecification of the data distribution is demonstrated numerically in \cite{hivert2024running}.

In our setting,
obtaining theoretical guarantees (finite-sample or asymptotic) with an estimated covariance matrix or a data-dependent tuning parameter is of course an important problem for future work. 
Note that relaxing the assumption of known covariance matrix can yield identifiability issues, because the mean vector $\bbeta$ is unrestricted (see also the discussion of misspecified clustering problems in Section~\ref{subsection:benefits:test}). These identifiability issues boil down to the fact that multiple pairs of mean vector and covariance matrix can ``explain'' the same dataset. Studying which minimal assumptions circumvent these identifiability issues is thus an important problem in the prospect of extending this work to an estimated covariance matrix. 

\subsection{Choice of the $\ell_1$ norm in the multi-dimensional convex clustering problem~\eqref{eq:conv-clust:multi-dim}}
\label{subsection:choice:lone:norm}

Our test procedure and its statistical guarantees for the multi-dimensional case rely on aggregating one-dimensional clusterings. As discussed in Remark \ref{rem:clustering-aggregation}, solving Problem \eqref{eq:conv-clust:multi-dim} with the multi-dimensional $\ell_1$ norm penalization boils down to one such aggregation. Hence, our procedure and guarantees apply to multi-dimensional convex clustering with $\ell_1$ penalization. 

One can see that our arguments, and crucially the proof of Theorem \ref{theorem:equivalence:polyhedral:one:d}, cannot be applied directly to convex clusterings obtained by replacing the $\ell_1$ penalization by a more general $\ell_q$ one, $q >0$, and especially by the $\ell_2$ one. In fact, we view the following question as an important open problem: is it possible to characterize the set of observation matrices $\Y$, such that Problem \eqref{eq:conv-clust:multi-dim}, with the $\ell_1$ penalization replaced by the $\ell_q$ one, yields a given clustering, with polyhedral sets or other tractable sets? 

Nevertheless, we note that the $\ell_1$ penalization in Problem \eqref{eq:conv-clust:multi-dim} has computational benefits. Indeed, the problem is separable, and for each subproblem, we have obtained an exact regularization path in Section \ref{subsection:regularization:path:one-dimensional} that stops after a maximal number of iterations known in advance. To our knowledge, such a favorable regularization path is not available for a general $\ell_q$ penalization. In agreement with this, the reference \cite{hocking2011} (from 2011) concludes that Problem \eqref{eq:conv-clust:multi-dim} can be readily solved for thousands of data points, while if the $\ell_1$ penalization is replaced by the $\ell_q$ one, this is the case for (only) hundreds of data points.

\subsection{Comparison with data thinning strategies}

For the problem of post-clustering inference, data thinning/data fission strategies \cite{leiner2023data-fission,dharamshi2024generalized,neufeld2024data,neufeld2023negative}
consist in splitting each observation of the dataset into two stochastically independent observations, keeping the same indexing of individuals as the original dataset (there exist extensions splitting into more than one observations).
%For the problem of post-clustering inference, data splitting (or data fission, data thinning) strategies \cite{dharamshi2024generalized,neufeld2024data,zhang2019}
%consist in separating the dataset into two stochastically independent ones, keeping the same indexing of individuals as the original dataset. 
Then, a clustering can be computed from the first dataset and then applied to the second dataset. By independence, the distribution of a post-clustering statistic of interest (for instance the difference of average between two classes for a variable, in view of studying \eqref{eq:example:comparison:signal:multiD}) on the second dataset remains simple. 
For instance if the original dataset is Gaussian, this distribution remains Gaussian conditionally to the clustering. Hence, a benefit of these strategies
compared to our approach is a simplicity of implementation. Furthermore, any clustering procedure can be used. 

On the other hand, with data thinning, conclusions are provided for a clustering computed on a dataset that differs from the original one. Hence, the conclusions of data thinning approaches might be more difficult to interpret for practitioners, compared to those of the present work, since these conclusions do not apply to the clustering that they would compute on the original dataset.  

Note also that data thinning and our approach share two similar difficulties. First, they share hyperparameters that should not be data-driven for the statistical guarantees. Indeed, 
with data thinning we need to fix the splitting mechanism to generate the two datasets above. Similarly, we fix the regularization parameter $\lambda$ in \eqref{eq:conv-clust:multi-dim}. Second, considering Gaussian data, the covariance matrix should be known for data thinning and our approach, as already discussed in Section \ref{sec:discussion-known-gamma}.

\subsection{On conditioning by the orders}
\label{section:cond:by:order}

Let us consider the one-dimensional setting (Section \ref{section:one:dimensional case}) for simplicity of exposition. A similar discussion could be made for the multi-dimensional case as well. 
Our test procedure is valid conditionally to both the clustering and the order of observations, see Proposition \ref{prop:level:conditional}, and our discussion at the beginning of Section \ref{subsection:polyhedral:caracterization:}. Being valid conditionally to the clustering can be considered as a desirable statistical feature, since the clustering is an object of interest in itself (see also the discussion before Proposition \ref{prop:level:conditional}). However, being valid conditionally to the order is more a by-product of our approach than a desirable statistical feature. Indeed, in order to obtain a polyhedral set with a tractable number of linear pieces ($2(n-1)$) in Theorem \ref{theorem:equivalence:polyhedral:one:d}, it was necessary in the proof to condition by the observation order. Importantly, the constraint \eqref{eq:equi:conditions:deux} is not a linear constraint on the observation vector if the order is not fixed. 

If a test procedure could be derived by only conditioning by the clustering, this test could have more power than the one we obtain in Section \ref{subsection:test:one:dimensional}, which is an interesting perspective for future work.
In other words, is it possible that we pay a price when conditioning by the order of observations? 
In the related regression context, a similar phenomenon occurs in \cite{Lee16}. There, a first test procedure is obtained by conditioning by the selected variables and a second one is obtained by conditioning by the selected variables and the signs of the coefficients. The first procedure has a computational cost that is exponential in the number of variables, but is more powerful. The second procedure has a small computational cost. In Section 6 of \cite{Lee16}, it is written on this point that ``one may be willing to sacrifice statistical efficiency for computational efficiency''.

%\begin{supplement}
%\stitle{Proofs and additional material}
%\sdescription{
%	The Supplementary Material contains the proofs, additional material regarding the computational aspects of convex clustering and additional numerical illustrations.}
%\end{supplement}

%%%%%%%%%%%%%%%%%%%%%%%%%%%%%%%%
%%%%%%%%%%%%%%%%%%%%%%%%%%%%%%%%
\appendix

\section{Technical lemmas and their proofs}
\label{section:technical:lemmas:and:proofs}

\begin{lmm}
	\label{lm:increasing-hat-B:least:square}
	Consider a fixed $\x = (x_1 , \ldots , x_n) \in \mathbb{R}^n$. Let, for $\B = (B_1,\ldots,B_n) \in \R^{n }$,
	\[
	R(\B) = 	||\B - \x||_2^2.
	\]
	Then, for $i,i'\in [|n|], i \neq i'$ such that $x_i = x_{i'}$, if $\B$ is such that $B_i \neq B_{i'}$, replacing $B_i$ and $B_{i'}$ by $(B_i + B_{i'}) /2$ strictly decreases $R(\B)$. Furthermore,  for $i,i'\in [|n|], i \neq i'$ such that $x_i < x_{i'}$, if $B_i > B_{i'}$, exchanging $B_i$ and $B_{i'}$ in $\B$ strictly decreases $R(\B)$.
\end{lmm}

\begin{proof}[Proof of Lemma~\ref{lm:increasing-hat-B:least:square}]
	In the first case, we compute the change of $R(\B)$,
	\begin{align*}
		\text{before} - \text{after} 
		= (B_i - x_i)^2
		+ (B_{i'} - x_i)^2
		-2 \left( \frac{B_i+B_{i'}}{2} - x_i \right)^2 
	\end{align*}
	which is strictly positive by strict convexity and because $B_i \neq B_{i'}$.
	In the second case,
	\begin{align*}
		\text{before} - \text{after} 
		& = (B_i - x_i)^2
		+(B_{i'} - x_{i'})^2
		-	(B_i - x_{i'})^2
		- (B_{i'} - x_i)^2
		\\
		& =
		- 2 B_i  x_i
		-2B_{i'} x_{i'}
		+2	B_i  x_{i'}
		+2 B_{i'} x_i
		\\
		& = 
		2	(B_i - B_{i'})
		(x_{i'} - x_i)
		\\
		& >0.
	\end{align*}
\end{proof}

\begin{lmm} \label{lem:linearization:convex}
	Let $k \in \mathbb{N}$. Let $f : \R^k \to \R$ be convex and continuously differentiable. Let $g : \R^k \to \R$ be convex and continuous. Let $\x \in \R^k$. 
	For a continuously differentiable function $\psi : \R^k \to \R$, we let $\mathrm{Lin}_{\x} (\psi)$ be the function $\t \mapsto \nabla_{\psi}(\x)^\top (\t-\x) $, letting $\nabla_{\psi}(\x)$ be the gradient of $\psi$ at $\x$.
	Then $\x $ is a minimizer of $\mathrm{Lin}_{\x} (f) + g$ if and only if $\x $ is a minimizer of $f + g$.  
\end{lmm}

\begin{proof}[Proof of Lemma~\ref{lem:linearization:convex}]
	For a convex function $\phi$, $\x$ is a minimizer of $\phi$ if and only if, for any $\bv \in \mathbb{R}^k$,
	\[
	\lim_{\substack{u \to 0 \\ u >0}}
	\frac{\phi( \x+ u \bv ) - \phi(\x)}{u}
	\geq 0. 
	\]
	For any $\bv \in \R^k$, the above limit is identical when $\phi = f+g$ and when $\phi = \mathrm{Lin}_{\x} (f) + g$. Hence the above limit is non-negative  when $\phi = f+g$ if and only if  it is non-negative when $\phi = \mathrm{Lin}_{\x} (f) + g$. 
\end{proof}

\begin{lmm} \label{lemma:condition:obs:absolute:value}
	Let $n \in \mathbb{N}$ and $(a_1 , \ldots, a_n)\in \R^n$ with $\sum_{i=1}^n a_i = 0$. Then the function 
	\begin{align*}
		g :  \R^n & \to \R
		\\
		(u_1 , \ldots , u_n )
		& \mapsto 
		\sum_{i=1}^n a_i u_i 
		+
		\sum_{ \substack{ i,i'=1 \\ i < i' }}^n 
		|u_i - u_{i'}|
	\end{align*}
	is minimal at $0$ if and only if, with $a_{[1]} \leq \dots \leq a_{[n]}$ the ordered values of $a_1 , \ldots , a_n$, for $\ell \in [| n-1|]$, 
	\[
	\sum_{i=1}^\ell
	a_{[i]} 
	+
	\ell 
	(n - \ell)
	\geq 0.
	\]
\end{lmm}

\begin{proof}[Proof of Lemma~\ref{lemma:condition:obs:absolute:value}]
	We write $b_1 \leq \dots \leq b_n$ for the ordered values of $a_1 , \ldots , a_n$. Then $g$ is minimal at $0$ if and only if $h$ is minimal at $0$ with 
	\begin{align*}
		h :  \R^n & \to \R
		\\
		(u_1 , \ldots , u_n) 
		& \mapsto 
		\sum_{i=1}^n b_i u_i 
		+
		\sum_{ \substack{ i,i'=1 \\ i < i' }}^n 
		|u_i - u_{i'}|. 
	\end{align*}
	
	The minimum of $h$ is reached when $u_1 , \ldots, u_n$ satisfy $u_1 \geq \dots \geq u_n$. Indeed if there is $i<i'$ with $u_i <u_{i'}$, 
	we can swap $u_i$ and $u_{i'}$ which lets the sum of absolute values unchanged and changes the linear combination as
	\[
	\text{before} - \text{after} 
	= 
	b_i u_i + b_{i'} u_{i'} -b_i u_{i'} - b_{i'} u_i 
	=
	( b_i - b_{i'} ) ( u_i - u_{i'}  ) 
	\geq 0.
	\] 
	We can do this swap each time there is $i <i'$ with $u_i < u_{i'}$, until we have $u_1 \geq \dots \geq u_n$ and $h$ has not been increased.
	Hence, to minimize $h$ it is sufficient to consider $u_1 \geq \dots \geq u_n$. 
	
	Let $v_{\ell} = u_{\ell} - u_{\ell+1} \geq 0$ for $\ell \in [| n-1 |]$.
	% These $u_1 \geq \dots \geq u_k$ are of the form $u_k, u_k + v_{k-1} , \ldots , u_k + v_{k-1}+  \dots +v_1$, with $u_k \in \R$ and $v_1, \ldots, v_{k-1} \geq 0 $. 
	We have 
	
	\begin{align*}
		\sum_{i=1}^n b_i u_i 
		& = 
		u_n \sum_{i=1}^n b_i 
		+
		\sum_{\ell=1}^{n-1} 
		v_{\ell} 
		\left(
		\sum_{i=1}^{\ell} b_i
		\right)
		\\
		& = 
		\sum_{\ell=1}^{n-1} 
		v_{\ell} 
		\left(
		\sum_{i=1}^{\ell} b_i
		\right)
	\end{align*}
	since by assumption $\sum_{i=1}^n a_i = \sum_{i=1}^n b_i =0$. We also have
	\begin{align*}
		\sum_{\substack{ i,i'=1 \\ i < i' }}^n
		|u_i - u_{i'}|
		& = 
		\sum_{\substack{ i,i'=1 \\ i < i' }}^n
		\sum_{\ell=i}^{i'-1}v_{\ell}
		\\
		& =
		\sum_{\ell=1}^{n-1}v_{\ell} \ell (n-\ell).  
	\end{align*}
	Therefore, 
	\begin{align*}
		\sum_{i=1}^n b_i u_i 
		+
		\sum_{\substack{ i,i'=1 \\ i < i' }}^n
		|u_i - u_{i'}|
		& =  
		\sum_{\ell=1}^{n-1} 
		v_{\ell} 
		\left(
		\sum_{i=1}^{\ell} b_i + \ell (n-\ell)
		\right).
	\end{align*}
	Hence $h$ is minimal at $0$ if and only if, for $\ell \in [|n-1|]$, $\sum_{i=1}^\ell b_i + \ell (n-\ell) \geq 0$.
\end{proof}

\section{Proofs for Section \ref{section:one:dimensional case}} \label{section:proof:one:dimensional}

\begin{proof}[Proof of Lemma~\ref{lm:increasing-hat-B}]
	For the first part, let $i,i' \in [|n|], i \neq i'$ such that $x_i = x_{i'}$ and assume that $\widehat{B}_i \neq \widehat{B}_{i'}$. Let us consider the increment of the criterion in \eqref{eq:conv-clust}  when replacing $\widehat{B}_i$ and $\widehat{B}_{i'}$ by $(\widehat{B}_i + \widehat{B}_{i'})/2$. From Lemma \ref{lm:increasing-hat-B:least:square}, the quadratic part is strictly decreased. Let us show that the absolute value part is decreased. This will lead to a contradiction since there is a unique minimizer in \eqref{eq:conv-clust} by strict convexity.  The increment of the absolute value part is given by
	\begin{align*}
		\text{before} - \text{after}
		= 
		|  \widehat{B}_i  - \widehat{B}_{i'} |
		+
		\sum_{\substack{\iota=1 \\ \iota \not \in \{i,i'\} }}^n 
		\left( 
		\left| \widehat{B}_\iota - \widehat{B}_i  \right| 
		+
		\left| \widehat{B}_\iota - \widehat{B}_{i'} \right|
		-
		2
		\left| \widehat{B}_\iota - \frac{\widehat{B}_i + \widehat{B}_{i'}}{2} \right|
		\right).
	\end{align*}
	In the right-hand side above, $	|  \widehat{B}_i  - \widehat{B}_{i'} | >0$ and the second sum is non-negative by convexity. This conclude the proof of the first part.
	
	For the second part, let $i,i' \in [|n|], i \neq i'$ such that $x_i > x_{i'}$ and assume that $\widehat{B}_i < \widehat{B}_{i'}$. Let us consider again the increment of the criterion in \eqref{eq:conv-clust} obtained by exchanging $\widehat{B}_i$  and $\widehat{B}_{i'}$. From Lemma \ref{lm:increasing-hat-B:least:square}, the quadratic part is strictly decreased. The absolute value part is left unchanged and thus the criterion in \eqref{eq:conv-clust} is strictly decreased which is a contradiction as before.
\end{proof}

\vspace{0.3cm}

\begin{proof}[Proof of Theorem~\ref{theorem:equivalence:polyhedral:one:d}]~\\
	{\bf Proof that \eqref{eq:conditions:un} and \eqref{eq:conditions:deux} imply \eqref{eq:equi:conditions:un},\eqref{eq:equi:conditions:deux},\eqref{eq:equi:conditions:trois}} 
	
	By \eqref{eq:conditions:un}, for any $k \in [|K|]$, all the $\widehat{B}_{i}$ for $i \in \C_k$ are identical to a value that we denote by $\widehat{b}_k$, with $\widehat{b}_1 , \ldots , \widehat{b}_K$ two-by-two distinct. 
	By Definition \ref{def:clustering}, Lemma \ref{lm:increasing-hat-B} and \eqref{eq:conditions:deux}, we have $\widehat{b}_1 > \dots > \widehat{b}_K$.
	With this notation, the  vector $(\widehat{b}_k)_{k \in [|K|]}$ is locally solution of 
	\[
	\min_{(b_k)_{k}}
	~ ~
	\frac{1}{2}
	\sum_{k=1}^{K} 
	\sum_{ i \in \C_{k} } 
	\left(
	b_k - x_{i} 
	\right)^2
	+ \lambda 
	\sum_{\substack{k,k'=1 \\ k'>k}}^{K}
	n_k n_{k'} (b_{k} - b_{k'}).
	\]
	Indeed, in \eqref{eq:conv-clust} we can assign to all the $(B_{i})_{i \in  \C_k}$ the same new value $b_k$ close to $\widehat{b}_k$, and we have  $|B_{i} - B_{i'}| = b_{k} - b_{k'}$ for all $i \in \C_k, i' \in \C_{k'}$, $k < k'$. Canceling the gradient with respect to $b_1,\ldots,b_K$ at $\widehat{b}_1,\ldots,\widehat{b}_K$  then provides, for $k \in [|K|]$,
	\[
	\sum_{i \in \C_{k}} 
	\left( 
	\widehat{b}_k
	-
	x_{i}
	\right)
	-
	\lambda 
	\sum_{k'=1}^{k-1}
	n_{k} n_{k'} 
	+
	\lambda 
	\sum_{k'=k+1}^K
	n_k n_{k'} 
	=0.
	\]
	This provides 
	\begin{align} 
		\widehat{b}_k
		& =
		\frac{1}{n_k}
		\sum_{i \in \C_{k}} 
		x_{i}
		+
		\lambda 
		\sum_{k' = 1}^{k-1}
		n_{k'}
		- 
		\lambda 
		\sum_{k'  = k+1}^{K}
		n_{k'}  \label{eq:hatB} \\
		& =
		\frac{1}{n_k}
		\sum_{i \in \C_{k}} 
		x_{i}
		+
		\lambda 
		t_{k-1}
		- 
		\lambda 
		(t_K - t_k). \label{eq:hatB:deux}
	\end{align}
	Hence, we have for $k,k'\in [|K|], k < k'$:
	\begin{equation*} 
		\widehat{b}_{k} - \widehat{b}_{k'}
		=
		\frac{1}{n_{k}}
		\sum_{i \in \C_{k}}
		x_{i}
		-
		\frac{1}{n_{k'}}
		\sum_{i \in \C_{k'}}
		x_{i}
		+
		\lambda (t_{k-1} - t_{k'-1})
		+
		\lambda (t_{k} - t_{k'}),
	\end{equation*}
	so that \eqref{eq:equi:conditions:un} holds
	by the previous observation that $\widehat{b}_1 > \dots > \widehat{b}_K$ and taking $k' = k+1$.
	
	Now we fix $k \in [|K|]$. If we replace $\widehat{B}_{i} = \widehat{b}_{k}$ by $\widehat{b}_{k} + U_{i}$ for $i \in \C_{k}$ and we keep the $\widehat{B}_{i}$, $i \not \in \C_{k}$ unchanged, we increase the cost function in Problem \eqref{eq:conv-clust}. Hence the following function of $(U_{i})_{i \in \C_{k}}$
	$$
		\frac{1}{2}
		\sum_{i \in \C_{k}}
		\left( 
		\widehat{b}_{k} + U_{i}
		- x_{i}  
		\right)^2
		+ \lambda 
		\sum_{\substack{ k'=1 \\k' \neq k }}^{K}
		\sum_{i \in \C_{k}} 
		n_{k'} \ 
		\sign(k'-k)
		\left(
		\widehat{b}_{k}
		-
		\widehat{b}_{k'}
		+ U_{i}
		\right)
		+ \lambda 
		\sum_{ \substack{ i,i' \in \C_{k} \\ i < i'} } 
		|  U_{i} - U_{i'} |
	$$
	%\begin{align*}
	%	& \frac{1}{2}
	%	\sum_{i \in \C_{k}}
	%	\left( 
	%	\widehat{b}_{k} + U_{i}
	%	- x_{i}  
	%	\right)^2
	%	+ \lambda 
	%	\sum_{\substack{ k'=1 \\k' \neq k }}^{K}
	%	\sum_{i \in \C_{k}} 
	%	n_{k'} \ 
	%	\sign(k'-k)
	%	\left(
	%	\widehat{b}_{k}
	%	-
	%	\widehat{b}_{k'}
	%	+ U_{i}
	%	\right)
	%	\\
	%	&
	%	+ \lambda 
	%	\sum_{ \substack{ i,i' \in \C_{k} \\ i < i'} } 
	%	|  U_{i} - U_{i'} |
	%\end{align*}
	is minimal locally around $0$.
	Above, we let $\sign(t) = 1$ if $t >0$, $\sign(0)=0$ and $\sign(t) = -1$ if $t <0$.
	From Lemma \ref{lem:linearization:convex}, this implies that the function 
	\begin{equation} 
		\label{eq:the:function:of:U:before}
		\sum_{i \in \C_{k}}
		\left( 
		\widehat{b}_{k}
		-
		x_{i}  
		\right)
		U_{i}
		+ \lambda 
		\sum_{\substack{ k'=1 \\ k' \neq k }}^{K}
		\sum_{i \in \C_{k}} 
		n_{k'} \
		\sign(k'-k)
		U_{i}
		+ \lambda 
		\sum_{ \substack{ i,i' \in \C_{k} \\ i < i' } } 
		|  U_{i} - U_{i'} |  
	\end{equation}
	of $(U_{i})_{i \in \C_{k}}$ has a local minimum at zero. From \eqref{eq:hatB}, this function is
	\begin{align} \label{eq:the:function:of:U}
		\sum_{i \in \C_{k}} 
		\left(
		\left(
		\frac{1}{n_k}
		\sum_{i' \in \C_{k}}
		x_{i'}
		\right) 
		- x_{i}
		\right) 
		U_i
		+ \lambda 
		\sum_{ \substack{ i,i' \in \C_{k} \\ i < i' } } 
		|  U_{i} - U_{i'} |. 
	\end{align}
	
	If $n_k=1$ this function is 0. 
	Otherwise, because this function has a local minimum at zero, and because $a_{i}:= \frac{1}{n_k}
	\left(
	\sum_{i'\in \C_{k}}
	x_{i'}
	\right) 
	- x_{i}$ satisfies $a_{\sigma(t_{k-1}+1)}\leq \dots \leq a_{\sigma(t_k)}$  by \eqref{eq:conditions:deux} , Lemma \ref{lemma:condition:obs:absolute:value} implies that for all $\ell \in [| n_k -1|]$, 
	\begin{equation}
		\label{eq:triangle}
		\sum_{i=1}^{\ell}
		\left[
		\left(
		\frac{1}{n_k}
		\sum_{i'\in \C_{k}}
		x_{i'}
		\right) 
		- x_{\sigma(t_{k-1} + i)}
		\right]
		+
		\lambda 
		\ell
		(n_k - \ell)
		\geq 0
	\end{equation}
	so that \eqref{eq:equi:conditions:deux} holds. 
	Finally, \eqref{eq:equi:conditions:trois} holds, being identical to \eqref{eq:conditions:deux}. 
	
	\noindent {\bf Proof that
		\eqref{eq:equi:conditions:un},\eqref{eq:equi:conditions:deux},\eqref{eq:equi:conditions:trois} imply 
		\eqref{eq:conditions:un} and \eqref{eq:conditions:deux} }

	Let $\tilde{b}_k$ be given by the right hand side of \eqref{eq:hatB} for $k \in [|K|]$. Let $\tilde{B}_i = \tilde{b}_k$ for $k \in [|K|]$ and $i \in \C_k$. Let us show that $\widetilde{\B} = (\tilde{B}_1 ,\ldots , \tilde{B}_n)$ provides a minimum of \eqref{eq:conv-clust} (that is $\widetilde{\B} = \widehat{\B}$).
	Note that \eqref{eq:equi:conditions:un} and \eqref{eq:hatB:deux} provide $\tilde{b}_{k} > \tilde{b}_{k'}$ for $k < k'$. Then we can write
	the cost function at $\widetilde{B}_{i} + U_{i}, i\in [|n|]$, locally around $0$ for $U = (U_1 , \ldots , U_n) \in \mathbb{R}^n$, using :

	$$
	\frac{1}{2} 
	\sum_{k=1}^K
	\sum_{i \in \C_k}
	\left( 
	\widetilde{b}_{k}
	-
	x_{i}  
	+ U_{i}
	\right)^2
	+
	\lambda 
	\sum_{\substack{k,k'=1 \\k' > k}}^{K}
	\sum_{ \substack{i \in \C_{k} \\ i' \in \C_{k'} }}
	\left( 
	\widetilde{b}_{k} 
	-
	\widetilde{b}_{k'} 
	+ U_{i} - U_{i'} 
	\right)
	+
	\lambda 
	\sum_{k=1}^{K} 
	\sum_{ \substack{i,i' \in \C_{k} \\ i < i'}}
	| U_{i} - U_{i'} |.
	$$
	From Lemma \ref{lem:linearization:convex} a sufficient condition to have a local minimum at $0$ is to have a local minimum at $0$ of 	
	$$
	\sum_{k=1}^K
	\sum_{i \in \C_k} 
	\left( 
	\widetilde{b}_{k}
	-
	x_{i}  
	\right)
	U_{i}
	+
	\lambda 
	\sum_{\substack{k,k'=1 \\k' > k}}^{K}
	\sum_{ \substack{i \in \C_{k} \\ i'\in \C_{k'} }}
	\left( 
	U_{i} - U_{i'} 
	\right)
	+
	\lambda 
	\sum_{k=1}^{K}
	\sum_{ \substack{i,i' \in \C_{k} \\ i < i'}}
	| U_{i} - U_{i'} |.
	$$
	This is a sum of functions of $(U_{i})_{i \in \C_{k}}$, the sum being over $k$. Hence, it is enough that the following function of $(U_{i})_{i \in \C_{k}}$ is locally minimal at $0$, for $k\in [|K|]$,	
	$$
	\sum_{i \in \C_k} 
	\left( 
	\widetilde{b}_{k}
	-
	x_{i}  
	\right)
	U_{i}
	+
	\lambda 
	\sum_{i \in \C_k}
	\sum_{\substack{k'=1 \\  k' \neq k }}^{K}
	n_{k'} \ 
	\sign(k'-k)
	U_{i}  
	+
	\lambda 
	\sum_{ \substack{i,i' \in \C_{k} \\ i < i'}}
	| U_{i} - U_{i'} |.
	$$
	This is the same function as in \eqref{eq:the:function:of:U:before} and \eqref{eq:the:function:of:U}. It is 0 when $n_k=1$. Otherwise, since the weights of the linear combination of $(U_{i})_{i \in \C_{k}}$ have sum zero and with Condition \eqref{eq:equi:conditions:deux} we indeed have a minimum at $U_{i} = 0,i \in \C_{k}$ from Lemma \ref{lemma:condition:obs:absolute:value}.
	Hence $\widetilde{\B}$ as defined above is the global minimizer of \eqref{eq:conv-clust} (since it is a local minimizer). Hence since we have seen before that $\tilde{b}_{k} \neq \tilde{b}_{k'}$ for $k \neq k'$, then \eqref{eq:conditions:un} is satisfied. Finally, \eqref{eq:conditions:deux} holds, being identical to \eqref{eq:equi:conditions:trois}.  
	
	\noindent {\bf Proof of \eqref{eq:hat:B:optimum}:}
	This equation was established in \eqref{eq:hatB}.
\end{proof}
\vspace{0.3cm}

\begin{proof}[ Proof and full expressions for Lemma \ref{corPolyh}]~\\
	\noindent Condition \eqref{eq:equi:conditions:trois}: $ x_{\sigma(1)} \geq x_{\sigma(2)} \geq \dots \geq x_{\sigma(n)} $ is equivalent to 
	$\bM_1 \bP_\sigma \x \leq \lambda \m_1$ with
	\begin{equation}
		\label{eq:exprM1}
		\bM_1=
		\left(\begin{array}{c c c c c c}
			-1 & 1 & 0 & 0 & \ldots & 0 \\
			0 & -1 & 1 & 0 & \ldots & 0\\
			0 & 0 & -1 & 1 & \ldots & 0\\
			& \ddots & \ddots & \ddots&\ddots&\\
			0 & 0 & 0& 0 & -1& 1
		\end{array}\right)\in\R^{n-1 \times n}
	\end{equation}
	and $\m_1=\bzero_{n-1}$.

	\noindent Condition \eqref{eq:equi:conditions:un}: For $k \in [|K-1|]$, 
	$$
	\frac{1}{n_k}
	\sum_{i=1}^{n_k} x_{\sigma(t_{k-1} + i)} - 
	\frac{1}{n_{k+1}}
	\sum_{i=1}^{n_{k+1}} x_{\sigma(t_{k} + i)} 
	> 
	\lambda (t_{k+1} - t_{k-1})
	$$
	is equivalent to $\bM_2(\t) \bP_\sigma \x <  \lambda
	\m_2(\t)$, where $\bM_2(\t)\in\R^{K-1 \times n}$
	and $\m_2(\t)\in\R^{K-1}$ are defined by
	
	\begin{equation}
		\label{eq:exprm2}
		\m_2 (\t)
		%= -\lambda \left(  (t_{l}-t_{k}+t_{l-1}-t_{k-1})_{k,l \in [|K|], k < l}\right) 
		%= -\left(\begin{array}{c}
		%t_2  - t_0\\
		%t_3  - t_1\\
		%\vdots\\
		%t_K -t_{K-2}
		%\end{array}\right) 
		= -\left(t_2  - t_0,t_3  - t_1,\ldots,t_K -t_{K-2}\right)^\top 
	\end{equation}
	and 
%	\scriptsize
	\begin{equation}
		\label{eq:exprM2}
		\scriptsize
		\bM_2(\t) = \left(
		\begin{array}{*{16}{c}}
			-\frac{1}{n_1} & \ldots & -\frac{1}{n_1}  &  \frac{1}{n_2} & \ldots & \frac{1}{n_2}  & 0 & \ldots & 0 & \ldots & 0 & \ldots & 0 & 0 & \ldots & 0 \\
			0 & \ldots & 0 &-\frac{1}{n_2} & \ldots & -\frac{1}{n_2}  &  \frac{1}{n_3} & \ldots & \frac{1}{n_3}  & \ldots & 0 & \ldots & 0 & 0 & \ldots & 0 \\
			\vdots & \vdots & \vdots & \vdots & \vdots & \vdots & \vdots & \vdots &\vdots &\vdots &\vdots &\vdots &\vdots &\vdots &\vdots &\vdots \\
			0 & \ldots & 0 & 0 & \ldots & 0 & 0 & \ldots & 0 & \ldots & - \frac{1}{n_{K-1}} & \ldots & -\frac{1}{n_{K-1}}  & \frac{1}{n_{K}} & \ldots & \frac{1}{n_{K}}  \\
		\end{array}
		\right),
		\normalsize
	\end{equation}
%	\normalsize
	where the number of repetitions of each $ \pm 1/n_k$ in each line is $n_k$.
	
	\noindent Condition \eqref{eq:equi:conditions:deux} : For $k \in [|K|]$ such that $n_k \geq 2$, for $\ell \in [| n_k-1 |]$,
	$$
	\frac{1}{n_k}
	\sum_{i=1}^{n_k} x_{\sigma(t_{k-1} + i)} 
	-
	\frac{1}{\ell}
	\sum_{i=1}^{\ell} x_{\sigma(t_{k-1} + i)} 
	\geq  
	\lambda ( \ell - n_k)
	$$
	is equivalent to
	$
	\bM_3(\t) \bP_\sigma \x \leq \lambda \m_3(\t)$ where $\bM_3(\t)\in\R^{n-K \times n}$ and 
	$\m_3(\t)\in\R^{n-K}$ are as follows. 
	We have
	\begin{equation}
		\label{eq:exprm3}
		\m_3(\t) =  \left(n_1 - 1,n_1-2,\ldots,1,n_2-1, \ldots,1,\ldots,n_K-1,\ldots,1\right)^\top
	\end{equation}
	with the convention that $(n_k - 1,n_k-2,\ldots,1)$ is empty when $n_k=1$, 
	and  $\bM_3(\t) = \mbox{diag}(\bM_3^{(1)},\ldots, \linebreak[1] \bM_3^{(K)})$ with
	\begin{equation}
		\label{eq:exprM3}
		\bM_3^{(k)} = \left(\begin{array}{*{6}{c}}
			1 & 0 & 0 & \ldots & 0 & 0 \\
			\frac 1 2  & \frac 1 2 & 0 & \ldots & 0 & 0\\
			\frac 1 3 & \frac 1 3 & \frac 1 3 & \ldots & 0 &0 \\
			\vdots & \vdots & \vdots & \ddots & \vdots & \vdots \\
			\frac{1}{n_{k}-1} &  \frac{1}{n_{k}-1}  &  \frac{1}{n_{k}-1} & \ldots &  \frac{1}{n_{k}-1}  & 0
		\end{array}\right)
		- 
		\frac{1}{n_k} \bone_{n_k -1 \times n_k},
	\end{equation}
	with the convention that $\bM_3^{(k)}$ is $0 \times 0$ when $n_k=1$ and where $\bone_{n_k -1 \times n_k}$ is the $n_k -1 \times n_k$ matrix composed of ones.
\end{proof}

\begin{proof}[Proof of Proposition \ref{prop:polyhedral:lemma}]
	The proposition is obtained from Lemma 5.1 in \cite{Lee16}. We will only show the last claim that the probability that $ \V^{-}(\Z) =  \V^{+}(\Z)$ is zero, conditionally to the event in \eqref{eq:conditioning:set}. We have, letting $\1_{\eqref{eq:conditioning:set}}$ denote the indicator function that the event in \eqref{eq:conditioning:set} holds,
	\begin{align*}
		\E \left[ 
		\1_{\eqref{eq:conditioning:set}} 
		\1_{\V^{-}(\Z) =  \V^{+}(\Z)} 
		\right]
		& =
		\E \left[
		\1_{	\V^{-}(\Z) =  \V^{+}(\Z) }
		\E
		\left[
		\left.
		\1_{\eqref{eq:conditioning:set}} 
		\right| 
		\V^{-}(\Z) =  \V^{+}(\Z)
		\right]
		\right].
	\end{align*}
	The above conditional expectation is zero from \eqref{eq:conditioning:set}, because $\boeta^\top \X$ is independent from $\Z$ and has non-zero variance because $\bSigma$ is invertible and $\boeta$ is non-zero.
\end{proof}

\begin{proof}[Proof of Lemma \ref{lemma:expression:pvalue}]
For convenience, we let $\V^-(\z_0) = \V^-$, $\V^+(\z_0) = \V^+$. Recall that $\Phi$ denotes the cdf of the Gaussian distribution $\N(0,\boeta^\top \bSigma \boeta)$. Recall also that $W$ follows the Gaussian distribution $\N(0,\boeta^\top \bSigma \boeta)$ truncated to $[\V^-,\V^+]$. Recall finally that $\boeta^\top \x \in [\V^-,\V^+]$. 

In Case 1.1, since $|\boeta^\top \x| >  \min (|\V^-| , |\V^+| )$ with $|\V^-| > |\V^+|$, then necessarily $\boeta^\top \x \in [\V^- , - \
| \V^+|]$ and thus $\boeta^\top \x \leq 0$. We have
\[ \mathbb{P}_{W} \left( 
| W |
\ge 
| \boeta^\top \x |
 \right)
=
\mathbb{P}_{W} \left(
W \in [ \V^- , \boeta^\top \x ]
\right)
=
\frac{\Phi(\boeta^\top \x)
 - \Phi(\V^-)
 }{\Phi(\V^+)
-
\Phi(\V^-)
}.
\]

In Case 1.2,
since $|\boeta^\top \x| >  \min (|\V^-| , |\V^+| )$ with $|\V^+| > |\V^-|$, then necessarily $\boeta^\top \x \in [|\V^-| ,   \V^+]$ and thus $\boeta^\top \x \geq 0$. We have
\[
\mathbb{P}_{W} \left( 
| W |
\ge 
| \boeta^\top \x |
 \right)
=
\mathbb{P}_{W} \left(
W \in [ \boeta^\top \x , \V^+ ]
\right)
=
\frac{
\Phi(\V^+)
-
\Phi(\boeta^\top \x)
}{
\Phi(\V^+)
-
\Phi(\V^-)
}.
\]

In Case 2, since $|\boeta^\top \x| \leq  \min (|\V^-| , |\V^+| )$, then necessarily $\V^- \leq 0$ and $\V^+ \geq 0$.  We have
\begin{align*}
\mathbb{P}_{W} \left( 
| W |
\ge 
| \boeta^\top \x |
 \right)
= &
\mathbb{P}_{W} \left(
W \in [ \V^-, -|\boeta^\top \x|]
\right)
+
\mathbb{P}_{W} \left(
W \in [ |\boeta^\top \x| , \V^+ ]
\right)
\\
= &
\frac{
\Phi(-|\boeta^\top \x|)
-
\Phi(\V^-)
}{
\Phi(\V^+)
-
\Phi(\V^-)
}
+
\frac{
\Phi(\V^+)
- \Phi(|\boeta^\top \x|)
}{
\Phi(\V^+)
-
\Phi(\V^-)
}.
\qedhere
\end{align*}
\end{proof}

\begin{proof}[Proof of Proposition \ref{prop:level:conditional}] 
	The proof is adapted from that of Theorem 5.2 in \cite{Lee16}.	
	Fix $t \in [0,1]$. Remark that in Lemma \ref{corPolyh}, all the lines of $\bM_2(\t)$ are non-zero.
	Furthermore, $\bSigma$ is invertible.
	This provides, from Theorem \ref{theorem:equivalence:polyhedral:one:d} and Lemma \ref{corPolyh} that the events $E_{\t,\sigma}$ and $\left\{\bM \bP_\sigma \X \leq \lambda\ \m\right\}$ have their symmetric difference of probability zero. 
	Hence we have 
	\[
	\mathbb{P} 
	\left( 
	\left.
	\pval( \X, \t,\sigma)  \leq t \right| E_{\t,\sigma}
	\right) 
	= 
	\mathbb{P} 
	\left( 
	\left.
	\pval( \X, \t,\sigma)  \leq t \right| 
	\bM \bP_\sigma \X \leq \lambda\ \m
	\right). 
	\]
	We then have
	\begin{align} \label{in:proof:level}
		& \mathbb{P} 
		\left( 
		\left.
		\pval( \X, \t,\sigma)  \leq t \right| E_{\t,\sigma}
		\right) \notag
		\\
		& = 
		\int_{ \mathbb{R}^{n} } 
		\mathbb{P} 
		\left( 
		\left.
		\pval( \X, \t,\sigma) 
		\leq 
		t 
		\right| 
		\bM \bP_\sigma \X \leq \lambda\ \m
		,
		\Z
		= \z_0
		\right)
		d \mathbb{P}_{|\bM \bP_\sigma \X \leq \lambda\ \m }  
		(\z_0),
	\end{align}
	where $d \mathbb{P}_{|\bM \bP_\sigma \X \leq \lambda\ \m } 
	(\z_0)$ denotes the law of $ \Z $ conditionally to $\bM \bP_\sigma \X \leq \lambda\ \m$.

	Consider $\z_0$ in the support of $\mathbb{P}_{|\bM \bP_\sigma \X \leq \lambda\ \m}$ such that $\mathcal{V}^-(\z_0) < \mathcal{V}^+(\z_0)$, which holds with $\mathbb{P}_{|\bM \bP_\sigma \X \leq \lambda\ \m} $-probability one from Proposition \ref{prop:polyhedral:lemma}.
	Then, as discussed in Section \ref{subsection:construction:test} and shown in \cite{Lee16}, conditionally to $\bM \bP_\sigma \X \leq \lambda\ \m$  and  $\Z=\z_0$, under the null hypothesis $\boeta^\top \bmu=0$,  $\boeta^\top  \X$ has cdf $F_{0, \boeta^\top \bSigma \boeta}^{[\V^{-}(\z_0), \V^+(\z_0)]}$.
 Let $\widetilde{F}$ be the cdf of $|W|$ in \eqref{eq:cond-p-value-1d}. Then $\pval( \X, \t,\sigma)$ can be written as
 \[
 \pval( \X, \t,\sigma)
 =
 1 - \widetilde{F}( |\boeta^\top  \X| ).
 \]
 Let $C \subseteq [0 , \infty)$ be the closed segment equal to the image of $[\V^-(\z_0),\V^+(\z_0) ]$ by the function $|\cdot|$. Let $\mathring{C}$ be the interior of $C$.
Because $\mathcal{V}^-(\z_0) < \mathcal{V}^+(\z_0)$ and $\boeta^\top \bSigma \boeta >0$, 
$\mathring{C}$ is non-empty and
the cdf $\widetilde{F}$ is continuous and strictly increasing, thus bijective, from $\mathring{C}$ to $(0,1)$. Furthermore, conditionally to $\bM \bP_\sigma \X \leq \lambda\ \m$  and  $\Z=\z_0$, $|\boeta^\top  \X|$ has cdf  $\widetilde{F}$. Hence conditionally to $\bM \bP_\sigma \X \leq \lambda\ \m$  and  $\Z=\z_0$, $\widetilde{F}( |\boeta^\top  \X| )$ is uniformly distributed on $[0,1]$.
	We thus obtain from \eqref{in:proof:level},  

	\begin{eqnarray*}
		\mathbb{P} \left( \left.\pval( \X, \t,\sigma)  \leq t \right| E_{\t,\sigma}\right)
		& = &  \int_{ \mathbb{R}^{n} }   t \   d \mathbb{P}_{|\bM \bP_\sigma \X \leq \lambda\ \m }  (\z_0) 
		\\
		& =  & t.
	\end{eqnarray*}
	This concludes the proof.
\end{proof}

\begin{proof}[Proof of Proposition \ref{proposition:unconditional:level}]
	Fix $t \in [0,1]$. We have
	\begin{align*}
		&	\mathbb{P}
		\left( 
		\left.
		\pval( \X, \T,S)
		\leq 
		t 
		\right|
		(\T,S) \in \mathcal{E},
		\boeta(\T,S)^\top \bmu
		= 0
		\right) 
		\\
		& = 
		\sum_{ \substack{ (\t,\sigma) \in \mathcal{E}
				\\
				\boeta(\t,\sigma)^\top \bmu = 0
				\\
				\mathbb{P} \left(  
				(\T,S) = (\t,\sigma)
				\right) >0 
				\\
		}  } 
		\\
		& ~ ~
		\mathbb{P} \left( 
		\left. 
		(\T,S) = (\t,\sigma)
		\right|
		(\T,S) \in \mathcal{E},
		\boeta(\T,S)^\top \bmu = 0
		\right)
		\mathbb{P}
		\left( 
		\left.
		\pval( \X, \t,\sigma)
		\leq 
		t   
		\right| 
		(\T,S) = (\t,\sigma) 
		\right). 
	\end{align*}
	In the conditional probability
	$	\mathbb{P}
	\left( 
	\left.
	\pval( \X, \t,\sigma)
	\leq 
	t   
	\right| 
	(\T,S) = (\t,\sigma) 
	\right)$
	of the above sum, conditionally to $(\T,S) = (\t,\sigma)$, one can check that all the conditions of Proposition \ref{prop:level:conditional} hold. Hence, from this proposition, we have 
	\begin{align*}
		&	\mathbb{P}
		\left( 
		\left.
		\pval( \X, \T,S)
		\leq 
		t 
		\right|
		(\T,S) \in \mathcal{E},
		\boeta(\T,S)^\top \bmu
		= 0
		\right) 
		\\
		& =
		\sum_{ \substack{ (\t,\sigma) \in \mathcal{E}
				\\
				\boeta(\t,\sigma)^\top \bmu = 0 \\
				\mathbb{P} \left(  
				(\T,S) = (\t,\sigma) 
				\right) >0 
		}  }
		\mathbb{P} \left( 
		\left. 
		(\T,S) = (\t,\sigma)
		\right|
		(\T,S) \in \mathcal{E},
		\boeta(\T,S)^\top \bmu = 0
		\right) \times t
		\\
		& = 
		t. 
	\end{align*}
	This concludes the proof.
\end{proof}

\begin{proof}[Proof of Theorem \ref{thm:regularization:path}]~\\
	We will show that, for the successive values of $r$, for $k \in [|K^{(r)}|]$, for $\lambda \geq  \lambda^{(r)}$,
	\begin{align} \label{eq:hat:b:deux:algo}
		\hat{b}_k^{(r)}(\lambda) 
		=
		\frac{1}{n^{(r)}_k}
		\sum_{i \in \C^{(r)}_k} 
		x_{i}
		+
		\lambda 
		\sum_{k' = 1}^{k-1}
		n^{(r)}_{k'}
		- 
		\lambda 
		\sum_{k'  = k+1}^{K^{(r)}}
		n^{(r)}_{k'}.
	\end{align}
	We will also show that for the successive values of $r$,
	\begin{align} \label{eq:lambda:equal:inf:deux}
		\lambda^{(r+1)} 
		=
		\inf 
		\left\{ 
		\lambda \geq \lambda^{(r)};
		~\text{there exists}~
		k \in [| K^{(r)}-1 |]
		~\text{such that}~
		\hat{b}_k^{(r)}(\lambda) 
		=
		\hat{b}_{k+1}^{(r)}(\lambda) 
		\right\}.
	\end{align}
	
	We will prove by induction that the following properties $\mathcal{O}^{(r)}$, $\mathcal{P}^{(r)}$, $\mathcal{Q}^{(r)}$ and $\mathcal{R}^{(r)}$  hold for $r=0,1,\ldots$ and as long as $K^{(r)} \geq 2$:
	\begin{align*}
		\mathcal{O}^{(r)} ~ & = ~
		\text{``\eqref{eq:hat:b:deux:algo} holds for $k \in [|K^{(r)}|]$ and $\lambda \geq  \lambda^{(r)}$''}, \\
		\mathcal{P}^{(r)} ~ & = ~
		\text{``the set in \eqref{eq:lambda:equal:inf:deux} is non-empty and  \eqref{eq:lambda:equal:inf:deux} holds''}, \\
		\mathcal{Q}^{(r)} ~ & = ~
		\text{``for $\lambda \in [\lambda^{(r)} ,\lambda^{(r+1)} )$, we have $\hat{b}^{(r)}_1(\lambda) > \dots > \hat{b}^{(r)}_{K^{(r)}}(\lambda)$''}, \\
		\mathcal{R}^{(r)} ~ & = ~
		\text{``for $\lambda \in [ \lambda^{(r)} , \lambda^{(r+1)} )$, $(\hat{B}_i^{(r)}(\lambda))_{i \in [|n|]}$ minimizes Problem \eqref{eq:conv-clust}''}.
	\end{align*}
	Along proving these properties by induction, we will show that $K^{(r)} > K^{(r+1)}$. Doing this, and discussing the case $r = r_{\max}$ at the end, will conclude the proof.
	
	\textbf{Initialization: $r=0$.}
	When $r=0$, we have $\lambda^{(0)} = 0$ and, for $k \in [|K^{(0)}|]$, 
	\begin{equation} \label{eq:hatbzero:equal}
		\hat{b}^{(0)}_k(\lambda^{(0)}) = \tilde{x}_k
		=
		\frac{1}{n^{(0)}_k}
		\sum_{i \in \C^{(0)}_k} 
		x_{i}
	\end{equation}
	and thus the right-hand sides of \eqref{eq:hat:b:algo} and \eqref{eq:hat:b:deux:algo} are equal and so $\mathcal{O}^{(0)}$  holds. 
	Furthermore, from \eqref{eq:hatbzero:equal}, $\hat{b}^{(0)}_1(0)> \dots > \hat{b}^{(0)}_{K^{(0)}}(0)$. We have, for $k \in [|K^{(0)} - 1|]$, using \eqref{eq:hat:b:deux:algo}, 
	\begin{align}
		\hat{b}^{(0)}_k(\lambda)
		-
		\hat{b}^{(0)}_{k+1}(\lambda)
		& = 
		\tilde{x}_k
		+
		\lambda 
		\sum_{k' = 1}^{k-1}
		n^{(0)}_{k'}
		- 
		\lambda 
		\sum_{k'  = k+1}^{K^{(0)}}
		n^{(0)}_{k'}
		-
		\tilde{x}_{k+1}
		-
		\lambda 
		\sum_{k' = 1}^{k}
		n^{(0)}_{k'}
		+ 
		\lambda 
		\sum_{k'  = k+2}^{K^{(0)}}
		n^{(0)}_{k'}
		\notag\\
		& = 
		\underbrace{
			\tilde{x}_k
			- \tilde{x}_{k+1} }_{>0}
		- \lambda 
		\underbrace{	
			\left(
			n^{(0)}_k
			+
			n^{(0)}_{k+1}
			\right)	
		}_{>0}.
		\label{eq:init-algo}
	\end{align}
	Hence, we see that indeed  the set in \eqref{eq:lambda:equal:inf:deux} is non-empty.
	
	Let $\tilde{\lambda}^{(1)}$ be given by the right-hand side of \eqref{eq:lambda:equal:inf:deux}. 
	The values of $\hat{b}^{(0)}_k(\lambda)$, $k \in [|K^{(0)}|]$, are continuous in $\lambda$ and thus by definition of $\tilde{\lambda}^{(1)}$ they remain two-by-two distinct and in the same order on $[0,\tilde{\lambda}^{(1)} )$. 
	Furthermore , from \eqref{eq:init-algo},
	\[
	\tilde{\lambda}^{(1)}
	=
	\min_{k \in [|  K^{(0)}-1 |] }
	\frac{  
		\tilde{x}_k
		-
		\tilde{x}_{k+1}
	}{
		n^{(0)}_k
		+ 
		n^{(0)}_{k+1}
	}
	=
	\lambda^{(0)} 
	+
	\min_{k \in [|  K^{(0)}-1 |] }
	\frac{
		\hat{b}^{(0)}_k(\lambda^{(0)})
		-
		\hat{b}^{(0)}_{k+1}(\lambda^{(0)})
	}{
		n^{(0)}_{k} +
		n^{(0)}_{k+1}}
	=
	\lambda^{(1)}.
	\]	
	Hence indeed \eqref{eq:lambda:equal:inf:deux} holds and thus $\mathcal{P}^{(0)}$ holds. 
	Since $\lambda^{(1)}$ is given by \eqref{eq:lambda:equal:inf:deux}, then also $\mathcal{Q}^{(0)}$ holds.
	
	Let us now show $\mathcal{R}^{(0)}$. Let $\lambda \in [\lambda^{(0)} ,\lambda^{(1)} )$.
	We will apply Theorem \ref{theorem:equivalence:polyhedral:one:d}, with $\sigma$ there being a permutation such that $x_{\sigma(1)} \geq \dots \geq x_{\sigma(n)}$ and $\C$ being the clustering $
	\C^{(0)}$. 
	For $k \in [|K^{(0)} - 1|]$, since 	$\hat{b}^{(0)}_k(\lambda)
	-
	\hat{b}^{(0)}_{k+1}(\lambda) > 0$ as seen above, we obtain 
	from \eqref{eq:hat:b:deux:algo}
	that \eqref{eq:equi:conditions:un} holds, using \eqref{eq:hatB} and \eqref{eq:hatB:deux}. It is immediate that \eqref{eq:equi:conditions:deux} holds because the left-hand term is zero and the right-hand term is non-positive. Hence from \eqref{eq:hat:B:optimum} in Theorem \ref{theorem:equivalence:polyhedral:one:d}, $\mathcal{R}^{(0)}$ indeed holds, also from \eqref{eq:hat:b:deux:algo}. Also, $K^{(1)} < K^{(0)}$ because, by definition of $\lambda^{(1)}$ in \eqref{eq:lambda:equal:inf:deux}, the values $\hat{b}_1^{(0)}(\lambda^{(1)}), \ldots,
	\hat{b}_{K^{(0)}}^{(0)}(\lambda^{(1)})$ are not two-by-two distinct.

	\textbf{Induction: from $r$ to $r+1$.}
	Let now $r \in \mathbb{N}$ such that $K^{(r+1)} \geq 2$. 
	Assume that $\mathcal{O}^{(r)}$, $\mathcal{P}^{(r)}$, $\mathcal{Q}^{(r)}$ and $\mathcal{R}^{(r)}$ hold. 
	For any $B \in \mathbb{R}^n$, from $\mathcal{R}^{(r)}$, for  $\lambda \in [\lambda^{(r)} , \lambda^{(r+1)})$, 
	\[
	\frac{1}{2} \sum_{i=1}^n 
	\left( \hat{B}_i^{(r)}(\lambda) - x_i \right)^2
	+
	\lambda 
	\sum_{ \substack{ i,i'=1 \\ i < i' }}^n
	\left|  \hat{B}_i^{(r)}(\lambda) - \hat{B}_{i'}^{(r)}(\lambda)  \right|
	\leq 
	\frac{1}{2} \sum_{i=1}^n 
	\left( B_i - x_i \right)^2
	+
	\lambda 
	\sum_{ \substack{ i,i' =1 \\ i < i'}}^n
	\left|  B_i - B_{i'}  \right|.
	\] 
	As $\lambda \to \lambda^{(r+1)}$, this yields
	\begin{align*}
		&		\frac{1}{2} \sum_{i=1}^n 
		\left( \hat{B}_i^{(r)}(\lambda^{(r+1)}) - x_i \right)^2
		+
		\lambda^{(r+1)}
		\sum_{ \substack{ i,i'=1 \\ i < i' }}^n
		\left|  \hat{B}_i^{(r)}(\lambda^{(r+1)}) - \hat{B}_{i'}^{(r)}(\lambda^{(r+1)})  \right| 
		\notag\\
		&		\leq 
		\frac{1}{2} \sum_{i=1}^n 
		\left( B_i - x_i \right)^2
		+
		\lambda^{(r+1)}
		\sum_{ \substack{ i,i'=1 \\ i < i' }}^n
		\left|  B_i - B_{i'} \right|.
	\end{align*}
	Hence, the minimizer of \eqref{eq:conv-clust} for $\lambda = \lambda^{(r+1)}$ is equal to   $\hat{B}_i^{(r)}(\lambda^{(r+1)})_{i \in [|n|]}$. Hence, $\C^{(r+1)}$ is the clustering obtained by minimizing \eqref{eq:conv-clust}. We can see from the successive definitions of 
	\[
	(\hat{b}_k^{(r)}( \lambda^{(r+1)} )  )_{k \in [|K^{(r)}|]},
	~ ~ ~ 
	(\hat{b}_k^{(r+1)}( \lambda^{(r+1)} ))_{k \in [|K^{(r+1)}|]}
	~ ~ ~ \text{and} 
	~ ~ ~ 
	(\hat{B}_i^{(r+1)} (\lambda^{(r+1)}))_{i \in [|n|]} 
	\]
	that 
	\[
	(\hat{B}_i^{(r)}(\lambda^{(r+1)}))_{i \in [|n|]} = (\hat{B}_i^{(r+1)}(\lambda^{(r+1)}))_{i \in [|n|]}.
	\]
	
	Hence, from \eqref{eq:hat:B:optimum} in Theorem \ref{theorem:equivalence:polyhedral:one:d},
	the right-hand side of \eqref{eq:hat:b:deux:algo}, with ($r,\lambda)$ there replaced by $(r+1,\lambda^{(r+1)})$
	and for $k \in [|K^{(r+1)}|]$, is equal to $\hat{b}_k^{(r+1)}(\lambda^{(r+1)})$. Hence \eqref{eq:hat:b:deux:algo} holds at step $r+1$ for $\lambda = \lambda^{(r+1)}$. Hence \eqref{eq:hat:b:deux:algo}  holds for $\lambda \geq \lambda^{(r+1)}$ since the right-hand-sides of \eqref{eq:hat:b:algo} and \eqref{eq:hat:b:deux:algo} have the same slope w.r.t. $\lambda$. Thus $\mathcal{O}^{(r+1)}$ is proved.
	The properties $\mathcal{P}^{(r+1)}$ and $\mathcal{Q}^{(r+1)}$ are shown similarly as in the initialization step. 
	
	Let us finally show $\mathcal{R}^{(r+1)}$. Let $\lambda \in [\lambda^{(r+1)} ,\lambda^{(r+2)} )$.
	Similarly as for the initialization step, we will apply Theorem \ref{theorem:equivalence:polyhedral:one:d}, with the same permutation $\sigma$ and with $\C$ being the clustering $
	\C^{(r+1)}$. 
	Equation~\eqref{eq:equi:conditions:un} is shown to hold similarly as before, using \eqref{eq:hat:b:deux:algo}.
	From $\mathcal{R}^{(r)}$ and with the above, we obtain that $\left( \hat{B}_i^{(r)}(\lambda^{(r+1)}) \right)_{i \in [|n|]} = \left( \hat{B}_i^{(r+1)}(\lambda^{(r+1)}) \right)_{i \in [|n|]}$ minimizes \eqref{eq:conv-clust} when $\lambda = \lambda^{(r+1)}$. Hence
	Equation~\eqref{eq:equi:conditions:deux} holds when $\lambda = \lambda^{(r+1)}$ from Theorem \ref{theorem:equivalence:polyhedral:one:d}. 
	For $\lambda \in [\lambda^{(r+1)} ,\lambda^{(r+2)} )$, the clustering is the same as when $\lambda = \lambda^{(r+1)}$ so the left-hand-side of \eqref{eq:equi:conditions:deux} is unchanged compared to when $\lambda = \lambda^{(r+1)}$. On the other hand, the right-hand side is decreased. Hence,  \eqref{eq:equi:conditions:deux} also holds for $\lambda \in [\lambda^{(r+1)} ,\lambda^{(r+2)} )$. Hence from \eqref{eq:hat:B:optimum} in Theorem \ref{theorem:equivalence:polyhedral:one:d}, and  \eqref{eq:hat:b:deux:algo}, $\mathcal{R}^{(r+1)}$ indeed holds.

	\textbf{When $r = r_{\max}$.}
	As before, we show that $\mathcal{O}^{(r_{\max}-1)}$ implies $\mathcal{O}^{(r_{\max})}$. Then using \eqref{eq:hat:b:deux:algo} for $r_{\max}$ we obtain, by the same arguments as when showing $\mathcal{R}^{(r+1)}$ above, that for $\lambda \geq \lambda^{(r_{\max})}$, $(\hat{B}_i^{(r_{\max})}(\lambda))_{i \in [|n|]}$ minimizes \eqref{eq:conv-clust}. Note that the right-hand-side of \eqref{eq:hat:b:deux:algo} is constant in $\lambda$ now and there is a single class. Hence the common value of $
	(\hat{B}_i^{(r_{\max})})_{i \in [|n|]}$ minimizes \eqref{eq:conv-clust} as $\lambda \to \infty$, so this value is $\sum_{i=1}^n x_i/n$.
\end{proof}

\section{Proofs for Section \ref{section:p:dimensional case}}
\label{appendix:proof:multiD}

\begin{proof}[Proof of Proposition \ref{proposition:example:p:dim}]
	Computing the $p$-value
	as described in Section \ref{subsubsubsection:construction:p:dim}
	requires computing
	$\vec(\obZ) := [ \bI_{np} - \vec(\obc)  \bkappa^\top] \vec(\Y)$ with $ \vec(\obc) := (\bI_p \otimes \bSigma) \bkappa (\bkappa^\top  (\bI_p \otimes \bSigma) \bkappa)^{-1}$, 
	similarly as $\Z$ and $\bc$ in Proposition~\ref{prop:polyhedral:lemma}.  Using the properties of Kronecker products, we have, letting $\be_{j_0}$ be the $j_0$th base column vector in $\mathbb{R}^p$,
	\begin{align*}
		\vec(\obc) 
		= &
		(\bI_p \otimes \bSigma) ( \be_{j_0} \otimes \boeta ) ( \bkappa^\top  (\bI_p \otimes \bSigma) \bkappa )^{-1}
		\\ 
		= &
		(\be_{j_0} \otimes  \bSigma \boeta ) ( \boeta^\top  \bSigma \boeta )^{-1} 
		\\ 
		= &
		\be_{j_0} \otimes \bc,
	\end{align*}
	where $\bc =  \bSigma \boeta ( \boeta^\top  \bSigma \boeta )^{-1}$ is as defined for the one-dimensional case in Proposition \ref{prop:polyhedral:lemma}.
	Hence, $\vec(\obc) $ is a $np \times 1$ vector where the subvector corresponding to the variable $j_0$ is equal to $\bc$ and the subvectors corresponding to the other variables are zero.
	Then, 
	\begin{align*}
		\vec(\obZ)
		= &
		\vec(\Y) - \vec(\obc)  \bkappa^\top \vec(\Y) \\ 
		= &
		\vec(\Y) - 
		(\be_{j_0} \otimes  \bc ) 
		( \be_{j_0}^\top \otimes \boeta^\top ) \vec(\Y) 
		\\
		= &
		\vec(\Y) - 
		\left(
		(\be_{j_0} \be_{j_0}^\top ) \otimes  (\bc \boeta^\top )
		\right)
		\vec(\Y) 
		\\
		= &
		\vec(\Y) - 
		\be_{j_0} 
		\otimes 
		\left( 
		\bc \boeta^\top 
		\Y_{.j_0}
		\right) 
		\\ 
		= &
		\be_{j_0} 
		\otimes 
		\Z
		+
		\vec( \Y_{.-j_0} ),
	\end{align*}
	where $\Z = [ \bI_n - \bc  \boeta^\top] \Y_{.j_0}$ is defined as in Proposition \ref{prop:polyhedral:lemma}
	and $\Y_{.-j_0}$ is defined by replacing the column $j_0$ of $\Y$ by zero. 
	Hence, $\vec(\obZ)$ is a $n p \times 1$ vector which subvector corresponding to the variable $j_0$ is $\Z$ which is computed as in the one-dimensional case. 
	
	The next step for obtaining the $p$-value is to compute
	\[
	\V^{-}( \vec(\obZ) ) := \underset{l:(\scrM \bD_\sigma \vec(\obc))_l<0}{\max}\ \frac{\lambda \mathscr{m}_l - (\scrM \bD_\sigma \vec(\obZ) )_l}{(\scrM \bD_\sigma \vec(\obc))_l}
	\]
	and
	\[
	\V^{+}( \vec(\obZ) ) := \underset{l:(\scrM \bD_\sigma \vec(\obc))_l>0}{\min}\ \frac{\lambda \mathscr{m}_l - (\scrM \bD_\sigma \vec(\obZ) )_l}{(\scrM \bD_\sigma \vec(\obc))_l}.
	\]
	In $\V^{-}( \vec(\obZ) ) $ the set of indices $l$ is the disjoint union of $p$ sets of cardinality $2(n-1)$ each, corresponding to the $p$ variables. Consider $l $ in the set corresponding to a variable $j \neq j_0$. Then in $(\scrM \bD_\sigma \vec(\obc))_l$, the row $l$ of $\scrM \bD_\sigma$, of size $np$ has non-zero components only for the indices corresponding to the variable $j$. On the other hand, as seen above, $\vec(\obc)$ has non-zero components only for the indices corresponding to the variable $j_0$. Hence, taking the inner product, $(\scrM \bD_\sigma \vec(\obc))_l = 0$. Hence the maximum in $\V^{-}( \vec(\obZ) )$ can simply be taken with the indices $l$ corresponding to the variable $j_0$. This, together with the expressions of $\vec(\obc) $ and $\vec(\obZ)$ above yields 
	\[
	\V^{-}( \vec(\obZ) )
	= 
	\underset{
		l:(\bM(\t^{(j_0)}) \bP_{\sigma^{(j_0)}} \bc)_l<0
	}{\max}\
	\frac{\lambda \m(\t^{(j_0)})_l - (\bM(\t^{(j_0)}) \bP_{\sigma^{(j_0)}} \Z )_l
	}{
		(\bM(\t^{(j_0)}) \bP_{\sigma^{(j_0)}} \bc)_l}
	:= 
	\V^{-}( \Z ),
	\]
	where $\V^{-}( \Z )$ has the same expression as in Proposition \ref{prop:polyhedral:lemma} for the one-dimensional case.
	We obtain similarly 
	\[
	\V^{+}( \vec(\obZ) )
	= 
	\underset{
		l:(\bM(\t^{(j_0)}) \bP_{\sigma^{(j_0)}} \bc)_l>0
	}{\min}\
	\frac{\lambda \m(\t^{(j_0)})_l - (\bM(\t^{(j_0)}) \bP_{\sigma^{(j_0)}} \Z )_l
	}{
		(\bM(\t^{(j_0)}) \bP_{\sigma^{(j_0)}} \bc)_l}
	:= 
	\V^{+}( \Z ).
	\]
The $p$-value is thus computed as in Lemma \ref{lemma:expression:pvalue} 
with the only difference that $\boeta^\top \x$ is replaced by $\boeta^\top \Y_{.j_0}$.
	This concludes the proof.
\end{proof}

\begin{proof}[Proof of Proposition \ref{prop:level:conditional:multidim}]
	
	From Theorem \ref{theorem:equivalence:polyhedral:one:d} and Lemma \ref{corPolyh}, for $j \in [|p|]$, the event 
	\[\left\{
	\text{	$\C^{(j)}$ is the clustering given by $\widehat{\B}_{.j}$ and
		$Y_{\sigma^{(j)}(1)j}  \geq \dots \geq Y_{\sigma^{(j)}(n)j}$ }\right\}
	\]
	is equal to the event 	$\left\{\bM(\t^{(j)}) \bP_{\sigma^{(j)}} \Y_{.j} \leq \lambda \m(\t^{(j)}) \right\}$, up to a symmetric difference of $\Y$-probability $0$ (because the rows of $\bM(\t^{(j)})$ are non-zero and the covariance matrix of $\Y_{.j}$ is invertible). 
	Hence, up to a symmetric difference of $\Y$-probability $0$, the event $E$ is equal to the event $\left\{\scrM \bD_{\sigma} \vec(\Y) \leq \lambda \scrm \right\}$, with the construction of Section \ref{subsubsubsection:construction:p:dim}. The rest of the proof is the same as the proof of Proposition~\ref{prop:level:conditional}. 
\end{proof}

\begin{proof}[Proof of Proposition \ref{proposition:unconditional:level:multi-dim}]
	The proof is the same as for Proposition \ref{proposition:unconditional:level}.
\end{proof}

\section{Time complexity of convex clustering}
\label{sec:complexity}

\subsection{Benchmarking existing implementations of convex clustering}
\label{sec:benchmark-convex-clust}
\begin{figure}
	\includegraphics[width=0.9\textwidth]{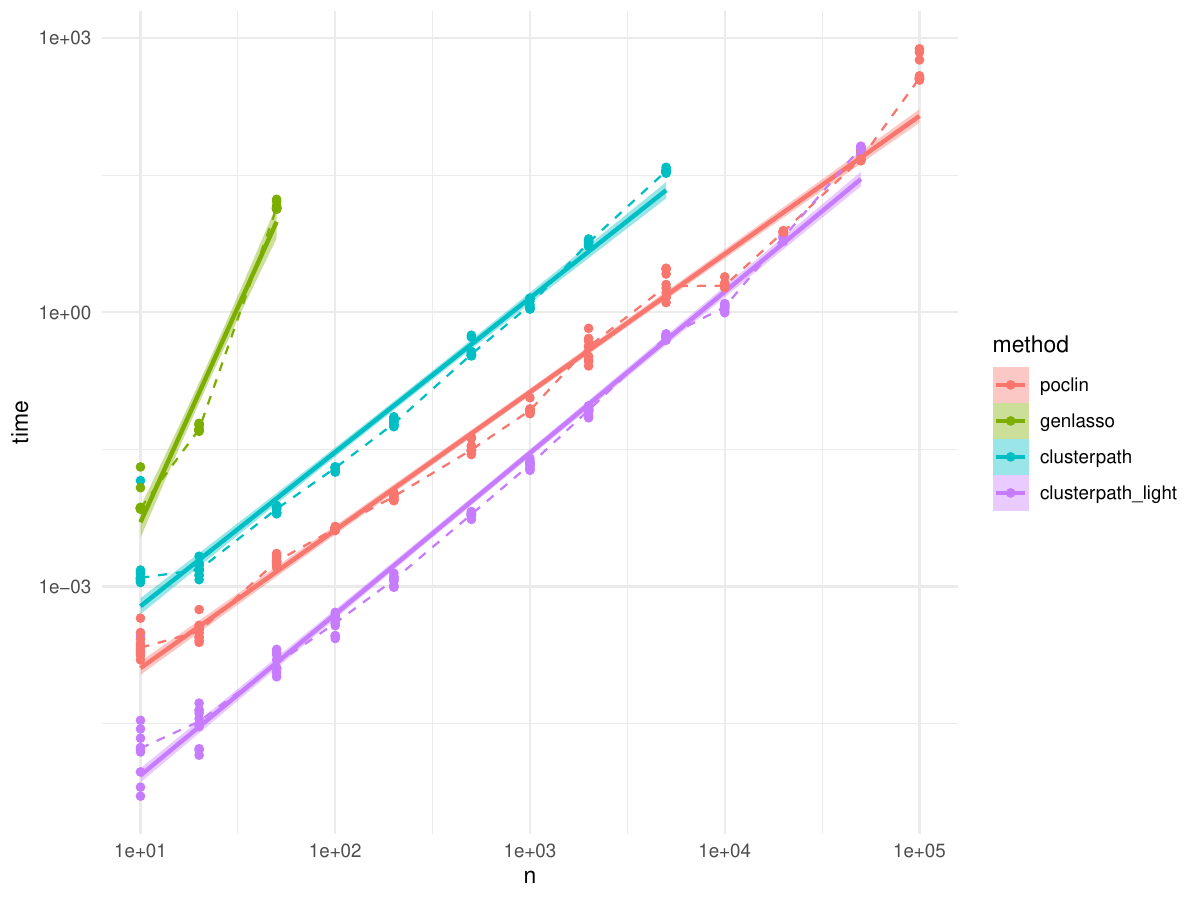}
	\caption{Comparison of the empirical complexities of existing \texttt{R} implementations of convex clustering. The axes are on a logarithmic scale. Each dot represents one experimental run. For a given method, the median computation times for a given problem size are connected by dashed lines. The solid lines have been obtained by a linear regression of time against problem size (on the log-log scale).}
	\label{fig:computational-time}
\end{figure}

In this section we compare the observed time complexities of existing implementations of one-dimensional convex clustering in the \texttt{R} language:
\begin{itemize}
	\item the \texttt{convex\_clustering\_1d} method in our \texttt{R} package \texttt{poclin}, which is available from \url{https://plmlab.math.cnrs.fr/pneuvial/poclin};
	\item the \texttt{genlasso} function in the \texttt{R} package \texttt{genlasso}, which is available from CRAN at \url{https://CRAN.R-project.org/package=genlasso};
	\item the \texttt{clusterpath.l1.id} function in the \texttt{R} package \texttt{clusterpath}, which is available from \texttt{R}-forge at \url{https://clusterpath.r-forge.r-project.org/}. The core functions of this package are implemented in \texttt{C}. 
\end{itemize}
We have used the \texttt{R} package \texttt{microbenchmark} to compare the execution time of these implementations on standard Gaussian signals of size $n \in \{10, 20, 50, 100, 200, 500, 1000, 2000, 5000, 10000, \linebreak[1] 20000,  50000, 100000\}$.
The results are displayed in Figure~\ref{fig:computational-time} on the log-log scale.
Each dot represents one experimental run.
For a given method, the median computation times for given problem sizes are connected by dashed lines.
The solid lines have been obtained by a linear regression of time against problem size (on the log scale).

The computation time of \texttt{genlasso} is much larger than for the other implementations; we were not able to get results for $n\geq 100$ with this method. 
This is explained by the fact that \texttt{genlasso} is a generic implementation where the constraints are stored in a $n(n-1)/2 \times n$ matrix.
In contrast, the other \texttt{clusterpath} and \texttt{poclin}  implementations are quite efficient.
For \texttt{clusterpath}, we report two computational times, which are labeled as 'clusterpath' and 'clusterpath\_light' in Figure~\ref{fig:computational-time}, respectively:
\begin{itemize}
	\item 'clusterpath' corresponds to the direct application of the \texttt{clusterpath.l1.id} function.
	We were not able to include 'clusterpath' for $n \geq 10000$ due to memory issues -- one run of this function for $n = 5000$ takes up 6.6 Gb of RAM;
	\item 'clusterpath\_light' directly calls the underlying \texttt{C} function \texttt{join\_clusters\_convert} in the \texttt{clusterpath} package, thereby avoiding some computational overhead.
	Thanks to this modification, we were able to run 'clusterpath\_light' for $n \leq 50000$. However, it was not possible to run it for $n \geq 100000$ because of memory issues. 
\end{itemize}
In contrast, since the space complexity of \texttt{poclin} is linear, we were able to run \texttt{poclin} without any memory issue for $n \leq 100000$.
The computational time of \texttt{poclin} is slightly higher than that of 'clusterpath\_light' for $n \leq 20000$.
However, we note that the implementation in \texttt{poclin} only uses \texttt{R} code, while 'clusterpath\_light' only uses \texttt{C} code:
we expect that a \texttt{C} implementation of \texttt{poclin} would lead to improved computational times.
More interestingly, a linear regression in the log/log space showed that the slope of the \texttt{poclin} curve is approximately 1.5, while that of 'clusterpath\_light' and 'clusterpath' are approximately 1.9.
This implies that the empirical complexity of \texttt{poclin} is of the order of $\mathcal{O}(n^{1.5})$, while that of \texttt{clusterpath} is of the order of $\mathcal{O}(n^{1.9})$.
%
% note: 1 run of poclin with n=10^5 => 510 seconds (memory usage negligible)

\subsection{Further reducing the time complexity of Algorithm~\ref{alg:regularization:path}}
\label{sec:binary-heaps}
The complexity of Algorithm~\ref{alg:regularization:path} can be reduced to the order $\mathcal{O}( n \log(n))$ without compromising the linear memory complexity.
This section gives an informal description of the main idea for this reduction.
We consider the pairs of consecutive clusters, associated to consecutive values of $\hat{b}$ in  Algorithm~\ref{alg:regularization:path}.
Let us define the ``merging distance'' of each of these pairs as the value of $\lambda$ for which the corresponding values of $\hat{b}$ become equal, that is, where this pair of clusters should be merged into one.
If two clusters are merged, the merging distances are updated \textit{only for these two clusters and the one or two adjacent ones}.
This property could be exploited in the implementation of Algorithm \ref{alg:regularization:path}, by storing these merging distances in a min heap binary tree~\cite{williams1964algorithm}. 
Indeed, the minimal element of a min heap (here, corresponding to the next merge), is obtained in constant time ($\mathcal{O}(1)$) as the root of the tree, while the cost of inserting an element in the heap is logarithmic ($\mathcal{O}(\log(n))$), corresponding to the depth of the binary heap.
Exploiting the  binary min-heap tree, we can keep a $\mathcal{O}(\log(n))$ cost at each step when two clusters are merged.
This yields a total computational complexity of $\mathcal{O}(n \log(n))$ for $\mathcal{O}(n)$ steps.

Note also that if more than two clusters are merged, then the computational cost of the corresponding step can be higher, but the total number of steps is more reduced. 
We eschew a full description of an implementation of Algorithm \ref{alg:regularization:path} with a binary min-heap tree for the sake of concision and to promote explicit formulas such as \eqref{eq:hat:b:algo} and \eqref{eq:lambda:equal:inf}.

\section{Additional illustrations}
\label{apx:addit-illustr}

\subsection{Calibration of the regularization parameter}
\label{sec:calibration-lambda}
We describe the procedure used in the numerical experiments to calibrate the value of the regularization parameter $\lambda$.
As explained in the main text, the goal of this procedure is to ensure that with high probability, the one-dimensional convex clustering finds at least two clusters under the null hypothesis.

\begin{figure}
	\includegraphics[width=0.8\textwidth]{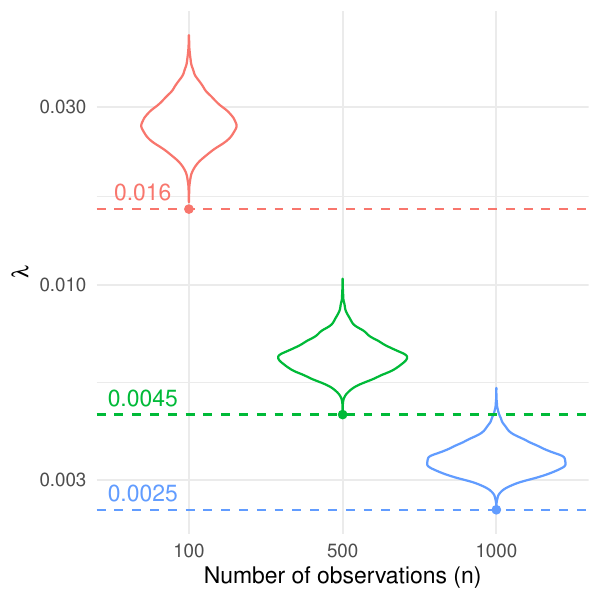}
	\caption{Empirical distribution of $\lambda^{(r_{\max})}$ across $10000$ replications of the procedure to choose $\lambda$ (see main text for details). The dots correspond to the chosen values.}
	\label{fig:lambda-vs-n}
\end{figure}

We generate $B=10000$ replicated null datasets $\Z_b \sim \mathcal{N}(\bzero_n, \bSigma), b \in [|B|]$. For each of these datasets, we calculate $\lambda^{(r_{\max})}_b$, the smallest value of the regularization parameter $\lambda$ for which the convex clustering yields a single cluster.
For a given input dataset, this value is obtained analytically in $\mathcal{O}(n \log(n))$ time using~\eqref{eq:lambda-max}.
Finally, we set $\lambda$ to $q_{0.01}(\boldsymbol{\lambda})-\mathrm{sd}(\boldsymbol{\lambda})$, where $\boldsymbol{\lambda} = (\lambda^{(r_{\max})}_b)_{b \in [|B|]}$, $q_{0.01}(\bz)$ is the first percentile of the vector $\bz$ and $\mathrm{sd}(\bz)$ is the standard deviation of the vector $\bz$.
The result of this procedure for $\bSigma = \bI_n$ is illustrated by Figure~\ref{fig:lambda-vs-n}.
For $n \in \{100, 500, 1000\}$, the empirical distribution of $\lambda^{(r_{\max})}$ is summarized by a violin plot (mirrored kernel density estimate) and the obtained value of $\lambda$ is represented by a dot and a dashed horizontal line.

\subsection{Level of the test in the one-dimensional case}
\label{apx:numerical-experiments-1d-level}
\begin{figure}[htp!]
	\includegraphics[width=0.7\textwidth]{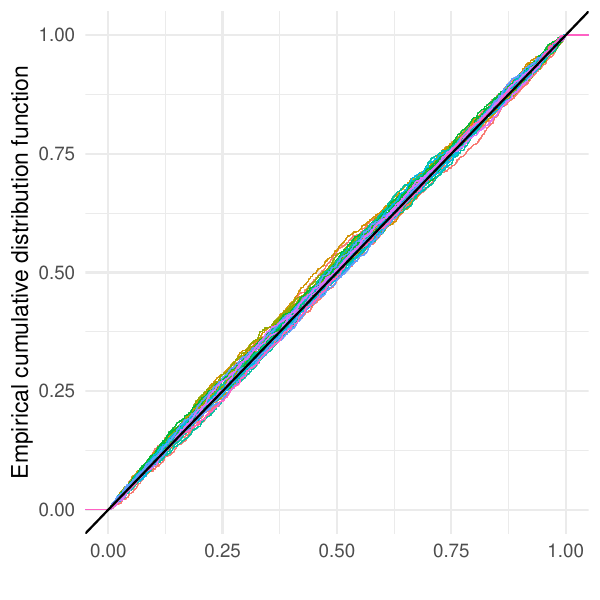}
	\caption{Experiments under the null hypothesis: empirical cumulative distribution functions of the $p$-value of the test of equality of means of all pairs of clusters. Each curve corresponds to a specific pair of clusters.}
	\label{fig:empirical-cdf-H0}
\end{figure}

We set a Gaussian sample $\X=(X_1,\ldots,X_n)$ with mean vector $\bmu= \bzero_n$ and known covariance matrix $\bSigma=\bI_n$.
For the given fixed value $\lambda = 0.0025$, we use our test in order to compare the means of cluster $\C_k$ and $\C_{k'}$ for $1 \leq k < k' \leq K_0=10$.
This corresponds to the test of the $K_0(K_0-1)/2$ null hypotheses $\boeta^{[kk']^\top} \bmu = 0$, where $\boeta^{[kk']} \in \R^n$ is defined by $\boeta^{[kk']}_i =  1/n_k\1_{i \in \C_k}  -  1/n_{k'}\1_{i \in \C_{k'}}$.
We retain $N=1000$ numerical experiments such that the clustering $\C(\X)$ associated to $\lambda$ verifies $K(\X) \geq K_0$.
The result is summarized in Figure \ref{fig:empirical-cdf-H0} by the empirical cumulative distribution of the conditional $p$-value \eqref{eq:cond-p-value-1d} for each pair of clusters.
Figure~\ref{fig:empirical-cdf-H0} illustrates the uniformity of the distribution of each of these $p$-values, for $n=1000$.

\subsection{The $p$-dimensional case}
\label{apx:p:dimensional case}

\begin{figure}[htp!]
	\includegraphics[width=0.9\textwidth]{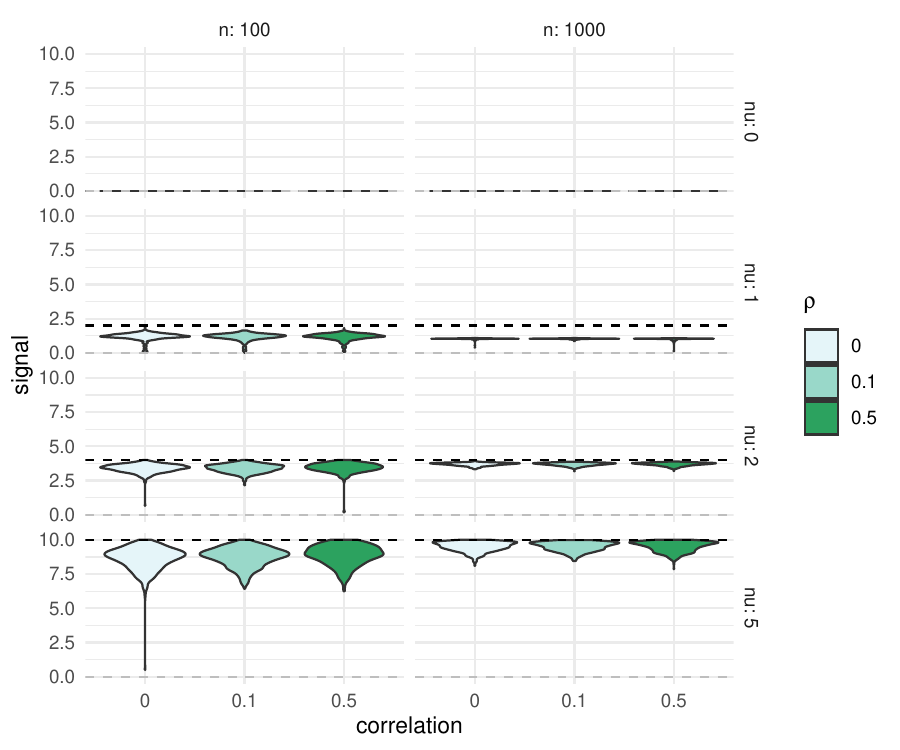}
	\caption{Empirical distribution (across 500 experiments) of the absolute value of the difference between the true means of the estimated clusters. Dashed horizontal lines are drawn for reference at $y=2\nu$ (in black) and $y=0$ (in gray).}
	\label{fig:signal-scenario1}
\end{figure}

In Figure~\ref{fig:signal-scenario1} we plot the empirical distribution (across 500 numerical experiments described in Section~\ref{sec:numer-exper-p}) of the absolute value of the difference between the true means of the estimated clusters for $n=100$ and $n=1000$, for the variable $\Y_{.1}$.
By construction, under this simulation scenario, this quantity is bounded by $2 \nu$, as indicated by a dashed horizontal line.
The other variables, $\Y_{.2}$ and $\Y_{.3}$, are not displayed because they do not carry any signal. 
This indicates that the convex clustering procedure works reasonably well: indeed, the difference between the true means of the estimated clusters is close to the difference between the true means of the true clusters.
Moreover, the variability is greater for $n=100$ than $n=1000$, consistently with the available information to solve the clustering problem.

\subsection{On the choice of the $p$-value \eqref{eq:cond-p-value-1d}}

The $p$-value in \eqref{eq:cond-p-value-1d} is
\begin{equation*}
	\pval( \x, \t,\sigma)  = 
 \mathbb{P}_{W} \left( 
| W |
\ge 
| \boeta^\top \x |
 \right),
\end{equation*}
and we can see that the $p$-value is a decreasing function of $| \boeta^\top \x |$. Hence, for $\alpha \in (0,1)$, the set of $\x$ that lead to $	\pval( \x, \t,\sigma) \le \alpha$ (that is to a rejection of the null hypothesis for the level $\alpha$) is of the form $\{ \x ; | \boeta^\top \x | \ge \mathrm{threshold} \}$ for a real  $\mathrm{threshold}$ that depends on the cdf of $W$ in \eqref{eq:cond-p-value-1d} (but not on $\x$). Hence the rejection region is composed of large values of $| \boeta^\top \x |$ which is natural when the null hypothesis is $\boeta^\top \bmu  = 0$ and we have in mind a symmetric alternative $\boeta^\top \bmu  \neq 0$. 

An alternative $p$-value construction could be to let
\begin{equation} \label{eq:T}
	T( \x, \t,\sigma):=F_{0, \boeta^\top \bSigma \boeta}^{[\V^{-}(\z_0), \V^+(\z_0)]} (\boeta^\top \x), 
\end{equation}
recalling that
$F_{0, \boeta^\top \bSigma \boeta}^{[\V^{-}(\z_0), \V^+(\z_0)]}$ is the cdf of $W$ in \eqref{eq:cond-p-value-1d}, and then to define the    $p$-value as
\begin{equation} \label{eq:tildepval}
\widetilde{\pval}( \x, \t,\sigma)  = 2 \min \left[T( \x, \t,\sigma) ,1-T( \x, \t,\sigma)\right].
\end{equation}
Since $W$ has cdf $F_{0, \boeta^\top \bSigma \boeta}^{[\V^{-}(\z_0), \V^+(\z_0)]}$, one can see that $\widetilde{\pval}( \X, \t,\sigma) $ is uniformly distributed on $[0,1]$ under the null hypothesis, similarly as $	\pval( \X, \t,\sigma)$. Note that the constructions \eqref{eq:T} and \eqref{eq:tildepval}  are made in \cite[Theorems 5.2 and 6.1]{Lee16}.   
For $\alpha \in (0,1)$, the set of $\x$ such that $\widetilde{\pval}( \x, \t,\sigma) \le \alpha$ is not necessarily of the form $\{ \x ; | \boeta^\top \x | \ge \mathrm{threshold} \}$. For instance, if $- \infty < \V^{-}(\z_0) < \V^+(\z_0) < 0$, then this set is composed of two disjoint segments, both included in $(- \infty , 0)$. Hence, for this alternative $p$-value $\widetilde{\pval}( \x, \t,\sigma)$, the rejection region can be considered as less natural than for $\pval( \x, \t,\sigma)$.

\begin{figure}[htp!]
	\includegraphics[width=10cm]{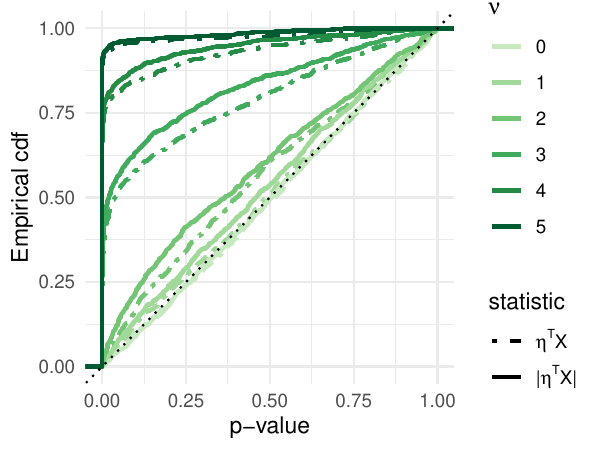}
	\caption{Empirical cumulative distribution functions of the two $p$-values defined of the test of equality between the means of two clusters. The statistic $|\boeta^T \X|$ corresponds to $\pval( \x, \t,\sigma)$ and $\boeta^T \X$ to $\widetilde{\pval}( \x, \t,\sigma)$.}
	\label{fig:num-exp-1d-H1-compare-stats}
\end{figure}

We have performed a numerical comparison of these two alternative definitions for the $p$-value, in the exact same settings as Section~\ref{sec:numerical-experiments-1d}. The results are summarized in Figure~\ref{fig:num-exp-1d-H1-compare-stats}, where we compare the empirical cdf of  $\pval( \X, \t,\sigma)$ and $\widetilde{\pval}( \X, \t,\sigma)$. We conclude that the $p$-value  $\pval( \x, \t,\sigma)$ yields more powerful tests than $\widetilde{\pval}( \x, \t,\sigma)$.

~ \\

\subsection{Comparison to generalized Lasso}
\label{sec:comp-gen-lasso}
\begin{figure}[htp!]
	\includegraphics[width=16cm]{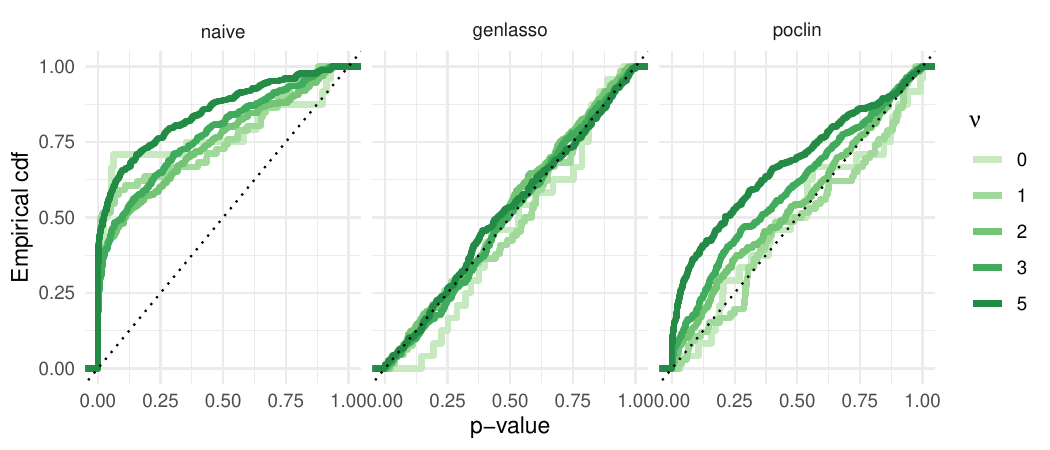}
	\caption{Empirical cumulative distribution functions of the $p$-values for three methods (in columns): the naive method (ignoring the randomness induced by the clustering step), the generalized Lasso method of \cite{hyun2018exact} (genlasso), and our proposed method as implemented in the \texttt{poclin} package.}
	\label{fig:compare-genlasso}
\end{figure}

%As discussed in Section \ref{subsection:regularization:path:one-dimensional} and Appendix \ref{sec:complexity}, it is not possible to solve the convex clustering problem using a generic implementation of the generalized lasso when the number of observations is large. 
%
In this section, we consider the same simulation setting as in Section \ref{sec:numerical-experiments-1d}, but set $n=40$. 
The generalized Lasso-based approaches proposed by \cite{chen2023more,hyun2018exact} are computationally demanding not only to solve the convex clustering problem (as discussed in Section \ref{subsection:regularization:path:one-dimensional} and illustrated in Section \ref{sec:complexity}), but also for the computation of the conditional $p$-value after convex clustering. 
In fact, we were not able to get the result of a single run of the procedure of \cite{chen2023more} for $n=40$, because the algorithm was still running after 5 days of computation. Hence it appears that this procedure is not suited to dense constraint matrices for the generalized Lasso.
Moreover, while the approach introduced by \cite{le2022more} is in theory applicable to any generalized Lasso problem, to the best of our knowledge it has only been implemented for the Lasso and fused Lasso problems\footnote{See \url{https://github.com/vonguyenleduy/parametric_generalized_lasso_selective_inference}.}. Therefore, we have only been able to compare our method to the one introduced by \cite{hyun2018exact}\footnote{We have adapted the code available in the \texttt{GFLassoInference} package, which provides an implementation of \cite{hyun2018exact}.}. We have performed $1000$ experiments for each value of the signal parameter $\nu \in \{0, 1, 2, 3, 5\}$, for a fixed value of the regularization parameter $\lambda=0.05$ of the convex clustering problem. For each of these runs, the inference procedure of \cite{hyun2018exact} took approximately 100 seconds, while our procedure returned the results instantly.
The results are summarized in Figure \ref{fig:compare-genlasso}. 
These results demonstrate that the generalized Lasso-based approach is not powerful enough to detect the signal, while the proposed approach is well-calibrated under the null hypothesis ($\nu=0$) and increasingly powerful for larger values of $\nu$. Qualitatively similar conclusions were obtained for other fixed values of $\lambda$.

~ \\

\paragraph{\textbf{Acknowledgements}}
The authors are grateful to two anonymous referees for their constructive comments that lead to an improvement of the paper.

%%-----------------------------
%%      your bibliography
%%-----------------------------

\bibliographystyle{abbrv}
\bibliography{biblio}

\end{document}